\documentclass[aps,rmp,reprint,longbibliography]{revtex4-2}

\usepackage{amsmath, amsthm, amsfonts} 
\usepackage{amssymb}
\usepackage{bm}          
\usepackage{xcolor}

\usepackage[defaultcolor=orange]{changes} 

\usepackage{csquotes}

\usepackage{soul} 
\usepackage{mathtools} 
\usepackage{tikz} 
\usepackage{cancel} 
\pdfoutput=1

\usepackage[unicode=true,pdfusetitle,
 bookmarks=true,bookmarksnumbered=false,bookmarksopen=false,
 breaklinks=true,pdfborder={0 0 0},backref=false,colorlinks=true,citecolor=blue]{hyperref}

\usepackage{graphicx}   
\usepackage[caption=false]{subfig}       
\usepackage{etoolbox} 
\usepackage{svg} 
\usepackage[nodisplayskipstretch]{setspace}
\usepackage{float}
\usepackage{epstopdf}

\usepackage{array}

\usepackage{calc}
\usepackage{tikz}
\usepackage{xspace}
\usepackage{upgreek}

\makeatother
\def\be{\begin{eqnarray}}
\def\ee{\end{eqnarray}}
\def\R{{\bf R}}
\def\r{{\bf r}}
\def\q{{\bf q}}

\def\E{{\bf E}}
\def\Eh{{\bf E}_\text{h}}
\def\Eb{{\bf E}_\text{b}}

\def\j{{\bf j}}

\def\P{{\bf P}}

\def\k{{\bf k}}
\def\u{{\bf u}}
\def\e{{\bf e}}

\def\T{{\bf T}}
\def\im{{\rm i}}

\def\G{{{\bf G}}}
\def\gh{\bm{\it g}_\text{h}}
\def\gb{\bm{\it g}_\text{b}}
\def\Gcalb{{\bm{\mathcal{G}}}_\text{b}}
\def\Gcalh{{\bm{\mathcal{G}}}_\text{h}}

\def\TGcalb{\tilde{\bm{\mathcal{G}}}_\text{b}}
\def\TGcalh{\tilde{\bm{\mathcal{G}}}_\text{h}}

\def\Hcal{\hat{\bm{\mathcal{G}}}_\text{b}}
\def\Tgcal{{\bm{\mathcal{T}}}}
\def\Xgcal{{\bm{\mathcal{X}}}}
\def\Sigmaop{\bm{\mathit{\Sigma}}}

\renewcommand{\bm}[1]{\boldsymbol{\mathbf{#1}}}
\newcommand{\ud}{\mathrm{d}}
\newcommand{\bra}{\left\langle}
\newcommand{\ket}{\right\rangle}

\newcommand{\imx}{\operatorname{Im}}
\newcommand{\rex}{\operatorname{Re}}

\newcommand{\Gammaop}{\bm{\mathit{\Gamma}}}

\newcommand{\Gg}{{\bf G}}

\newcommand{\Egav}{\langle {\bf E} \rangle}

\newcommand{\Ig}{\bm{1}}

\newcommand{\Tg}{{\bf T}}
\newcommand{\rg}{{\bf r}}

\newcommand{\Rg}{{\bf R}}

\newcommand{\sinc}{{\mathrm{sinc}}}

\newcommand{\dd}{\mathrm{d}}

\newcommand{\eps}{\epsilon}

\newcommand{\lcor}{\ell_\text{c}}

\begin{document}

\title{Light in correlated disordered media}

\author{Kevin Vynck}
\email{kevin.vynck@univ-lyon1.fr}
\affiliation{Univ. Bordeaux, Institut d'Optique Graduate School, CNRS, Laboratoire Photonique Num\'erique et Nanosciences (LP2N), F-33400 Talence, France}
\affiliation{Univ. Claude Bernard Lyon 1, CNRS, Institut Lumi\`ere Mati\`ere (iLM), F-69622 Villeurbanne, France}

\author{Romain Pierrat}
\affiliation{ESPCI Paris, PSL University, CNRS, Institut Langevin, 
F-75005 Paris, France}

\author{R\'emi Carminati}
\email{remi.carminati@espci.psl.eu}
\affiliation{ESPCI Paris, PSL University, CNRS, Institut Langevin, 
F-75005 Paris, France}

\author{Luis S. Froufe-P\'erez}
\affiliation{Physics Department, University of Fribourg, 
CH-1700 Fribourg, Switzerland}

\author{Frank Scheffold}
\email{frank.scheffold@unifr.ch}
\affiliation{Physics Department, University of Fribourg, 
CH-1700 Fribourg, Switzerland}

\author{Riccardo Sapienza}
\affiliation{Imperial College London, Blackett Laboratory, 
London, SW7 2AZ, United Kingdom}

\author{Silvia Vignolini}
\affiliation{Department of Chemistry, University of Cambridge, Lensfield Road, Cambridge, CB2 1EW, United Kingdom}

\author{Juan Jos\'e S\'aenz}
\affiliation{Donostia International Physics Center (DIPC), 
20018 Donostia-San Sebastian, Spain}
\affiliation{IKERBASQUE, Basque Foundation for Science, 48013 Bilbao, Spain}





\begin{abstract}
The optics of correlated disordered media is a fascinating research topic emerging at the interface between the physics of waves in complex media and nanophotonics. Inspired by photonic structures in nature and enabled by advances in nanofabrication processes, recent investigations have unveiled how the design of structural correlations down to the subwavelength scale could be exploited to control the scattering, transport and localization of light in matter. From optical transparency to superdiffusive light transport to photonic gaps, the optics of correlated disordered media challenges our physical intuition and offers new perspectives for applications.
This article reviews the theoretical foundations, state-of-the-art experimental techniques and major achievements in the study of light interaction with correlated disorder, covering a wide range of systems -- from short-range correlated photonic liquids, to L\'evy glasses containing fractal heterogeneities, to hyperuniform disordered photonic materials. The mechanisms underlying light scattering and transport phenomena are elucidated on the basis of rigorous theoretical arguments. We overview the exciting ongoing research on mesoscopic phenomena, such as transport phase transitions and speckle statistics, and the current development of disorder engineering for applications such as light-energy management and visual appearance design. Special efforts are finally made to identify the main theoretical and experimental challenges to address in the near future.

\end{abstract}
\date{\today}


\maketitle

\tableofcontents{}


\section{Introduction} \label{sec:1}

Correlated disordered media are non-crystalline heterogeneous materials exhibiting pronounced spatial correlations in their structure and morphology. The topic has bloomed in the context of optics and photonics, gradually unveiling the considerable impact of structural correlations on the scattering, transport, and localization of light in matter.
In essence, correlations engender constructive and destructive interferences that survive configurational average. This leads not only to more pronounced spectral and angular features at the single scattering level, but also to profound modifications of the radiation properties of quantum emitters and the macroscopic diffusion of photons by intricate near-field and multiple wave scattering phenomena.
Recent findings let us envision novel types of materials with unprecedented optical functionalities and raise a number of challenges in theoretical modelling, material fabrication and optical spectroscopy. This article aims to provide an overview of this emerging research field, starting from the basic principles of light interaction with heterogeneous media to the most recent and still actively debated topics.

The scattering of light by heterogeneous media has a long and venerable history, which started more than a century ago with pioneering studies on the refractive index of fluids of atoms or molecules~\cite{lorenz1880ueber, lorentz1880ueber}, the electromagnetic scattering by particles~\cite{rayleigh1899xxxiv, maxwellgarnett1904colours, mie1908beitrage}, and the phenomenon of critical opalescence in binary fluid mixtures~\cite{Smoluchowski1908molekular, einstein1910theorie, ornstein1914accidental}. The foundations of a rigorous theoretical treatment of multiple light scattering were built in the 1930s~\cite{kirkwood1936theory, yvon1937recherches} to take its full dimension a few decades later with various important contributions~\cite{foldy1945multiple, lax1951multiple, lax1952multiple, twersky1964propagation, keller1964stochastic}. These early works already pointed out the key role played by structural correlations on light scattering, a nice illustration of this being the transparency of the cornea resulting from short-range correlations in ensembles of discrete scatterers~\cite{maurice1957structure, hart1969light, benedek1971theory, twersky1975transparency}.

A new branch of research exploiting light waves to study mesoscopic phenomena in disordered systems emerged in the 1980s, prompted by experimental demonstrations of weak localization~\cite{tsang1984backscattering, van1985observation, wolf1985weak} and theoretical predictions for the three-dimensional Anderson localization of light~\cite{john1984electromagnetic, anderson1985question}. The advent of photonic crystals~\cite{yablonovitch1987inhibited, john1987strong}, wherein photonic band gaps are created by a periodic modulation of the refractive index in two or three dimensions, gave additional momentum to research by greatly stimulating the development of nanofabrication techniques for high-index dielectrics~\cite{lopez2003materials}. The following decade witnessed a flourishment of studies on periodic dielectric nanostructures~\cite{joannopoulos2011photonic} and disordered media made of resonant (Mie) scatterers~\cite{van1991speed, busch1995transport, lagendijk1996resonant} from two overlapping communities~\cite{soukoulis2012photonic}.

\begin{figure*}[htbp]
\centering
\includegraphics[width=0.9\textwidth]{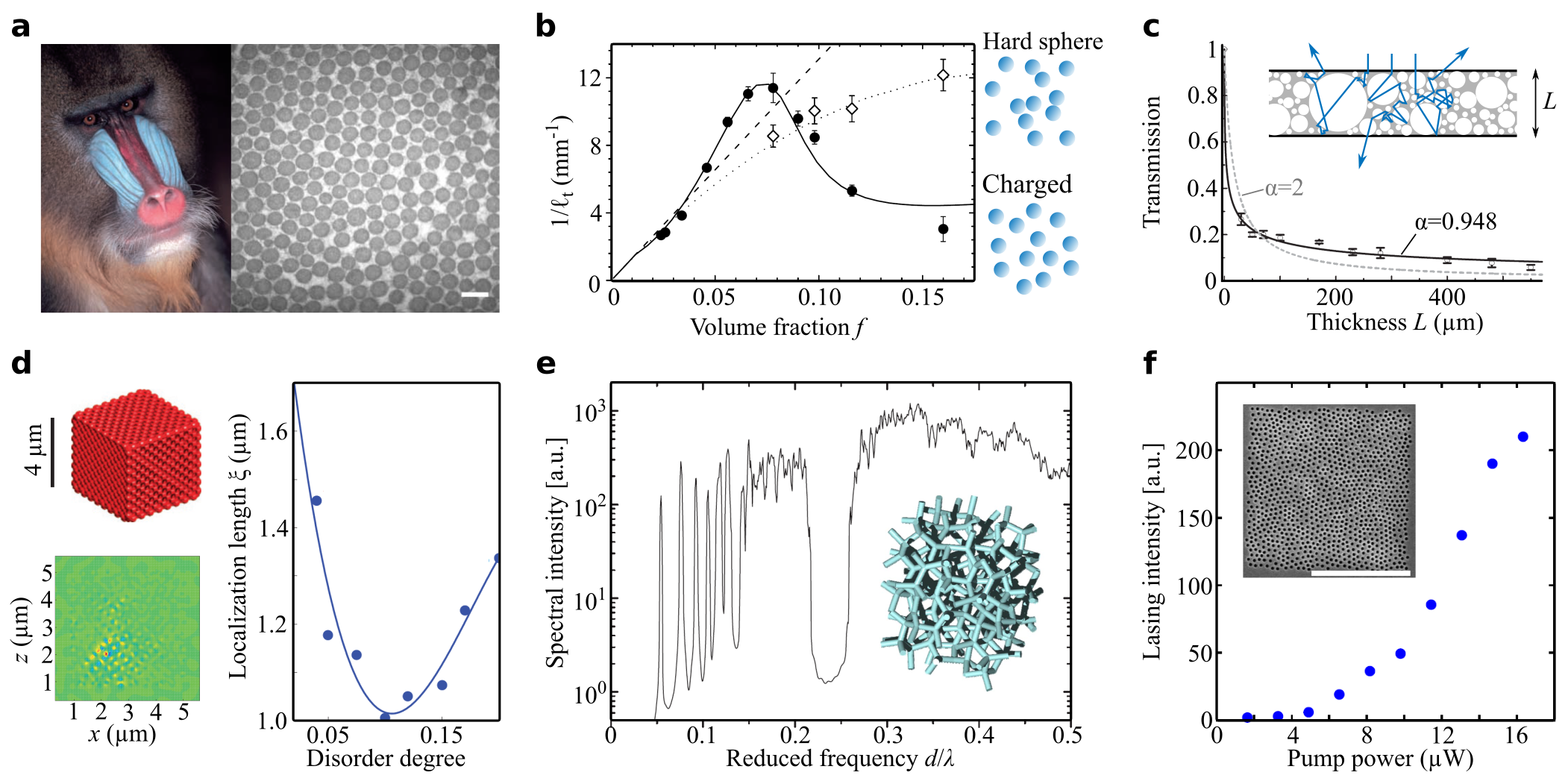}
\caption{\label{fig:Introduction} (Color online) Early achievements and applications of correlated disordered media in optics and photonics. (a) Male mandrill (Mandrillus sphinx) blue facial skin and cross-section of its dermis in the structurally colored area that reveals parallel collagen fibres organised in a correlated array. Adapted with permission from \cite{prum2004structural}. (b) Modified light transport (described by the inverse transport mean free path $\ell_\text{t}$) by engineering short-range structural correlations thanks to Coulomb repulsion between charged particles (filled symbols and solid line). Hard sphere systems (open symbols and dotted line) exhibit weaker correlations. The dashed line is a model neglecting completely structural correlations. Adapted with permission from \cite{rojas2004photonic}. (c) Anomalous light transport in L\'evy glasses. A fractal heterogeneity is engineered by adding transparent spheres with sizes varying over orders of magnitude in a host matrix. Transport can be modelled by a truncated L\'evy walk. On finite size samples, this leads to an anomalous scaling of the total transmittance $T \sim L^{\alpha/2}$ (experiments shown as symbols, lines are fits). Adapted with permission from \cite{barthelemy2008levy}. (d) Light localization in randomly perturbed inverse opal photonic crystals (upper left inset). Simulations reveal spatially-localized modes near the photonic band edge (lower left inset). Their typical spatial extent (the localization length $\xi$) depends strongly on the degree of disorder. Adapted with permission from \cite{conti2008dynamic}. (e) Existence of photonic gaps in amorphous photonic materials. Simulations of the spectral density - a quantity proportional to the density of states - in a connected amorphous diamond structure exhibiting short-range order shows a photonic gap near $d/\lambda \simeq 0.23$, where $d$ is the average bond length. Adapted with permission from \cite{edagawa2008photonic}. (f) Random lasing in two-dimensional photonic structures with correlated disorder. Short-range correlations are shown to increase the lasing efficiency at certain frequencies due to enhanced optical confinement. Adapted with permission from \cite{noh2011control}.}
\end{figure*}

The importance of short-range structural correlations on light transport in disordered systems~\cite{fraden1990multiple, saulnier1990scatterer} and of random imperfections on light propagation in periodic systems~\cite{sigalas1996localization, asatryan1999effects, vlasov2000manifestation} was recognized quite early. Research on correlated disordered media in optics however really took off in the mid-2000s with experimental studies showing that disorder could be \textit{engineered} to harness light transport~\cite{rojas2004photonic, garcia2007photonic, barthelemy2008levy}. The surprising observation of photonic gaps in disordered structures with short-range correlations~\cite{edagawa2008photonic, liew2011photonic}, reports of mesoscopic phenomena in imperfect photonic crystals~\cite{conti2008dynamic, toninelli2008exceptional, garcia2012nonuniversal} and the prospects of new generations of photonic devices like random lasers~\cite{gottardo2008resonance}, thin-film solar cells~\cite{vynck2012photon, oskooi2012partially, martins2013deterministic} and integrated spectrometers~\cite{redding2013compact}, greatly contributed to the emergence of the research field. Figure~\ref{fig:Introduction} presents some of the early achievements and applications of correlated disordered media in optics and photonics. 

Important efforts have been made in recent years to elucidate the role of structural correlations on the emergence of photonic gaps and Anderson localization of light in two-dimensional~\cite{conley2014light, froufe2016role, froufe2017band, monsarrat2022pseudogap} and three-dimensional disordered systems~\cite{ricouvier2019foam, klatt2019phoamtonic, aubry2020experimental, haberko2020transition, scheffold2022transport}. Near-field interaction and light polarization considerably complicate theoretical modelling~\cite{cherroret2016induced, vynck2016multiple, vantiggelen2021longitudinal}, explaining the widespread use of full-wave numerical methods to address this issue, alongside phenomenological models~\cite{naraghi2015near}. The so-called hyperuniform disordered structures~\cite{torquato2003local}, introduced in photonics by \citet{florescu2009designer}, have received considerable attention in this context, leading to a wider exploration of their optical properties~\cite{leseur2016high, froufe2017band, bigourdan2019enhanced, gorsky2019engineered, sheremet2020absorption, rohfritsch2020impact, torquato2021nonlocal, piechulla2021tailored} and advances on top-down and bottom-up fabrication techniques~\cite{man2013isotropic, weijs2015emergent, muller2017photonic, ricouvier2017optimizing, maimouni2020micrometric, chehadi2021scalable, piechulla2022toward}.

In a different context, the interplay of order and disorder appeared quite early as an essential ingredient to explain the colored appearance of certain plants and animals~\cite{kinoshita2005structural}. Research on natural photonic structures continued at a fast pace with important findings, such as the ubiquity of correlated disorder in animals exhibiting vivid diffuse blue colors~\cite{noh2010noniridescent, yin2012amorphous, magkiriadou2012disordered, johansen2017photonics, moyroud2017disorder}, the use of short-range correlations to reduce light reflectance~\cite{deparis2009assessment, siddique2015role, pomerantz2021developmental} or structural anisotropy to enhance whiteness~\cite{burresi2014bright}. Efforts are nowadays made to realize artificial materials exhibiting correlated disorder to create materials with versatile visual appearances~\cite{forster2010biomimetic, takeoka2012angle, park2014full, shang2018photonic, schertel2019structural, chan2019visual, goerlitzer2018bioinspired, salameh2020origin, jacucci2021light}.

In this review, we will introduce the key concepts and techniques in the study of light in correlated disordered media, assess the current state of knowledge on the topic, and define the main challenges that lie ahead of us. Compared to existing reviews on correlated disorder and disorder engineering in optics and photonics~\cite{wiersma2013disordered, shi2013amorphous, yu2020engineered, wang2020dependent, cao2022harnessing}, we provide here a broader view on the field and sufficient technical details for the readers who wish to dive into it, be it from the theoretical or experimental side. This article is also an attempt to bridge the gap between different research fields for which excellent textbooks already exist, namely on random heterogeneous materials~\cite{torquato2013random}, multiple light scattering in complex media~\cite{tsang2001scattering, sheng2006introduction, akkermans2007mesoscopic, carminati2021principles} and periodic photonic crystals~\cite{joannopoulos2011photonic}, and which may serve as complementary literature. We focus on two and three-dimensional dielectric materials, intentionally leaving aside one-dimensional dielectric structures (i.e., layered media)~\cite{izrailev2012anomalous} and metallic nanostructures~\cite{shalaev2002optical}. 
Quasicrystalline media, which are non-periodic, yet deterministic structures, are not discussed explicitly here, despite many conceptual overlaps discussed at length, for instance, by \citet{dalnegro2022waves}.
We also do not discuss the very fertile fields of metamaterials and metasurfaces, which show some apparent similarities with the present topic in terms of theoretical models and concepts~\cite{mackay2020modern}, yet with different scopes of application. Finally, certain concepts discussed here relate to transport theory in correlated, \textit{stochastic} media, where spatial correlations take place on scales larger than the wavelength and do not give rise to interferences. This article only covers a tiny portion of the vast literature on the topic, which has been instead meticulously reviewed by~\citet{deon2022hitchhiker}.

The remainder of the article is structured as follows. Section~\ref{sec:2} introduces the basic concepts and important quantities for light scattering and transport in correlated disordered media, namely the extinction, scattering and transport mean free paths. We derive mathematically explicit results, as a function of the degree of structural correlations, from rigorous multiple scattering theories for both continuous permittivity media and discrete particulate media, emphasizing conceptual similarities between these two viewpoints. Section~\ref{sec:3} addresses the statistical description of the structural properties of correlated disordered media. Different classes of correlated systems are discussed, together with numerical and experimental techniques to realize and characterize them. Section~\ref{sec:5} reviews experimental and theoretical studies wherein structural correlations yield substantial variations of light transport parameters, including enhanced scattering in colloidal suspensions of particles, optical transparency in hyperuniform media and anomalous diffusion in materials with large-scale (fractal) heterogeneities. Section~\ref{sec:6} is concerned with emergent mesoscopic phenomena relying on an interplay of order and disorder, most of which are not yet fully understood. This includes the formation of photonic gaps and localized states in disordered systems, and the statistical properties of near-field speckles and local density of states. Section~\ref{sec:7} describes various applications of correlated disordered media in optics and photonics, namely light trapping for enhanced absorption, random lasing and visual appearance design. Section~\ref{sec:8} concludes the review with a discussion on some open challenges in the field.

\section{Theory of multiple light scattering by correlated disordered media} \label{sec:2}

The theoretical study of light propagation in disordered media is a notoriously difficult problem that has experienced many developments for more than a century. In this section, we introduce the basic concepts of multiple light scattering by heterogeneities with the aim to give a solid theoretical ground to the role of structural correlations in light scattering and transport.

In Sec.~\ref{sec:theory_basics}, we first focus on the ``constitutive'' linear relation between the average electric field and the average polarization density in disordered media, which allows us to introduce the concepts of effective permittivity tensor and extinction mean free path. Many of the derived results have been used in the study of the effective optical response and homogeneization processes of periodic and amorphous dielectrics~\cite{van1977v, mackay2020modern}. We derive the main equations that govern the propagation of the average intensity and introduce the scattering and transport mean free paths, two experimentally measurable quantities that form the backbone of radiative transfer theory~\cite{chandrasekhar1960radiative}.

\begin{figure}[htbp]
\centering
\includegraphics[width=8cm]{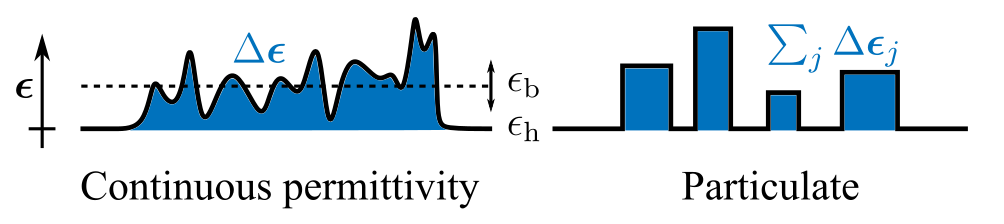}
\caption{\label{fig:continuous_particulate} (Color online) Disordered media may be described by a continuous permittivity model (left) in the most general case, or by a particulate model (right) in the case where the permittivity variation is compact.}
\end{figure}

Within this unique theoretical framework, we then address the light scattering problem for non-absorbing media described by either a continuous permittivity that fluctuates in space [Sec.~\ref{subsec:random}] or discrete particles correlated in their position [Sec.~\ref{subsec:particulate}], see Fig.~\ref{fig:continuous_particulate}. We derive analytical expressions for the characteristic lengths, allowing us to show, on rigorous grounds, how structural correlations impact scattering and transport. Interestingly, we find that the choice of a specific effective medium model does not affect the form of the expressions, thereby demonstrating their generality.

The main outcomes of the theoretical analysis are summarized in Sec.~\ref{subsec:theory-summary} for the readers who prefer to skip the mathematical details. Table~\ref{tab:mfp_expressions} in this section provides the final expressions for the scattering and transport lengths (mean free paths), to be used in practical situations.

\subsection{General framework} \label{sec:theory_basics}

\subsubsection{Average field and self-energy}

We consider a region of space filled with a non-magnetic, isotropic material (relative permeability $\mu(\mathbf{r}) = 1$), described by a scalar spatially-varying relative permittivity $\epsilon(\r)$ in a uniform \textit{host} medium with relative permittivity $\epsilon_\text{h}$. Throughout this review, we consider harmonic fields at frequency $\omega$ with the $e^{-\im \omega t}$ convention and drop the explicit dependence on $\omega$ in the permittivities, fields, etc. In absence of charges and currents, the electric field $\E(\r)$ at frequency $\omega$ satisfies the propagation equation
\be
\bm{\nabla} \times \bm{\nabla} \times \E(\r) - k_0^2 \epsilon_\text{h} \E(\r) =  k_0^2 \P(\r)/\epsilon_0,\label{waveequation} 
\ee
where $k_0 =\omega/c$ is the vacuum wave number and $\P(\r) = \epsilon_0 (\epsilon(\r)-\epsilon_\text{h}) \E(\r)$ is the polarization density (electric dipole moment per unit volume). The permittivity variation $\Delta \epsilon(\r) = \epsilon(\r)-\epsilon_\text{h}$ readily appears as the source of light scattering in the system.

Upon statistical average over an ensemble of realizations of disorder, denoted here as $\langle \dots \rangle$, Eq.~\eqref{waveequation} describes the propagation of the average field $\langle \E(\r) \rangle$ with the average polarization density $\langle \P(\r) \rangle$ as a source term. The difficulty to solve this general problem essentially comes from the fact that the permittivity variation and the electric field are not statistically-independent, i.e., $\langle \Delta \epsilon(\r) \E(\r) \rangle \ne  \langle \Delta \epsilon(\r) \rangle \langle \E(\r)\rangle$. The standard approach to solve this problem is to make an \textit{ansatz} about the effective permittivity ${\bm{\epsilon}}_{\text{eff}}$ of the medium, thereby establishing a constitutive linear relation between the average field and the average polarization density, and then calculate it perturbatively from the spatial permittivity fluctuations.


Let us then rewrite Eq.~\eqref{waveequation} as~\cite{ryzhov1965spatial}
\be
	\bm{\nabla} \times \bm{\nabla} \times \E (\r) - k_0^2 \epsilon_\text{b} \E(\r) =  \Xgcal(\r) \E(\r), \label{waveequationb}
\ee
where $\epsilon_\text{b}$ is a constant, auxiliary, background permittivity that can differ from the permittivity of the host medium $\epsilon_\text{h}$, and $\Xgcal$ is an effective scattering potential (or electric susceptibility) defined as
\be
	\Xgcal(\r) = k_0^2 (\epsilon(\r) - \epsilon_\text{b}) \Ig, \label{eq:scatt_potential}
\ee
with $\Ig$ the unit tensor. The average field $\Egav$ then fulfills the vector wave equation
\be
	\bm{\nabla} \times \bm{\nabla} \times \langle \E(\r)\rangle - k_\text{b}^2 \langle \E(\r)\rangle = \langle \Xgcal(\r) \E(\r)\rangle, \label{waveequationav}
\ee
where $k_\text{b}^2 = k_0^2 \epsilon_\text{b}$. We now introduce the susceptibility tensor ${\bm{\Sigma}}$, commonly known as the ``self-energy'' or ``mass operator'' following the language of many-body scattering theory~\cite{dyson1949radiation}, as
\be
	\langle \Xgcal(\r) \E(\r)\rangle \equiv \int \bm{\Sigma}(\r,\r') \langle \E(\r')\rangle d\r', \label{self-energy}
\ee
with
\be
	\bm{\Sigma}(\r,\r') = k_0^2 \left({\bm{\epsilon}}_{\text{eff}}(\r,\r') - \epsilon_\text{b} \Ig \delta(\r-\r')\right). \label{eq:Sigma_rr}
\ee	
The self-energy depends on the non-local effective permittivity tensor that results from multiple scattering in the disordered medium. Hereafter, we assume that the system has a proper thermodynamic limit in which it becomes spatially homogeneous and translationally invariant on average (i.e., ${\bm{\epsilon}}_{\text{eff}}(\r,\r') = {\bm{\epsilon}}_{\text{eff}}(\r-\r')$). Aspects related to the effective medium description in large but finite systems and the deep connection with the Ewald-Oseen extinction theorem~\cite{hynne1987scattering,van1977v} will thus not be discussed here. In Fourier space, the self-energy is given by
\be
	\bm{\Sigma}(\k,\k') &=& \iint e^{-\im  \k \cdot \r} \bm{\Sigma}(\r-\r')  e^{\im  \k' \cdot \r'} d\r d\r' \\
	&=& (2\pi)^3 {\bm{\Sigma}}(\k) \delta(\k-\k'). \label{Sigma_k}
\ee
It is further convenient to decompose ${\bm{\Sigma}}(\k)$ into its transverse ($\perp$) and longitudinal ($\parallel$) components as
\be
	{\bm{\Sigma}}(\k) = {\Sigma}_{\perp}(\k) \left(\Ig -\u \otimes \u \right) + {\Sigma}_{\parallel}(\k) \u \otimes \u, \label{SigmaTransLong}
\ee
where $\u = \k/|\k|$ and $\u \otimes \u$ is the outer tensor product between $\u$ and itself. Defining $\e$ as the unit polarization vector with $\e \cdot \u = 0$, the transverse component of the self-energy reads
\be
    {\Sigma}_{\perp}(\k) = \e \cdot {\bm{\Sigma}}(\k) \e. \label{eq:sigma_perp}
\ee

\subsubsection{Refractive index and extinction mean free path}

To understand the role played by the self-energy in wave propagation and scattering, we can seek for transverse solutions of the vector wave propagation equation of the form
\be
 \langle \E(\r) \rangle &=& E_0 \e  e^{\im \k_{\text{eff}} \cdot \r}, \label{pwt}
\ee
with $ \k_{\text{eff}} = k_0 n_{\text{eff}} \u$ the wavevector describing propagation in a homogeneous medium with effective refractive index $n_\text{eff}$. Assuming a statistically isotropic system, substituting Eq.~\eqref{pwt} in Eq.~\eqref{waveequationav} and making use of Eqs.~\eqref{self-energy},\eqref{SigmaTransLong},\eqref{eq:sigma_perp} leads to a transcendental equation for the effective wave number
\be
	k_\text{eff} = k_0 n_{\text{eff}} &=& \sqrt{k_\text{b}^2 + {\Sigma}_{\perp}(k_{\text{eff}})}, \nonumber \\
	&\equiv& k_\text{r} + \im \frac{1}{2 \ell_\text{e}}. \label{keff}
\ee
The real part of the effective index $\text{Re} [n_\text{eff}] = k_\text{r}/k_0$ describes the phase velocity of the average field (often called ``coherent'' or ``ballistic'' component) in the material, while the imaginary part $\text{Im} [n_\text{eff}] = (2 k_0 \ell_\text{e})^{-1}$ describes its exponential decay with propagation due to absorption and/or scattering on a characteristic length scale that is the extinction mean free path, $\ell_\text{e}$. In the weak extinction regime (i.e., $\text{Im} {\Sigma}_{\perp}(k_{\text{eff}}) \ll k_\text{b}^2 + \text{Re} {\Sigma}_{\perp}(k_{\text{eff}})$ and $k_\text{r} \ell_\text{e} \gg 1$), Eq.~\eqref{keff} leads to
\be
	\frac{1}{\ell_\text{e}} \simeq \frac{\text{Im} \Sigma_{\perp}(k_{\text{r}})}{k_\text{r}}. \label{extmfp}
\ee
In non-absorbing dielectric materials, extinction is purely driven by scattering ($\ell_\text{e} = \ell_\text{s}$ with $\ell_\text{s}$ the scattering mean free path). The problem of light scattering by correlated disordered media can therefore be apprehended by determining the self-energy of the system.

\subsubsection{Multiple-scattering expansion}

A key ingredient in solving multiple light scattering problems is the electromagnetic Green tensor $\Gg_\text{b}(\r,\r')$, which is the solution of the wave equation in a homogeneous medium with permittivity $\epsilon_\text{b}$ [Eq.~\eqref{waveequationb}] with a point source,
\be
	\bm{\nabla} \times \bm{\nabla} \times \Gg_\text{b}(\r,\r') - k_\text{b}^2 \Gg_\text{b}(\r,\r') = \delta(\r-\r') \Ig.
\ee
Physically, it corresponds to the electric field produced at a point $\r$ by a radiating point electric dipole at $\r'$, and is given by
\begin{multline}
	\Gg_\text{b}(\r,\r') =  -\frac{\Ig}{3k_\text{b}^2} \delta(\r-\r') \\
	+ \lim_{a \rightarrow 0} \Theta(|\r-\r'|-a) \left\{ \left( \Ig +\frac{\bm{\nabla} \otimes \bm{\nabla} }{k_\text{b}^2} \right) \frac{e^{\im k_\text{b}|\r-\r'|}}{4 \pi |\r-\r'|} \right\} , \label{GFsing}
\end{multline}
where $\Theta$ is the Heaviside step function. The Dirac delta function in the right hand side gives the well-known singularity in the source region while the second term corresponds to the non-singular, {\em principal value} of the Green function~\cite{yaghjian1980electric, van1991singular}.
The exclusion volume defining the source region is chosen here to be spherical, but nonspherical (e.g., spheroidal, cubic, etc.) regions may also be used~\cite{yaghjian1980electric, tsang1981scattering, torquato2021nonlocal}. The choice of a nonspherical geometry can be particularly adapted to certain microstructures, for instance, with anisotropic correlation functions. Mathematically, the geometry of the source region affects the values of integrals involving the individual singular or non-singular contributions of the Green function.

The Green function enables writing a general solution of the wave equation in the form
%
\be
	\E(\r) = \Eb(\r)  +  \int { \Gg_\text{b}(\r,\r')  \Xgcal(\r') \E(\r')  d\r'}, \label{LS-integral}
\ee
where $\Eb(\r)$ is the solution of the homogeneous problem, which can be seen as a background (incident) field with wave number $k_\text{b}$. Equation~\eqref{LS-integral}, known as the Lippmann-Schwinger equation, can conveniently be written in operator form as
\be
	\E = \Eb + \Gcalb  \Xgcal \E, \label{LS}
\ee
where $\Gcalb$ is an integral operator. Equation~\eqref{LS} can be formally solved by successive iterations, leading to a multiple-scattering expansion on orders of $\Xgcal$ (i.e., single scattering, double scattering, etc.). Eventually, all multiple-scattering orders are taken into account by defining the transition operator $\Tgcal$ relating the polarization induced in the medium to the \textit{background} field, as
\be
	\E = \Eb + \Gcalb \Tgcal \Eb, \label{LS-T}
\ee
with
\be
	\Tgcal &=& \Xgcal + \Xgcal \Gcalb \Xgcal + \cdots \nonumber \\
	&=& \left[ \Ig - \Xgcal \Gcalb \right]^{-1} \Xgcal. \label{T-operator}
\ee
Keeping only the lowest order in the expansion, $\Tgcal = \Xgcal$, is known as the Born approximation, which corresponds to single scattering.
In the general case of multiple scattering, the transition operator is spatially non-local, $\Tgcal \E_\text{b} \equiv \int{\Tg(\r,\r') \Eb(\r') d\r'}$.

Upon statistical average of Eqs.~\eqref{LS} and \eqref{LS-T}, and having Eq.~\eqref{self-energy}, we finally reach a general expression for the average field as a function of the self-energy operator $\Sigmaop$, known as the Dyson equation~\cite{yvon1937recherches,dyson1949radiation,dyson1949s,rytov1989principles}
\be
	\Egav &=& \Eb + \Gcalb \Sigmaop \Egav, \label{eq:th_dyson}
\ee
with
\be
	\Sigmaop = \langle \Tgcal \rangle  \left[ \Ig +  \Gcalb  \langle \Tgcal \rangle \right]^{-1}. \label{Sigmaexact}
\ee

In summary, the disordered medium is described as a permittivity that fluctuates around an auxiliary background permittivity $\epsilon_\text{b}$ via the scattering potential $\Xgcal$ [Eq.~\eqref{eq:scatt_potential}]. The field propagates from fluctuation to fluctuation via the Green tensor $\Gcalb$ in the homogeneous background with wave number $k_\text{b}$ [Eq.~\eqref{GFsing}]. The multiple scattering process on the scattering potential $\Xgcal$ is described (to infinite order) via the transition operator $\Tgcal$ [Eq.~\eqref{T-operator}]. The average transition operator $\langle \Tgcal \rangle$ finally defines the self-energy $\Sigmaop$ [Eq.~\eqref{Sigmaexact}], which describes the propagation of the average field $\Egav$ in the disordered medium, and leads to the extinction mean free path in the medium [Eq.~\eqref{extmfp}].

\subsubsection{Average intensity and four-point irreducible vertex}

Light transport, that is the propagation of the energy, is formally described by the average intensity $\bra |\E(\r)|^2 \ket$, which we will now consider. First and foremost, let us remark that the average intensity can be decomposed into two components with distinct physical meaning. Indeed, writing the field as the sum of its average value and a fluctuating part, $\E = \bra \E \ket + \Delta \E $ with $\bra \Delta \E \ket = 0$ by definition, straighforwardly leads to 
\be
 \bra |\E(\r)|^2 \ket = |\bra \E (\r) \ket |^2 + \bra |\Delta \E(\r)|^2 \ket.
\ee
The first term, $|\bra \E \ket |^2$, corresponds to the so-called ballistic (or coherent) intensity, which describes the part of the intensity that propagates in the direction of the incident light and is attenuated (exponentially) by scattering and absorption. Its behavior is fully determined by the theory for the average field presented in the previous section.
Our attention here should be given instead to the second term, $\bra |\Delta \E |^2 \ket$, which corresponds to the so-called diffuse (or incoherent) intensity, describing the part of the intensity that spreads throughout the volume of the medium by successive scattering events. The diffuse intensity will lead to the definition of the scattering and transport mean free paths, two additional length scales at the heart of light propagation in disordered media~\cite{apresyan1996radiation, van1999multiple, rytov1989principles}.

Let us then consider the spatial correlation function of the electric field, or ``coherence matrix''~\cite{mandel1995optical}, $\bm{C}(\r,\r') \equiv \bra \E(\r) \otimes \E^*(\r')\ket$, with ${ }^*$ denoting complex conjugate. Starting from the Lippmann-Schwinger equation [Eq.~\eqref{LS}], one can easily show that $\bm{C}$ depends on the correlator of the polarization density in the effective scattering potential $\bra (\Xgcal(\r) \E(\r)) \otimes (\Xgcal^*(\r') \E^*(\r'))\ket$. Similarly to the self-energy that allowed us to relate the average polarization to the average field [Eq.~\eqref{self-energy}] eventually leading to the Dyson equation [Eq.~\eqref{eq:th_dyson}], we can introduce here an operator $\bm{\Gamma}$ known as the ``four-point irreducible vertex'' -- or intensity vertex -- that relates the effective polarization density correlation to the electric field correlation. This leads to a closed-form equation, known as the Bethe-Salpeter equation~\cite{salpeter1951relativistic}, which reads
\begin{multline}
   \bm{C}(\r,\r') = \bra \bm{E}(\bm{r}) \ket \otimes \bra\bm{E}^*(\r')\ket
      + \int \bra \bm{G}(\bm{r},\bm{r}_1)\ket \otimes \bra\bm{G}^*(\r',\r_1')\ket
   \\
      \cdot\bm{\Gamma}(\r_1,\r_2,\r_1',\r_2') \cdot\bm{C}(\r_2,\r_2')
      \ud\r_1 \ud\r_1' \ud\r_2 \ud\r_2', \label{eq:th_bethe_salpeter_full_vector}
\end{multline}
where the average Green function $\bra \bm{G} \ket$ is given by the Dyson equation [Eq.~\eqref{eq:th_dyson}],
\be
   \bra\bm{\mathcal{G}}\ket = \Gcalb + \Gcalb \Sigmaop \bra\bm{\mathcal{G}}\ket. \label{eq:av_Green_dyson}
\ee
The symbol $\cdot$ denotes here a tensor contraction, defined such that $(\bm{A}\otimes\bm{B})\cdot(\bm{e}\otimes\bm{f}) = (\bm{A}\bm{e})\otimes (\bm{B}\bm{f})$ and $(\bm{A}\otimes\bm{B})\cdot(\bm{C}\otimes\bm{D}) = (\bm{A}\bm{C})\otimes (\bm{B}\bm{D})$, where $\bm{e},\bm{f}$ are vectors and $\bm{A},\bm{B},\bm{C},\bm{D}$ second-rank tensors.

The first term in Eq.~\eqref{eq:th_bethe_salpeter_full_vector} is the correlation function on the average field that leads to the coherent intensity. The second term expresses the field correlation as a multiple scattering process, wherein the propagation is described by the average Green tensors and scattering by the vertex $\bm{\Gamma}$ that connects two pairs of points (for the field and the complex conjugate). Following similar steps as those leading to Eq.~\eqref{Sigmaexact}, we obtain the following general expression for $\Gammaop$~\cite{carminati2021principles}
\be
   \Gammaop &=& \left[ \Gcalb\Gcalb^*\right ]^{-1} \big[ \left( \Ig + \Gcalb \bra \Tgcal \ket + \Gcalb^* \bra \Tgcal^* \ket + \bra \Tgcal \ket \Gcalb \Gcalb^* \bra \Tgcal^* \ket \right)^{-1} \nonumber \\
   &-& \left( \Ig + \Gcalb \bra \Tgcal \ket + \Gcalb^* \bra \Tgcal^* \ket + \bra \Tgcal \Gcalb \Gcalb^* \Tgcal^* \ket \right)^{-1} \big]. \label{eq:gamma}
\ee

To proceed further, it is convenient to rewrite the Bethe-Salpeter equation in Fourier space. Assuming that the scattering events take place on distances larger than the wavelength, the average Green tensor can be approximated by its transverse component,
\be
   \bra \bm{G}(\k) \ket &=& \left[k^2\bm{P}(\u)-k_\text{b}^2\Ig-\bm{\Sigma}(\k)\right]^{-1}, \\
   & \simeq & \bra G_\perp(\k) \ket \bm{P}(\u),
\ee
where $\bm{P}(\u) = \Ig - \u \otimes \u$ is the transverse projection operator and $\bra G_\perp(\k) \ket = \left[k^2-k_\text{b}^2-\Sigma_\perp(\k)\right]^{-1}$ is the (scalar) transverse component. After some algebra provided in details in Appendix~\ref{App:BS-eq-Fourier}, we find that
\begin{widetext}
\begin{multline}
   \left[\left(\k-\frac{\q}{2}\right)^2-\left(\k+\frac{\q}{2}\right)^2
      -\Sigma_{\perp}^*\left(\k-\frac{\q}{2}\right)
      +\Sigma_{\perp}\left(\k+\frac{\q}{2}\right)\right] \bm{L}_{\perp}(\k,\q)
\\
    =\left[\bra G_{\perp}\left(\k+\frac{\q}{2}\right) \ket -\bra G_{\perp}^*\left(\k-\frac{\q}{2}\right)\ket\right]
      \int \bar{\bm{\Gamma}}_{\perp}\left(\k+\frac{\q}{2},\k'+\frac{\q}{2},\k-\frac{\q}{2},\k'-\frac{\q}{2}\right)
      \cdot\bm{L}_{\perp}\left(\k',\q \right) \frac{\ud \bm{k}'}{(2\pi)^3}, \label{eq:th_bethe_salpeter_qk}
\end{multline}
\end{widetext}
where we have assumed statistical homogeneity and translational invariance of the medium and neglected the exponentially small coherent intensity. The field correlation, described by a new function
\be
	\bm{L}_{\perp}(\k,\q) \equiv \bm{C}_{\perp}\left(\k+\frac{\q}{2},\k-\frac{\q}{2}\right),
\ee
with
\be
   \bm{C}_{\perp}\left(\bm{k},\bm{k}'\right)
      =\bm{P}(\u) \otimes \bm{P}(\u') \cdot\bm{C}\left(\bm{k},\bm{k}'\right),
\ee
depends only on the transverse part of the intensity vertex, which is given by
\begin{multline}
 \bm{\Gamma}_{\perp}(\bm{k},\bm{\upkappa},\bm{k}',\bm{\upkappa}') = \bm{P}(\u) \otimes \bm{P}(\u')
    \cdot\bm{\Gamma}(\bm{k},\bm{\upkappa},\bm{k}',\bm{\upkappa}')
\\
   =(2\pi)^3\delta(\k-\bm{\upkappa}-\k'-\bm{\upkappa}')
      \bar{\bm{\Gamma}}_{\perp}(\bm{k},\bm{\upkappa},\bm{k}',\bm{\upkappa}').
\end{multline}
Equation~\eqref{eq:th_bethe_salpeter_qk} is very general, as it considers all multiple scattering events within the medium and does not make any explicit assumption on the kind of disorder. Note however that neglecting the longitudinal component of the Green tensor implicitly excludes near-field interactions between scattering centers, that might be important, for example, in dense packings of high-index resonant particles.

\subsubsection{Radiative transfer limit and scattering mean free path}

Further approximations are required to obtain an explicit transport equation for the average intensity. First, we take the large-scale approximation $|\q| \ll \{|\k|,|\k'|\}$, also known as the radiative transfer limit~\cite{barabanenkov1968radiation, ryzhik1996transport}, which assumes that the average intensity varies on length scales $2\pi/|\q|$ much larger than the wavelength in the medium $2\pi/k_\text{r}$. This amounts to assuming $k_\text{r} \ell_\text{e} \gg 1$, which corresponds to the weak extinction regime.
Equation~\eqref{eq:th_bethe_salpeter_qk} becomes
\begin{multline} \label{eq:th_bethe_salpeter_q}
	\left[-2\k\cdot\q+2i\imx\Sigma_{\perp} (\k)\right] \bm{L}_{\perp}(\k,\q) = 2i\imx \bra G_{\perp}(\k)\ket
\\
	\times \int \bar{\bm{\Gamma}}_{\perp}\left(\k,\k',\k,\k'\right) \cdot\bm{L}_{\perp}(\k',\q) \frac{\ud\k'}{(2\pi)^3}.
\end{multline}
The weak extinction regime also corresponds to $|\Sigma_{\perp}|\ll k_\text{b}^2$, see Eqs.~\eqref{keff},\eqref{extmfp}.
Using the relation
\be \label{eq:th_distribution_relation}
	\lim_{\epsilon\to 0^+}\frac{1}{x-x_0-i\epsilon} = \operatorname{PV}\left[\frac{1}{x-x_0}\right]+i\pi\delta(x-x_0),
\ee
where $\operatorname{PV}$ stands for the Cauchy principal value operator, the imaginary part of the average Green function reduces to
\be\label{eq:th_im_g}
	\imx \bra G_{\perp} (\k) \ket = \pi \delta\left[k^2-k_\text{b}^2-\rex\Sigma_{\perp}(\k)\right].
\ee
This relation fixes the real part of the effective wavevector $k_\text{r} = \rex k_\text{eff}$ to
\be\label{eq:th_real_effective_k}
   k_\text{r} = \sqrt{k_\text{b}^2+\rex \Sigma_{\perp}(k_\text{r})},
\ee
which is the so-called ``on-shell approximation''. Second, we assume that the field is fully depolarized, which is valid when the observation point is at a large distance from the source compared to the average distance between scattering events~\cite{bicout1992multiply, gorodnichev2014depolarization,  vynck2016multiple}. This means that
\be
   \bm{C}\left(\bm{k},\bm{k}'\right)=C\left(\bm{k},\bm{k}'\right)\Ig,
\ee
leading to
\be\label{eq:th_depolarized}
   \bm{L}_{\perp}(\k,\q)= L(\k,\q)\bm{P}(\u) \otimes \bm{P}(\u')\cdot\Ig.
\ee
An inverse Fourier transform of the trace of Eq.~\eqref{eq:th_bethe_salpeter_q} together with Eqs.~\eqref{eq:th_im_g} and~\eqref{eq:th_depolarized} eventually leads to the well-known Radiative Transfer Equation (RTE)~\cite{chandrasekhar1960radiative}
\be\label{eq:th_rte}
   \left[\u \cdot \bm{\nabla}_{\r}+\frac{1}{\ell_\text{e}}\right] I(\r,\u) = \frac{1}{\ell_\text{s}} \int p(\u,\u') I\left(\r,\u' \right) \ud \u',
\ee
where $\ud \u$ means an integration over the unit sphere or equivalently over the solid angle, and $I$ is the specific intensity, defined as
%
\be
   \delta(k-k_\text{r})I(\r,\u)=L(\r,\k).
\ee
The specific intensity can be interpreted as a local (at position $\bm{r}$) and directional (on direction $\u$) radiative flux. In the RTE, $\ell_\text{s}$ and $p(\u,\u')$ are the scattering mean free path and the phase function, describing respectively the average distance between two scattering events and the angular diagram for an incident planewave along $\u'$ scattered along direction $\u$. Both quantities are related to the intensity vertex via the relation
\begin{multline}\label{eq:l_s}
   \frac{1}{\ell_\text{s}} p(\u,\u') = \frac{1}{32\pi^2}
      \operatorname{Tr}\left[\bm{P}(\u) \otimes \bm{P}(\u) 
      \right.
\\
      \left.
         \cdot\bar{\bm{\Gamma}} (k_\text{r}\u,k_\text{r}\u',k_\text{r}\u,k_\text{r}\u')
         \cdot\bm{P}(\u') \otimes \bm{P}(\u')\cdot\Ig
      \right],
\end{multline}
and the phase function is normalized as
\be\label{eq:phase_function_normalization}
	\int p(\u,\u')\ud\u' = 1.
\ee
This normalization immediately shows that $1/\ell_\text{s}$ is given by the integral of the right hand side of Eq.~\eqref{eq:l_s} over $\u'$.
The trace appearing in Eq.~\eqref{eq:l_s} is a consequence of the assumption of a depolarized field.
The RTE [Eq.~\eqref{eq:th_rte}] can be seen as an energy balance~\cite{chandrasekhar1960radiative}. The spatial variation of the specific intensity (term involving the derivative) is due to the loss induced by extinction along the direction $\u$ [term involving $\ell_\text{e}$] and the gain from scattering from direction $\u'$ to direction $\u$ [term involving $\ell_\text{s}$ and $p(\u,\u')$]. Equations~\eqref{eq:l_s} and \eqref{eq:phase_function_normalization} show that $\ell_\text{s}$, the key quantity to describe the scattering strength of a medium, is obtained in the radiative transfer limit from the angular integral of the intensity vertex.

Previously, we showed that the extinction mean free path $\ell_\text{e}$ could be obtained from the self-energy $\bm{\Sigma}$ and noted that, in absence of absorption, we should have $\ell_\text{e} = \ell_\text{s}$, the latter being defined from the intensity vertex $\bm{\Gamma}$. It is worth emphasizing at this point that the two operators are indeed formally linked by the Ward identity~\cite{barabanenkov1995diffusion, cherroret2016induced}, which may be seen as a generalization of the extinction (optical) theorem and ensures energy conservation~\cite{lagendijk1996resonant, apresyan1996radiation, tsang2001scattering, sheng2006introduction, carminati2021principles}.

\subsubsection{Transport mean free path and diffusion approximation}\label{diffusion}

Many experiments on light in disordered media are performed in situations where light experiences not just a few but many scattering events on average.
In the deep multiple scattering regime, the RTE can be simplified into a diffusion equation. In this limit, light transport is driven by a new length scale, known as the transport mean free path $\ell_\text{t}$, which we will introduce here.

We start by taking the first moment of Eq.~\eqref{eq:th_rte} (i.e., multiplying both sides by $\u$ and integrating over
$\u$) which directly leads to
\be\label{eq:th_first_moment_rte}
	\int \left[\u \cdot \bm{\nabla}_{\r} I(\r,\u)\right] \u \ud\u + \frac{1}{\ell_\text{t}} \j(\r) = 0,
\ee
where $\j(\r) = \int I(\r,\u) \u \ud\u$ is the radiative flux vector and
\be\label{eq:l_t}
	\ell_\text{t} \equiv \frac{\ell_\text{s}}{1-g},
\ee
is the transport mean free path.
\be\label{eq:g}
	g = \int p(\u,\u') \u \cdot \u' \ud \u
\ee
is the average cosine of the scattering angle, or scattering anisotropy factor.
Structural correlations impact the transport mean free path via both the scattering mean free path $\ell_\text{s}$ and the scattering anisotropy described by $g$.

After a large number of scattering events, we can assume that the specific intensity becomes quasi-isotropic.
Expanding the specific intensity in the RTE [Eq.~\eqref{eq:th_rte}] on Legendre polynomials to first order in $\u$, which is known as the $P_1$-approximation~\cite{ishimaru1978wave}, leads to the diffusion equation. In the steady-state regime, it reads
\be\label{eq:th_diffusion_approximation}
	-\mathcal{D}\Delta u(\r) = s(\r)
\ee
where $u(\r)=v_\text{E}^{-1} \int I(\r,\u)\ud\u$ is the energy density with $v_\text{E}$ the energy velocity~\cite{lagendijk1996resonant}, $\mathcal{D}=v_\text{E}\ell_\text{t}/3$ is the diffusion constant and $s$ is a source term.
An analysis of the diffusion equation shows that it is valid on length scales large compared to $\ell_\text{t}$. This allows us to reinterpret $\ell_\text{t}$ as the distance after which the intensity distribution is quasi-isotropic~\cite{ishimaru1978wave, carminati2021principles}.

Resolving this equation in a slab geometry of thickness $L$ under planewave illumination at normal incidence gives the following asymptotic behavior for the total transmittance
\be
	T \sim \frac{5}{3}\frac{\ell_\text{t}}{L},
\ee
which is Ohm's law for light~\cite{van1999multiple}. Many transport observations in the diffusive limit depend directly on the transport mean free path, including the linewidth of the coherent backscattering cone~\cite{akkermans1986coherent, akkermans1988theoretical}, the time-resolved transmittance and reflectance~\cite{contini1997photon}, and long-range speckle correlations~\cite{shapiro1999new,scheffold1998universal,fayard2015intensity}.

In summary, the average transition operator of the medium $\bra \Tgcal \ket$ defines the four-point irreducible vertex $\Gammaop$ [Eq.~\eqref{eq:gamma}]. Neglecting near-field interaction between scattering elements, taking the radiative transfer limit and assuming fully depolarized light allow us to relate the transverse component of $\Gammaop$ to the scattering mean free path $\ell_\text{s}$ and phase function $p(\u,\u')$ [Eq.~\eqref{eq:l_s}]. In the diffusion approximation, the transport mean free path $\ell_\text{t}$ [Eq.~\eqref{eq:l_t}] drives the energy flux. It is related to the intensity vertex via Eqs.~\eqref{eq:l_s} and \eqref{eq:phase_function_normalization}.

The theoretical framework described here will now be used to get closed-form expressions for the different optical length scales in the cases of random media described by a continuous permittivity [Sec.~\ref{subsec:random}] or as an assembly of discrete particles [Sec.~\ref{subsec:particulate}].

\subsection{Media with fluctuating continuous permittivity} \label{subsec:random}

\subsubsection{Weak permittivity fluctuations}

We consider a statistically homogeneous and isotropic disordered medium, described by a spatially-dependent permittivity $\epsilon(\r) = \langle \epsilon \rangle + \Delta \epsilon (\r)$, where $\Delta \epsilon$ is the fluctuating part with statistics
\be
	\langle \Delta \epsilon(\r) \rangle &=& 0, \\
	\langle \Delta \epsilon(\r) \Delta \epsilon(\r') \rangle &=& \langle \epsilon \rangle^2 \delta_\epsilon^2 h_{\epsilon}(|\r-\r'|). \label{eq:statistics-random}
\ee
Here, $\delta_\epsilon^2 = \langle \Delta \epsilon^2 \rangle / \langle \epsilon \rangle^2$ is the normalized variance of $\epsilon$ and $h_\epsilon(|\r-\r'|) = \langle \Delta \epsilon(\r) \Delta \epsilon(\r') \rangle / \langle \Delta \epsilon^2 \rangle$ is the normalized permittivity-permittivity correlation function ($h_\epsilon(0) = 1$). Hereafter, we assume ergodicity, such that the ensemble average is equivalent to a volume average in the infinite-volume limit, and isotropic permittivity fluctuations, keeping in mind, however, that anisotropic fluctuations may take place even in isotropic materials~\cite{landau2013electrodynamics}.

The statistical properties of $\epsilon(\r)$ straightforwardly translate into statistical properties of $\Xgcal(\r)$ via Eq.~\eqref{eq:scatt_potential}. The self-energy $\Sigmaop$ can be expressed in terms of $\Xgcal$ by inserting the expression for the transition operator $\Tgcal$ given by Eq.~\eqref{T-operator} into Eq.~\eqref{Sigmaexact}, leading to
\be
	\Sigmaop = \bra \Xgcal \left[ \Ig - \Gcalb \Xgcal \right]^{-1} \ket \bra \left[ \Ig - \Gcalb \Xgcal \right]^{-1} \ket^{-1}. \label{eq:expansion-sigmaop-weak}
\ee
In the simplest approach, we proceed by assuming that the scattering potential $\Xgcal$ weakly fluctuates around its average value $\bra \Xgcal \ket$. Expanding the last expression near $\bra \Xgcal \ket$ leads to
\be
	\Sigmaop \sim \bra \Xgcal \ket + \bra \left (\Xgcal - \bra \Xgcal \ket \right) \Gcalb \left( \Xgcal - \bra \Xgcal \ket \right) \ket + \cdots
\ee

At this stage, we need to define explicitly the constant auxiliary background permittivity $\epsilon_\text{b}$, which describes the reference value around which the permittivity fluctuates. A reasonable choice is to set it to the average permittivity, $\epsilon_\text{b} = \bra \epsilon \ket \equiv \epsilon_\text{av}$, which, for a two-component medium with permittivities $\epsilon_\text{p}$ and $\epsilon_\text{h}$ at filling fractions $f$ and $1-f$, respectively, would simply be $\epsilon_\text{av} = f \epsilon_\text{p} + (1-f) \epsilon_\text{h}$. Having then $\bra \Xgcal \ket = 0$, the leading term for the self-energy becomes $\bra \Xgcal \Gcalb \Xgcal \ket$, such that
\be
	\bm{\Sigma}(\r-\r') &=& k_\text{av}^4 \delta_\epsilon^2 h_{\epsilon}(|\r-\r'|) \Gg_\text{av}(\r-\r'), \label{Sigma_r_av}
\ee
where $\Gg_\text{av}$ is the Green tensor in a homogeneous medium with permittivity $\epsilon_\text{av}$. Correlated permittivity fluctuations mutually interacting via $\Gg_\text{av}$ are readily responsible for the non-local character of the self-energy [Eq.~\eqref{eq:Sigma_rr}]. In Fourier space, Eq.~\eqref{Sigma_r_av} becomes
\be
	\bm{\Sigma}(\k) &=& k_\text{av}^4 \delta_\epsilon^2 \int h_{\epsilon}(|\k-\k'|) \Gg_\text{av}(\k') \frac{d\k'}{(2\pi)^3}. \label{Sigma_r_av_kspace}
\ee

The extinction mean free path $\ell_\text{e}$ can finally be determined using Eq.~\eqref{extmfp} with $k_\text{r} = k_\text{av}$. For non-absorbing media [$\epsilon (\r)$ real], the imaginary part of the Green tensor is given by
\be
    \text{Im} \Gg_\text{av} (\k) = \pi \delta \left( k^2 - k_\text{av}^2 \right) \P(\u),
\ee
with $k_\text{av} = k_0 \sqrt{\epsilon_\text{av}}$. Using Eq.~\eqref{eq:sigma_perp} for the transverse component with $\e$ the unit vector defining the polarization direction such that $\e \cdot \u = 0$, we find that
\be \label{eq:ext_mfp_weak_fluctuations_general}
    \frac{1}{\ell_\text{e}} & \simeq & \frac{\text{Im} \Sigma_\perp(k_\text{av})}{k_\text{av}}, \nonumber \\
    &=& \frac{k_\text{av}^4}{16 \pi^2} \delta_\epsilon^2 \int h_{\epsilon}(k_\text{av}|\u-\u'|) \left( \e \cdot \P(\u') \e \right) d\u'. \nonumber \\
\ee
Introducing the scattering wavenumber $q = k_\text{av} |\u - \u'|$, we eventually reach
\be
    \frac{1}{\ell_\text{e}} = \frac{k_0^4}{8 \pi k_\text{av}^2} \epsilon_\text{av}^2 \delta_\epsilon^2 \int_0^{2 k_\text{av}} P\left(\frac{q}{2 k_\text{av}}\right) h_{\epsilon}(q) q dq, \label{eq:ext_mfp_weak_fluctuations}
\ee
where
\be\label{eq:projector-pola}
    P(k) \equiv 1 - 2 k^2 + 2 k^4,
\ee
is specific to the vector nature of light. 

%
%

Equation~\eqref{eq:ext_mfp_weak_fluctuations}, obtained in non-absorbing disordered media with weak permittivity fluctuations, constitute the first analytical expression for the extinction mean free path in correlated media. Very importantly, it shows that besides the amplitude of the permittivity fluctuations, $\epsilon_\text{av}^2 \delta_\epsilon^2 = \bra \Delta \epsilon^2 \ket$, spatial correlations, described here by $h_{\epsilon}$, also play a crucial role in light scattering.

\subsubsection{Lorentz local fields: strong fluctuations}

The approximation of weak fluctuations is prohibitive in many realistic cases. Fortunately, this constraint can be eliminated by properly handling the singularity of the dyadic Green function at the origin [Eq.~\eqref{GFsing}], which constitutes the basis of a strong fluctuation theory~\cite{finkel1964mean, ryzhov1965spatial, tsang1981scattering}. Related approaches were introduced by \citet{bedeaux1973critical} and \citet{felderhof1974propagation} in the description of the optical response of non-polar fluids. Extensions to chiral and anisotropic media have also been proposed~\cite{ryzhov1970radiation, michel1995strong, mackay2020modern}, but will not be discussed here.
Noteworthy is the recent work by~\citet{torquato2021nonlocal}, which relies on the same theoretical grounds and proposes an analytical expression for the effective permittivity of two-phase composite media that takes into account structural correlations up to an arbitrary order $n$ (whereas we restrict the discussion to correlations of order $n=2$ in the present review).

The singularity of the Green function is handled by considering the scattering medium as being made of infinitesimal volume elements within which the polarization density $\P(\r)$ is constant. This physical viewpoint is the basis of the renowned discrete dipole approximation~\cite{draine1994discrete, lakhtakia1992general}. Let us then write the Green function as a sum of two contributions
\be
	\Gg_\text{b}(\r,\r') &=& \Theta(|\r-\r'|-a) \tilde{\Gg}_\text{b}(\r,\r') \nonumber \\
	&+& \Theta(a-|\r-\r'|) \bm{g}_\text{b}(\r,\r'), \label{Def_Lorentz_Green_1}
\ee
where $\bm{g}_\text{b}$ contains the singular part of the Green function and $\tilde{\Gg}_\text{b}$ is the so-called Lorentz propagator, which is purely non-local. The contributions are distinguished as belonging or not to a spherical region with radius $a$ and volume $v=4\pi a^3/3$ around $\r'$. Choosing $a$ such that $k_\text{b} a \ll 1$, the singular part reads
\be
	\left. \bm{{g}}_\text{b}(\r,\r')\right|_{k_\text{b} a \ll 1} & = & -\frac{1}{3k_\text{b}^2} \delta(\r-\r') \Ig + \im \frac{k_\text{b}}{6\pi} \Ig +\cdots \label{Def_singular_Green}
\ee
Keeping the lowest-order terms in the real and imaginary parts, the Lippmann-Schwinger equation [Eq.~\eqref{LS-integral}] can be rewritten as
\be
	\E(\r) &=& \Eb(\r) + \left( - \frac{1}{3k_\text{b}^2} +  \im \frac{k_\text{b} v}{6\pi} \right) \Xgcal(\r) \E(\r) \nonumber \\
	&+& \int \tilde{\Gg}_\text{b}(\r,\r') \Xgcal(\r') \E(\r') d\r'. \label{LS2}
\ee
The actual field at $\r$ is then given by the sum of the external field $\Eb(\r)$, the \textit{local} contributions, and the \textit{non-local} contributions coming from neighboring permittivity fluctuations (integral term).

In this framework, the field exciting a small volume element around $\r$, $\E_{\text{exc}}(\r)$, is the sum of the incident (background) field and the field scattered by other permittivity fluctuations. Using again operator notations, we thus reach an important set of equalities
\be
	\E_\text{exc} &=& \Eb + \TGcalb \Xgcal \E = \left[ \Ig - \gb \Xgcal \right] \E \nonumber \\
	&=& \Eb + \TGcalb \tilde{\Tgcal} \E_\text{exc} = \left[ \Ig + \TGcalb \Tgcal \right] \Eb. \label{eq:exciting-field}
\ee
We have introduced here a new quantity, $\tilde{\Tgcal}$, that plays the role of a \textit{local} transition operator. From Eqs.~\eqref{eq:exciting-field}, we straightforwardly obtain
\be
	\Tgcal = \tilde{\Tgcal} \left[ \Ig - \TGcalb \tilde{\Tgcal} \right]^{-1}. \label{polexpT}
\ee
The transition operator of the medium can be seen as a multiple-scattering expansion on independent scattering elements, connected via the (non-local) Lorentz propagator. Consistently with this picture, from Eqs.~\eqref{eq:exciting-field}, we also obtain
\be
	\tilde{\Tgcal} = \Xgcal \left[ \Ig - \gb \Xgcal \right]^{-1}.
\ee
For $k_\text{b} a \ll 1$, the (local) transition operator $\tilde{\Tg}(\r,\r')$ is directly proportional to the polarizability of the volume element,
\be
	\tilde{\Tg}(\r,\r') &=& k_\text{b}^2 \frac{\alpha(\r)}{v} \delta(\r-\r') \Ig, \label{T0}
\ee
with
\be
	\alpha(\r) = \frac{\alpha_0(\r)}{ 1 - \im \frac{k_\text{b}^3}{6\pi} \alpha_0(\r)}, \;
	\alpha_0(\r) = 3 v \frac{\epsilon(\r) -\epsilon_\text{b}} {\epsilon(\r) +2\epsilon_\text{b}}, \label{alpha0}
\ee
where $\alpha_0$ is the quasi-static polarizability. Note that $\alpha(\r)$ is a space-dependent local polarizability defined in a continuous medium.

\subsubsection{Average exciting field}

To determine the self-energy $\Sigmaop$ of the system, it is convenient to introduce a self-energy $\tilde{\Sigmaop}$ for the exciting field defined from a Dyson equation
\be
	\langle \E_\text{exc} \rangle = \Eb + {\TGcalb} \tilde{\Sigmaop} \langle \E_{\text{exc}} \rangle, \label{eq:th_dyson_ex} 
\ee
thereby leading to
\be
	\tilde{\Sigmaop} = \bra \tilde{\Tgcal} \left[ \Ig - \TGcalb \tilde{\Tgcal} \right]^{-1} \ket \bra \left[ \Ig - \TGcalb \tilde{\Tgcal} \right]^{-1} \ket^{-1}. \label{SigmaFormal3}
\ee
Note the similarity of this expression with Eq.~\eqref{eq:expansion-sigmaop-weak}, where the self-energy $\Sigmaop$ was expressed directly in terms of the scattering potential $\Xgcal$. From Eqs.~\eqref{Sigmaexact}, \eqref{polexpT} and \eqref{SigmaFormal3}, one shows that the two self-energies are related as
\be
	\Sigmaop = \tilde{\Sigmaop} \left[ \Ig + \gb \tilde{\Sigmaop} \right]^{-1}. \label{Sigma_Sigmatilde}
\ee

Let us then expand $\tilde{\Sigmaop}$ in Eq.~\eqref{SigmaFormal3} near $\langle \tilde{\Tgcal} \rangle = \tilde{\Tgcal} - \Delta \tilde{\Tgcal}$,
\be
	\tilde{\Sigmaop} \sim \bra \tilde{\Tgcal} \ket + \bra \Delta \tilde{\Tgcal} \Hcal \Delta \tilde{\Tgcal} \ket + \cdots \equiv \tilde{\Sigmaop}_1 + \tilde{\Sigmaop}_2 + \cdots. \label{eq:sigma_exc-expansion}
\ee
We have introduced here a new ``dressed'' propagator,
\be
	\Hcal = \left[ \Ig - \TGcalb \bra \tilde{\Tgcal} \ket \right]^{-1} \TGcalb. \label{Hcaldef}
\ee
Noting that $\langle \tilde{\Tgcal} \rangle$ corresponds to an average polarizability of the medium (for $k_\text{b} a \ll 1$, see Eqs.~\eqref{T0}), one understands that $\Hcal$ describes the fact that the fields propagate from fluctuation to fluctuation via a medium with a permittivity that can differ from the background permittivity $\epsilon_\text{b}$~\cite{bedeaux1973critical,felderhof1974propagation}.

The self-energy for the average exciting field, $\tilde{\Sigmaop}$ in Eq.~\eqref{eq:sigma_exc-expansion}, now explicitly depends on the spatial correlations of the fluctuations of $\tilde{\Tgcal}$ (i.e., of the polarizability of small volume elements). In most practical cases, the expansion is limited to second order, corresponding to the so-called ``bilocal'' approximation~\cite{tsang2001scattering}, due to the lack of information on higher-order correlation functions in real systems.

Expanding $\Sigmaop$ in Eq.~\eqref{Sigma_Sigmatilde} near $\tilde{\Sigmaop}_1$, we obtain
\be
	\Sigmaop &\sim& \tilde{\Sigmaop}_1 \left[ \Ig + \gb \tilde{\Sigmaop}_1 \right]^{-1} + \tilde{\Sigmaop}_2 \left[ \Ig + \gb \tilde{\Sigmaop}_1 \right]^{-2} + \cdots \nonumber \\
	&\equiv & \Sigmaop_1 + \Sigmaop_2 + \cdots \label{sigma_expansion}
\ee
The self-energy is now expressed in terms of the scattering properties of vanishingly small, individual scattering elements.

\subsubsection{Long-wavelength solutions: Bruggeman versus Maxwell-Garnett models}

As in the case of weakly fluctuating media discussed previously, we now need to give an explicit definition of the constant auxiliary background permittivity $\epsilon_\text{b}$, that describes the homogeneous effective medium in which the permittivity fluctuations scatter light. We will see that this sole parameter constitutes the essential difference between the two celebrated ``mixing rules'' due to~\citet{bruggeman1935berechnung} and \citet{maxwellgarnett1904colours}, presented here in a unique theoretical framework. Despite the arbitrariness in the choice of $\epsilon_\text{b}$, it is important to realize that all models would eventually be strictly equivalent when carried out to infinite order. The applicability of a model is thus mainly a question of accuracy at low orders and convergence.

A first possibility for $\epsilon_\text{b}$ is to set it such that $\langle \tilde{\Tgcal} \rangle = 0$. In the limit of small volume elements, this corresponds to having a zero average polarizability, see Eqs.~\eqref{T0}-\eqref{alpha0}. Considering a two-component medium with relative permittivities $\epsilon_\text{p}$ (at filling fraction $f$) and $\epsilon_\text{h}$ (at filling fraction $1-f$) in the quasi-static limit ($\alpha = \alpha_0$ in Eq.~\eqref{alpha0}), one obtains
\be
	\frac{\epsilon_\text{p} -\epsilon_\text{BG}} {\epsilon_\text{p} +2 \epsilon_\text{BG}} f + \frac{\epsilon_\text{h} -\epsilon_\text{BG}} {\epsilon_\text{h} +2 \epsilon_\text{BG}} (1-f) = 0
\ee
which is the Bruggeman mixing rule~\cite{bruggeman1935berechnung} with $\epsilon_\text{b} \equiv \epsilon_\text{BG}$. The generalization to $N$-component media is straightforward. Having $k_\text{b}^2 = k_\text{BG}^2 = k_0^2 \epsilon_\text{BG}$, $\hat{\bm{\mathcal{G}}}_\text{b} = \tilde{\bm{\mathcal{G}}}_\text{BG}$ and $\bm{\Sigma} \sim \tilde{\bm{\Sigma}}$ since $\tilde{\Sigmaop}_1=0$, we eventually find that
\be
	\bm{\Sigma}(\k) = k_\text{BG}^4 \delta_\alpha^2 \int h_\alpha(|\k-\k'|) \tilde{\G}_\text{BG}(\k') \frac{d\k'}{(2\pi)^3}, \label{Sigma_r_BG_kspace}
\ee
with $\delta_\alpha^2 = \langle \Delta \alpha^2 \rangle / v^2$ a normalized variance of the polarizability and $h_\alpha(|\k-\k'|)$ the Fourier transform of the normalized polarizability-polarizability correlation function $h_\alpha (|\r-\r'|) = \langle \Delta \alpha (\r) \Delta \alpha (\r') \rangle/\langle \Delta \alpha ^2 \rangle $. The function $h_\alpha$ plays the same role as $h_\epsilon$ in the weakly fluctuating permittivity model to describe structural correlations.

Assuming non-absorbing media and following the same steps as those leading to Eq.~\eqref{eq:ext_mfp_weak_fluctuations} with $\text{Im} \tilde{\G}_\text{BG} (\k) = \pi \delta \left( k^2 - k_\text{BG}^2 \right) \P(\u)$, we obtain
\be
    \frac{1}{\ell_\text{e}} = \frac{k_0^4}{8 \pi k_\text{BG}^2} \epsilon_\text{BG}^2 \delta_\alpha^2 \int_0^{2 k_\text{BG}} P\left(\frac{q}{2 k_\text{BG}}\right) h_{\alpha}(q) q dq, \nonumber \\
    \label{eq:ext_mfp_Bruggeman}
\ee
with $q = k_\text{BG} |\u - \u'|$. Equation~\eqref{eq:ext_mfp_Bruggeman} is strikingly similar to Eq.~\eqref{eq:ext_mfp_weak_fluctuations}, the essential differences being (i) the permittivity of the homogeneous effective medium and (ii) the description of the medium via a local polarizability instead of a local permittivity.

A second possibility for the choice of $\epsilon_\text{b}$ is to set it to the permittivity of the host medium (i.e., $\epsilon_\text{b} = \epsilon_\text{h}$), in which case $\langle \tilde{\Tgcal} \rangle \neq 0$. Considering again a two-component system in the quasi-static limit, we obtain
\be
	\tilde{\bm{\Sigma}}_1 (\r-\r') = k_\text{h}^2 \rho \alpha_0 \delta(\r-\r') \Ig, \label{Sigma_1_tilde_MG}
\ee
with $\rho=f/v$ the average number density of the small volume elements with permittivity $\epsilon_\text{p}$, and
\be
	\tilde{\bm{\Sigma}}_2 (\r-\r') =  k_\text{h}^4 \delta_\alpha^2 h_\alpha(|\r-\r'|) \hat{\G}_\text{h}(\r-\r'). \label{Sigma_2_tilde_MG}
\ee 
Using Eq.~\eqref{sigma_expansion}, we find the following expression for the self-energy in reciprocal space
\be
	\bm{\Sigma} (\k) &=& k_0^2 \left( \eps_\text{MG} -\eps_\text{h} \right) \Ig \nonumber \\
	&+& k_\text{h}^4 \delta_\alpha^2 f_\text{L} \int h_\alpha(|\k-\k'|) \hat{\G}_\text{h}(\k') \frac{d\k'}{(2\pi)^3}, \label{MG_sigma_continuous_kspace}
\ee
where $f_\text{L} = (\partial \epsilon_\text{MG}/\partial \rho)/(\epsilon_\text{h} \alpha_0)$ and
\be	
	\eps_\text{MG} = \eps_\text{h} + \eps_\text{h} \frac{\rho \alpha_0}{1-\rho \alpha_0 /3}, \label{MG_epsilon}
\ee
is the Maxwell-Garnett mixing rule~\cite{maxwellgarnett1904colours, markel2016introduction}. By contrast with the previous case, the lowest-order term now provides the renormalization of the wave number in the effective medium, structural correlations appearing at the next order. The factor $f_\text{L}$ is a local-field correction coming from the fact that the fluctuation of polarizability ($\Delta \alpha$ in $\delta_\alpha^2$) is evaluated with respect to the host medium. Following again the same steps as those leading to Eq.~\eqref{eq:ext_mfp_weak_fluctuations}, noting that $\eps_\text{MG}$ is real in the quasi-static limit for non-absorbing media, and using
\be \label{eq:Im-Gh}
    \text{Im} \hat{\G}_\text{h} (\k) = \pi f_\text{L} \delta \left( k^2 - k_\text{MG}^2 \right) \P(\u),
\ee
which is derived in Appendix~\ref{App:Dressed_Green}, we eventually obtain
\be
    \frac{1}{\ell_\text{e}} = \frac{k_0^4}{8 \pi k_\text{MG}^2} f_\text{L}^2 \epsilon_\text{h}^2 \delta_\alpha^2 \int_0^{2 k_\text{MG}} P\left(\frac{q}{2 k_\text{MG}}\right) h_{\alpha}(q) q dq, \nonumber \\
    \label{eq:ext_mfp_MaxwellGarnett}
\ee
with $q = k_\text{MG} |\u - \u'|$.
This expression for $\ell_\text{e}$ takes the same form as Eq.~\eqref{eq:ext_mfp_Bruggeman} with differences in the definition of the effective medium and a prefactor that accounts for local-field corrections.

All in all, the similarity between Eqs.~\eqref{eq:ext_mfp_weak_fluctuations}, \eqref{eq:ext_mfp_Bruggeman} and \eqref{eq:ext_mfp_MaxwellGarnett} demonstrates the deep physical implication of structural correlations for light scattering. The same functional structure is kept, whatever the approach used to define the effective medium.

\subsubsection{Expressions for the scattering and transport mean free paths from the average intensity}

We conclude this part on continuous permittivity media by deriving the expressions for $\ell_\text{s}$ and $\ell_\text{t}$ from the theory for the average intensity. A second-order expansion of the intensity vertex $\Gammaop$ in Eq.~\eqref{eq:gamma}, with $\Tgcal$ given by Eq.~\eqref{T-operator}, leads to
\be
   \Gammaop \sim \bra \Tgcal \Tgcal^* \ket - \bra \Tgcal \ket \bra \Tgcal^* \ket
   \sim \bra \Xgcal \Xgcal^* \ket - \bra \Xgcal \ket \bra \Xgcal^* \ket.
\ee
This expansion is valid in the weak extinction limit $k_\text{r} \ell_\text{e} \gg 1$. Similarly to the case of weakly
fluctuating media, we set the auxiliary background permittivity as $\epsilon_\text{b} = \bra \epsilon \ket \equiv
\epsilon_\text{av}$, such that $\bra \Xgcal \ket = 0$, and assume a non-absorbing material. This leads to
\begin{multline}
   \bar{\bm{\Gamma}} (k_\text{av}\u,k_\text{av}\u',k_\text{av}\u,k_\text{av}\u')
      =k_0^4\epsilon_\text{av}^2 \delta_\epsilon^2 h_{\epsilon}(k_\text{av}|\u-\u'|)
\\
         \times\Ig\otimes\Ig.
\end{multline}
The scattering mean free path is then obtained by integrating Eq.~\eqref{eq:l_s} over $\u'$, and making use of the equation above, we find that
\be\label{eq:th_scattering_length_q_continuous}
   \frac{1}{\ell_\text{s}}=\frac{k_0^4}{8\pi k_\text{av}^2} \epsilon_\text{av}^2 \delta_\epsilon^2
      \int_0^{2k_\text{av}} P\left(\frac{q}{2 k_\text{av}}\right)h_{\epsilon}(q) q \ud q,
\ee
with $q = k_\text{av} |\u-\u'|$. Similarly, the transport mean free is obtained by calculating the average cosine of Eq.~\eqref{eq:l_s} and using Eqs.~\eqref{eq:l_t}-\eqref{eq:g}, leading to
\be\label{eq:th_transport_length_q_continuous}
   \frac{1}{\ell_\text{t}}=\frac{k_0^4}{16\pi k_\text{av}^4} \epsilon_\text{av}^2 \delta_\epsilon^2
      \int_0^{2k_\text{av}} P\left(\frac{q}{2 k_\text{av}}\right)h_{\epsilon}(q) q^3 \ud q.
\ee

In absence of absorption, we expect $\ell_\text{s} = \ell_\text{e}$, which is actually found by comparing
Eqs.~\eqref{eq:th_scattering_length_q_continuous} and \eqref{eq:ext_mfp_weak_fluctuations}.


\subsection{Particulate media} \label{subsec:particulate}

\subsubsection{Expansion for identical scatterers}

Let us now consider a system whose morphology consists in localized (i.e., compact) permittivity variations in a uniform background. We take the most natural choice for the background permittivity $\epsilon_\text{b} = \epsilon_\text{h}$ from the start but the theory can also be developed with an arbitrary $\epsilon_\text{b}$. We also restrict the discussion to composite media made of \textit{identical} inclusions with relative permittivity $\epsilon_\text{p}$ confined to a volume $v$, centered at positions $\Rg = \left[\Rg_1,\Rg_2,\dots \Rg_N \right]$. The medium permittivity then reads
\be
	\epsilon(\r) = \sum_j \epsilon_\text{p}(\r-\Rg_j) \Theta(a-|\r-\Rg_j|).\label{eq:permittivity-particulate}
\ee
A configuration of the medium is described statistically by the probability distribution function $p(\Rg)$. Implicitly, we neglect here the possibility to have orientational correlations between particles (otherwise, the distribution should include orientational variables). Under the ergodic hypothesis, defining the statistical average as an average over all possible particle positions as $\langle \mathbf{f}(\Rg) \rangle = \int \mathbf{f}(\Rg) p(\Rg) d\Rg$, where $\mathbf{f}(\Rg)$ is an arbitrary tensor, the statistical properties of the medium can be described by $n$-particle probability density functions~\cite{tsang2004scattering, lebowitz1963integral}
\be
	\rho_n(\r_1, \cdots \r_n) = \big{\langle} &\sum_{j_1 \neq j_2 \cdots \neq j_n} & \delta(\r_1-\Rg_{j_1}) \nonumber \\
	& \cdots  & \delta(\r_n-\Rg_{j_n}) \big{\rangle},
\ee
or equivalently by $n$-particle correlation functions
\be
	g_n(\r_1, \cdots \r_n) = \frac{1}{\rho^n} \rho_n(\r_1, \cdots \r_n), \label{eq:n-particle-correlation}
\ee
where $\rho$ is the constant particle number density reached in the limit of infinite system size ($\rho = \lim_{N,V \rightarrow \infty} N/V$). In statistically homogeneous ensembles of impenetrable spheres, the particle correlation functions $g_n$ are formally related to the probability functions of finding $n$ points separated by given distances in the particle phase~\cite{torquato1982microstructure, torquato2013random}.

Similar to the case of random media described by a continuous permittivity, the first step is to derive an expression for the transition operator $\Tgcal$ of the medium. Having a discrete set of identical particles allows us to express the multiple-scattering problem in such a way as to separate the effects associated to (local) particle resonances and (non-local) structural correlations on light scattering. We start by rewritting the integral equation for the total field, Eq.~\eqref{LS}, for particulate media,
\be
	\E = \Eh + \sum_j \Gcalh \Xgcal_j \E, \label{eq:LS-particles}
\ee
with the effective scattering potential $\Xgcal_j(\r-\Rg_j) \equiv k_0^2 [\epsilon_\text{p}(\r-\Rg_j) \Theta(a-|\r-\Rg_j|) - \epsilon_\text{h}] \Ig$. We can then express the polarization induced in particle $j$ in terms of the polarization induced in all particles as
\be
	\Xgcal_j \E &=& \Xgcal_j \Eh + \Xgcal_j \sum_k \Gcalh \Xgcal_k \E, \\
	&=& \Xgcal_j \Eh + \Xgcal_j \Gcalh \Xgcal_j \E + \Xgcal_j \sum_{k \ne j} \Gcalh \Xgcal_k \E, \label{eq:polaj-total}\\
	&=& \Tgcal_j \Eh + \Tgcal_j \sum_{k \ne j} \Gcalh \Xgcal_k \E. \label{eq:polaj-particle}
\ee
In the second expression, we separated the self-contribution of particle $j$ from the contribution of all other particles. The last expression was obtained by introducing the transition operator $\Tgcal_j$ of an individual particle centered at $\Rg_j$ as
\be
	\Tgcal_j = \Xgcal_j \left[ \Ig - \Gcalh \Xgcal_j \right]^{-1}. \label{eq:Tj-particle}
\ee
Very importantly, $\Tgcal_j$ can be determined for particles of virtually any size, shape and composition, either analytically using Mie theory for simple geometries like spherical particles~\cite{bohren2008absorption} or numerically by any method for solving Maxwell's equations otherwise~\cite{mishchenko1999light}. This allows us to consider resonant particles exhibiting high-order multipolar resonances as the building blocks of the disordered medium.

Inserting Eq.~\eqref{eq:polaj-particle} into Eq.~\eqref{eq:LS-particles} and iterating over scattering sequences, we reach an expression for the transition operator $\Tgcal$ of the entire system [Eq.~\eqref{LS-T}] in terms of the transition operator of the individual particle, as
\be
	\Tgcal &=& \sum_j \Tgcal_j  + \sum_j \Tgcal_j \sum_{k \neq j} \Gcalh \Tgcal_k \nonumber \\
	&+& \sum_j \Tgcal_j \sum_{k \neq j} \Gcalh \Tgcal_k \sum_{l \neq k} \Gcalh \Tgcal_l + \cdots \label{eq:T-operator-particles}
\ee
This series expansion is the root of multiple scattering theory for particulate media and was introduced in the pioneering works of \citet{kirkwood1936theory} and \citet{yvon1937recherches} to determine the permittivity of molecular liquids. Similar multiple scattering equations were later discussed by \citet{foldy1945multiple} and \citet{lax1951multiple}. It is further interesting to note the occurrence of so-called ``recurrent scattering'', that is, scattering sequences that involve the same particle multiple times (for instance, $l$ can be equal to $j$ in the last displayed term).

The Green function in Eq.~\eqref{eq:Tj-particle} for the transition operator $\Tgcal_j$ of a specific particle $j$ always connects two points that belong to the same particle, whereas the Green function in Eq.~\eqref{eq:T-operator-particles} for the transition operator $\Tgcal$ of the entire medium always connects two points that belong to different particles. This is conceptually analogous to Eq.~\eqref{Def_Lorentz_Green_1} for continuous permittivity media where the Green function was split into local and non-local terms. To determine the self-energy $\Sigmaop$ in the Dyson equation [Eq.~\eqref{eq:th_dyson}], we may then follow the strategy used for continuous permittivity media and consider the exciting field $\E_\text{exc}$. Let us then write the Green function $\Gcalh$ as $\gh$ when connecting two points in the same particle and $\TGcalh$ otherwise. Removing the local contribution on the induced polarization in Eq.~\eqref{eq:polaj-total} and rewriting $\Xgcal_k \E$ in terms of the exciting field leads to
\be
	\Xgcal_j \E_\text{exc} = \Xgcal_j \Eh + \Xgcal_j \sum_{k} \TGcalh \tilde{\Tgcal}_k \E_\text{exc},
\ee
with $\tilde{\Tgcal}_j = \Xgcal_j \left[ \Ig - \gh \Xgcal_j \right]^{-1} = \Tgcal_j$, see Eq.~\eqref{eq:Tj-particle}. Note that the sum now runs over all particles $k$. Further defining $\tilde{\Tgcal} \equiv \sum_j \tilde{\Tgcal}_j$, we find
\be
	\Tgcal = \tilde{\Tgcal} \left[ \Ig - \TGcalh \tilde{\Tgcal} \right]^{-1}, \label{polexpT-2}
\ee
which agrees with Eq.~\eqref{polexpT} derived for continuous media. Expanding the self-energy $\tilde{\Sigmaop}$ for the average exciting field near $\langle \tilde{\Tgcal} \rangle = \tilde{\Tgcal} - \Delta \tilde{\Tgcal}$ then leads to Eq.~\eqref{eq:sigma_exc-expansion}, and expanding $\Sigmaop$ near $\tilde{\Sigmaop}_1$ to Eq.~\eqref{sigma_expansion}.

The problem of scattering by particulate media is thus described in a strictly similar manner to that of scattering by strongly fluctuating continuous permittivity media (i.e., including local-field corrections). The essential difference is that the volume elements composing the medium are no longer vanishingly small but actual finite-size scattering particles.

Finally, let us remined that Eq.~\eqref{eq:sigma_exc-expansion} is obtained by neglecting particle correlations beyond second order. Higher-order correlations may yet be taken into account by treating them as sequences of two-particle correlations (which is formally exact for crystalline media). This approach corresponds to the so-called quasicrystalline approximation (QCA) originally introduced by~\citet{lax1952multiple}, further developed by ~\citet{fikioris1964multiple} and~\citet{tsang1980multiple, tsang1982effective} among others, and used nowadays in various contexts~\cite{tsang2000dense, kristensson2015coherent, wang2018achieving, gower2018reflection}.

\subsubsection{Extinction mean free path and effective medium theories}

To get an expression for $\ell_\text{e}$, we need an expression for the self-energy $\tilde{\Sigmaop}$.
Using Eq.~\eqref{eq:sigma_exc-expansion} to the lowest order, we obtain
\be
	\tilde{\Sigmaop}_1 \equiv \bra \tilde{\Tgcal} \ket = \bra \sum_j \tilde{\Tgcal}_j \ket.
\ee
Writing the transition operator of an individual particle as $\Tgcal_j \equiv \T_0(\r-\R_j,\r'-\R_j)$, this can be rewritten as
\be
	\tilde{\bm{\Sigma}}_1 (\r-\r') = \rho \int \T_0(\r-\r_\text{p},\r'-\r_\text{p}) d\r_\text{p}. \label{eq:sigmatilde1}
\ee
%
%
Similarly, the second-order contribution is
\be
	\tilde{\Sigmaop}_2 &\equiv& \bra \Delta \tilde{\Tgcal} \hat{\bm{\mathcal{G}}}_\text{h} \Delta \tilde{\Tgcal} \ket \nonumber \\
	&=& \bra \tilde{\Tgcal}  \hat{\bm{\mathcal{G}}}_\text{h} \tilde{\Tgcal} \ket - \bra \tilde{\Tgcal} \ket  \hat{\bm{\mathcal{G}}}_\text{h} \bra \tilde{\Tgcal} \ket,
\ee
which leads to
\be
	\tilde{\bm{\Sigma}}_2 (\r-\r') &=& \rho \int \left[ \delta(|\r_\text{p}-\r_\text{q}|) + \rho h_2(|\r_\text{p}-\r_\text{q}|) \right] \nonumber \\
	& \times & \T_0(\r-\r_\text{p},\r''-\r_\text{p}) \hat{\G}_\text{h}(\r''-\r''') \nonumber \\
	& \times & \T_0(\r'''-\r_\text{q},\r'-\r_\text{q}) d\r'' d\r''' d\r_\text{p} d\r_\text{q}. \label{eq:sigmatilde2}
\ee
In this expression, we have defined the total pair correlation function
\be \label{eq:h2_r}
    h_2(\r)\equiv g_2 (\r)-1,
\ee
and $g_2$ is defined from Eq.~\eqref{eq:n-particle-correlation}.

To reach a general expression for the self-energy $\bm{\Sigmaop}$ via Eq.~\eqref{sigma_expansion}, we need to define the operator $\gh$ that describes the local propagation of radiation within each particle. As first pointed out by Sullivan and Deutch~\cite{sullivan1976local}, $\gh$ may be chosen to obtain different lowest order results for the effective permittivities and refractive index, such as the Maxwell-Garnett~\cite{bedeaux1973critical, felderhof1974propagation}, Onsager-B\"utcher \cite{onsager1936electric, bottcher1978theory, hynne1987scattering} or Wertheim \cite{wertheim1973dielectric} models. Differences vanish when all orders of the expansion are taken into account, but the rate of convergence of the different formulations is influenced by the particular choice of the Green function~\cite{geigenmuller1986effective, sullivan1976local, bedeaux1973critical, bedeaux1987effective}.

For pedagogical reasons, we restrict ourselves here to the Maxwell-Garnett result obtained in the long-wavelength limit. For some applications dealing with particles with complex shapes and relatively low scattering contrast, one can simplify the problem using the well-known Rayleigh-Gans approximation~\cite{bohren2008absorption} or subsequent generalizations~\cite{acquista1976light}. In this framework, the transition operator can be written as
\be	
	\T_0(\r-\R_j,\r'-\R_j) = k_h^2 \frac{\alpha_\text{p}}{v} \Theta\left[ a-|\r-\R_j| \right] \delta(\r-\r') \Ig \nonumber \\
\ee
with $\alpha_\text{p}$ the particle polarizability. Inserting this expression in Eqs.~\eqref{eq:sigmatilde1} and \eqref{eq:sigmatilde2} straightforwardly leads to an expression for the self-energy that reads in reciprocal space as
\be
	\bm{\Sigma} (\k) &=& k_0^2 \left( \eps_\text{MG} -\eps_\text{h} \right) \Ig + \rho \frac{k_\text{h}^4 \alpha_\text{p}^2}{\eps_\text{h} \alpha_\text{p}} \frac{\partial \epsilon_\text{MG}}{\partial \rho} \nonumber \\
	&\times& \int S(|\k-\k'|) \left| \frac{3 j_1(|\k-\k'|a)}{|\k-\k'|a} \right|^2 \hat{\G}_\text{h}(\k') \frac{d\k'}{(2\pi)^3}. \nonumber \\
\ee
In this expression, $j_n$ is the spherical Bessel function of the first kind and order $n$, $\eps_\text{MG}$ is the Maxwell-Garnett permittivity given by Eq.~\eqref{MG_epsilon} with $\alpha_0 \equiv \alpha_\text{p}$, and
\be
	S(\k) \equiv 1 + \rho h_2(\k),
\ee
is the \textit{static structure factor}, a key quantity for light scattering studies in correlated disordered media.

Following the same steps as those for continuous permittivity random media, assuming non-absorbing materials, we eventually reach a simple expression for the extinction mean free path,
\be \label{eq:extinction_mfp_particles_Sq}
	\frac{1}{\ell_\text{e}} = \frac{2\pi\rho}{k_\text{MG}^4} \int_0^{2k_\text{MG}} F(q) S(q) q \ud q,
\ee
with $q = k_\text{MG}|\u-\u'|$. We have defined here the form factor
\be \label{eq:form-factor}
	F(q) = k_\text{MG}^2 \frac{\ud \sigma}{\ud \Omega} \left( \frac{q}{2k_\text{MG}} \right) f_\text{L}^2 \left| \frac{3 j_1(qa)}{qa} \right|^2,
\ee
and the Rayleigh differential scattering cross-section
\be \label{eq:Rayleigh-diff-sc}
	\frac{\ud \sigma}{\ud \Omega} (k) = k_\text{h}^4 \frac{\alpha_\text{p}^2}{(4\pi)^2} P(k),
\ee
where $P(k)$ is given by Eq.~\eqref{eq:projector-pola}.

Remarkably, the respective contributions of the individual scattering elements, via the form factor $F(q)$, and of their spatial arrangement, via the structure factor $S(q)$, on the extinction (or scattering) strength of the medium are treated independently. Structural correlations act as a weighting function to the optical response of a random assembly of identical scatterers. Physically, the structure factor describes far-field interferences between fields scattered by pairs of particles.

Equation~\eqref{eq:extinction_mfp_particles_Sq} was obtained here in the long-wavelength limit, for small (non-resonant) particles. The more general situation of resonant particles is significantly more difficult to address within the theory for the average field. A heuristic extension of the Maxwell-Garnett approximation to resonant dipolar particles has been proposed by \citet{doyle1989optical} and further analyzed by \citet{grimes1991permeability, ruppin2000evaluation}. The approach provides some understanding to the spectral resonances observed in certain light scattering experiments, but it fails to fulfill a fundamental scaling law between the material and effective permittivities, as observed by~\citet{bohren2009extended}. A more rigorous framework is given by the so-called Coherent Potential Approximation (CPA)~\cite{tsang1980multiple, tsang2001scattering}, which may be seen as a generalization of the approach leading to the Bruggeman mixing rule for continuous permittivity media, as presented above, for particulate media. Considering scattering elements that are not only the particles but also the host medium, one looks for an auxiliary background permittivity $\epsilon_\text{b}$ such that the average transition operator vanishes $\bra \Tgcal \ket = 0$, or equivalently that the background Green function ${\bm{\mathcal{G}}}_\text{b}$ equals the actual averaged Green function $\langle \bm{\mathcal{G}} \rangle$. Different CPA-like models have been developed based on different self-consistent conditions~\cite{soukoulis1994propagation, busch1995transport}. The so-called Energy-density CPA (ECPA) introduced by \citet{busch1995transport}, in particular, stands out from classical effective medium approaches as it focuses on energy transport (described by the average intensity) rather than on wave propagation and attenuation (described by the average field).

\subsubsection{Scattering and transport mean free paths for resonant scatterers}

In this last part, we show that, in the presence of resonant particles, it is more convenient to use the theory for the average intensity.
To this aim, we assume that an effective index can be defined, such that $\text{Re} [n_\text{eff}] = k_\text{r}/k_0$ (but an explicit model for $n_\text{eff}$ is not needed). In the weak extinction limit (i.e., $k_\text{r}\ell_e\gg 1$) and using the expansion of the transition operator $\Tgcal$ in Eq.~\eqref{eq:T-operator-particles}, the intensity vertex given by Eq.~\eqref{eq:gamma} can be reduced to its lowest order terms as
\be
   \Gammaop &\sim& \bra\Tgcal\Tgcal^*\ket-\bra\Tgcal\ket\bra\Tgcal^*\ket \nonumber \\
   &\sim& \bra \sum_{i,j} \Tgcal_i \Tgcal_j^* \ket - \bra \sum_i \Tgcal_i \ket \bra \sum_j \Tgcal_j^* \ket. \label{eq:expansion-gammaop}
\ee
In Fourier space, this leads to
\begin{multline}\label{eq:th_gamma}
	\bar{\bm{\Gamma}} (k_\text{r}\u,k_\text{r}\u',k_\text{r}\u,k_\text{r}\u') =
	   \rho\T_0(k_\text{r}\u,k_\text{r}\u') \otimes \T_0^*(k_\text{r}\u,k_\text{r}\u')
\\\times
      S(k_\text{r}(\u-\u')).
\end{multline}
In the radiative transfer limit, taking the on-shell approximation and using Eqs.~\eqref{eq:l_s}-\eqref{eq:phase_function_normalization} leads to
\be \label{eq:th_scattering_length_correlations}
   \frac{1}{\ell_\text{s}} = \frac{\rho}{16\pi^2} \int \left| \P(\u) \T_0(k_\text{r}\u,k_\text{r}\u') \e' \right|^2 S(k_\text{r}(\u-\u')) \ud \u \nonumber \\
\ee
where $\e'$ is the polarization vector perpendicular to $\u'$. This expression is valid for a spherical particle of arbitrary size. Also note that $\T_0$ is the transition operator of the particle in the host medium, evaluated for the incident and scattered wavevectors in the \textit{effective} medium (i.e., at the wavenumber $k_\text{r}$). Equation~\eqref{eq:th_scattering_length_correlations} can eventually be reformulated using the form factor
\be
    \label{eq:form_factor_intensity}
	F(q)= k_\text{r}^2 \frac{\ud \sigma}{\ud \Omega}(q),
\ee
where $q = k_\text{r} |\u-\u'|$ and the differential scattering cross-section is now defined as
\be
   \frac{\ud \sigma}{\ud
   \Omega}(q) = \frac{1}{16\pi^2}\left|\P(\u) \T_0(k_\text{r} |\u-\u'|)\e'\right|^2.
\ee
This leads to
\be \label{eq:th_scattering_length_structure_factor}
	\frac{1}{\ell_\text{s}} = \frac{2\pi\rho}{k_\text{r}^4} \int_0^{2k_\text{r}} F(q) S(q) q \ud q.
\ee
We observe that the approaches based on the average field and the average intensity lead to the same final expressions [Eqs.~\eqref{eq:extinction_mfp_particles_Sq} and \eqref{eq:th_scattering_length_structure_factor},
respectively] for non-absorbing media.

A similar derivation using Eqs.~\eqref{eq:l_s}, \eqref{eq:l_t} and \eqref{eq:g} finally leads to a closed-form expression of the transport mean free path,
\be\label{eq:th_transport_length_structure_factor}
   \frac{1}{\ell_\text{t}} = \frac{\pi\rho}{k_{\text{r}}^6} \int_0^{2k_\text{r}} F(q) S(q) q^3 \ud q.
\ee
Equations~\eqref{eq:th_scattering_length_structure_factor} and \eqref{eq:th_transport_length_structure_factor} seem to be the most widely-used expressions in the literature on light scattering and transport in correlated disordered media~\cite{fraden1990multiple, saulnier1990scatterer, rojas2004photonic, reufer2007transport}. These two expressions, originally obtained from phenomenological arguments, have been derived here within a rigorous theoretical framework.

\subsection{Summary and further remarks}\label{subsec:theory-summary}

\begin{table*}[t]
\centering
\begin{tabular}{|>{\centering\arraybackslash} m{2.5cm} || >{\centering\arraybackslash}m{7cm} | >{\centering\arraybackslash}m{7cm} | } 
 \hline
  & Scattering (= Extinction) &  Transport \\ 
 \hline \hline
 Fluctuating permittivity media
 & \be \frac{1}{\ell_\text{s}}=\frac{k_0^4}{8\pi k_\text{r}^2} \epsilon_\text{b}^2 \Delta_x^2
      \int_0^{2k_\text{r}} P(q/2 k_\text{r}) h_{x}(q) q \ud q \nonumber \ee
 & \be \frac{1}{\ell_\text{t}} = \frac{k_0^4}{16\pi k_\text{r}^4} \epsilon_\text{b}^2 \Delta_x^2
      \int_0^{2k_\text{r}} P(q/2 k_\text{r}) h_{x}(q) q^3 \ud q \nonumber \ee \\
 \hline
 Monodisperse particulate media
 & \be \frac{1}{\ell_\text{s}} &=& \frac{2\pi\rho}{k_\text{r}^4} \int_0^{2k_\text{r}} F(q) S(q) q \ud q \nonumber \\
 	&=& \rho \int_{4\pi} \frac{\ud \sigma}{\ud \Omega}(\theta) S(\theta) \ud \Omega \nonumber \ee
 & \be \frac{1}{\ell_\text{t}} &=& \frac{\pi\rho}{k_{\text{r}}^6} \int_0^{2k_\text{r}} F(q) S(q) q^3 \ud q \nonumber \\
 	&=& \rho \int_{4\pi} \frac{\ud \sigma}{\ud \Omega}(\theta) S(\theta) (1- \cos \theta) \ud \Omega  \nonumber \ee \\ 
 \hline
\end{tabular}
\caption{Analytical expressions for the scattering and transport mean free paths, $\ell_\text{s}$ and $\ell_\text{t}$, in non-absorbing media with correlated disorder. These expressions were obtained from the average field and/or the average intensity, using different models (fluctuating permittivity or identical particles). $\Delta_x$ and $h_x$ describe the amplitude of the fluctuation and the two-point correlation function of 
material descriptor $x$, respectively, $\epsilon_\text{b}$ is the auxiliary background permittivity, and $k_\text{r}$ is wavenumber associated to the real part of the effective index. Hence, for weakly fluctuating media [Eqs.~\eqref{eq:ext_mfp_weak_fluctuations}, \eqref{eq:th_scattering_length_q_continuous} and \eqref{eq:th_transport_length_q_continuous}], one should set $\Delta_x = \delta_\epsilon$, $h_x = h_\epsilon$, $\epsilon_\text{b} = \epsilon_\text{av}$ and $k_\text{r} = k_0 \sqrt{\epsilon_\text{av}}$. For strongly fluctuating media, one should set $\Delta_x = \delta_\alpha$, $h_x = h_\alpha$, $\epsilon_\text{b} = \epsilon_\text{BG}$ and $k_\text{r} = k_0 \sqrt{\epsilon_\text{BG}}$, when using the Bruggeman mixing rule [Eq.~\eqref{eq:ext_mfp_Bruggeman}], or $\Delta_x = f_\text{L} \delta_\alpha$ with $f_\text{L} = (\partial \epsilon_\text{MG}/\partial \rho)/(\epsilon_\text{h} \alpha_0)$, $h_x = h_\alpha$, $\epsilon_\text{b} = \epsilon_\text{h}$ and $k_\text{r} = k_0 \sqrt{\epsilon_\text{MG}}$, when using the Maxwell-Garnett mixing rule [Eq.~\eqref{eq:ext_mfp_MaxwellGarnett}]. For monodisperse particulate media [Eqs.~\eqref{eq:extinction_mfp_particles_Sq}, \eqref{eq:th_scattering_length_structure_factor} and \eqref{eq:th_transport_length_structure_factor}], the form factor $F(q)$ can be expressed directly in terms of the differential scattering cross-section $\frac{\ud \sigma}{\ud \Omega}(q)$ [Eq.~\eqref{eq:form_factor_intensity}], leading to the simple expressions given on the last line, see also Eqs.~\eqref{eq:ls_theta} and \eqref{eq:lt_theta} in Sec.~\ref{sec:5-glasses-insight}. Remarkably, the fact to recover the same expressions with different approaches highlights a fundamental relation between structural correlations and light scattering and transport, that is independent of the choice of a specific effective medium approximation.}
\label{tab:mfp_expressions}
\end{table*}

The literature on multiple light scattering theory is vast and many approaches have been developed throughout the years. We adopted here a unique theoretical framework that can handle both families of systems described by a continuous permittivity that randomly fluctuates in space or by a set of identical particles that are randomly arranged in space. This has the great benefit of highlighting the main physical principles behind the role of spatial correlations on light scattering and transport, as well as the underlying approximations. Analytical expressions for the characteristic lengths were obtained in the weak extinction limit ($k_\text{r} \ell_\text{e} \gg 1$, with $k_\text{r}$ the effective wave number and $\ell_\text{e}$ the extinction mean free path), for statistically homogeneous, isotropic and non-absorbing media. These final expressions are provided in Table~\ref{tab:mfp_expressions} in their most general form.

The model for continuous permittivity media presented in Sec.~\ref{subsec:random} does not make any specific assumption on the size or shape of a particular scattering element, and may thus be applied to different types of microstructures. Several expressions for the scattering mean free path $\ell_\text{s}$ were derived from the average field or from the average intensity, assuming or not weak fluctuations, taking or not the long-wavelength limit [Eqs.~\eqref{eq:ext_mfp_weak_fluctuations}, \eqref{eq:ext_mfp_Bruggeman}, \eqref{eq:ext_mfp_MaxwellGarnett} and \eqref{eq:th_scattering_length_q_continuous}]. All expressions eventually take the same form, reported in Table~\ref{tab:mfp_expressions}, thereby unveiling the fundamental relation between structural correlations and light scattering and transport. In the case of strong fluctuations [Eqs.~\eqref{eq:ext_mfp_Bruggeman} and \eqref{eq:ext_mfp_MaxwellGarnett}], the expression to use depends on the choice of the ``reference'' homogeneous medium around which the permittivity fluctuates: the former is associated to the Bruggeman mixing rule and the latter to the Maxwell-Garnett mixing rule. The common ground that links these two approaches, as discussed for instance by~\citet{mackay2020modern}, is often overlooked.

The model for particulate media presented in Sec.~\ref{subsec:particulate} applies specifically to disordered assemblies of \textit{identical} particles, wherein the contributions of the individual particles (possibly exhibiting a resonant behavior) and of the particle spatial arrangement on light scattering can be formally separated. A first expression for the scattering mean free path $\ell_\text{s}$ was obtained from the average field in the long-wavelength limit [Eq.~\eqref{eq:extinction_mfp_particles_Sq}]. A comparison with Eq.~\eqref{eq:ext_mfp_MaxwellGarnett} unveils a fundamental concept, namely that a fluid of tiny identical particles with a fluctuating density (e.g., a fluid of molecules) can be assimilated to a continuous medium with a fluctuating permittivity (or polarizability).
Indeed, inserting Eqs.~\eqref{eq:form-factor} and \eqref{eq:Rayleigh-diff-sc}
 in Eq.~\eqref{eq:extinction_mfp_particles_Sq}, taking the limit $qa \rightarrow 0$ (very small particles) and comparing the resulting expression for $1/\ell_\text{e}$ with Eq.~\eqref{eq:ext_mfp_MaxwellGarnett} leads to the equivalence $\delta_\alpha^2 h_\alpha(q) \equiv \alpha_\text{p}^2 \rho S(q)$ in reciprocal space. In real space, using Eq.~\eqref{def_S_r} in Appendix~\ref{subsec:fluctNumber} leads to
\be
\bra \Delta \rho (\r) \Delta \rho (\r') \ket \alpha_\text{p}^2 \equiv \bra \Delta \alpha (\r) \Delta \alpha (\r') \ket/v^2,
\ee
with $\Delta \rho (\r) = \rho(\r) - \langle \rho (\r) \rangle$, the particle density fluctuation around the average density ($\langle \rho (\r) \rangle \equiv \rho$ using our notation). This is the reason why Rayleigh's and Einstein's quantitative results on the mean free paths (explaining the blue color of the sky) were essentially identical~\cite{rayleigh1899xxxiv, einstein1910theorie}.

A second yet identical expression for $\ell_\text{s}$ was obtained from the average intensity [Eq.~\eqref{eq:th_scattering_length_structure_factor}] under the assumption that a solution for the real part of the effective index exists. The correspondance between the two expressions highlights again how structural correlations impact light scattering, independently of the effective medium description. Importantly, this and the expression for the transport mean free path $\ell_\text{t}$ [Eq.~\eqref{eq:th_transport_length_structure_factor}] are not restricted to the long-wavelength limit, thereby explaining their popularity in studies on resonant scattering and transport in photonic liquids and glasses [Sec.~\ref{sec:5-glasses}].

To reach the expressions reported in Table~\ref{tab:mfp_expressions}, we assumed the disordered media to be non-absorbing everywhere in space, $\epsilon(\r) \in \mathbb{R}$. This makes that extinction is driven by scattering, leading to $1/\ell_\text{e} = 1/\ell_\text{s}$. To end this section, let us motivate this initial choice and explain how absorption might change the picture.

In practice, absorption reduces the number of scattering events a wave can experience in a disordered medium and hinders wave interference phenomena between multiply-scattered waves. Although the absorptance of a medium has been proposed as a means to identify Anderson localization of light~\cite{john1984electromagnetic}, the mesoscopic optics community has mostly been interested in the study of strongly scattering materials with negligible absorption, which has naturally led to considering high-index dielectrics like Si and Ge in the near-infrared range, and TiO$_2$ in the visible range. Many experiments have also been made with lower-index materials with negligible absorption in the visible range, including polymers (polystyrene, PMMA) -- well-suited to direct laser writing (see Sec.~\ref{sec:3-fabrication}) -- and SiO$_2$. Thus, our assumption of non-absorbing media is valid in most cases of interest.

The theoretical problem of multiple light scattering in presence of absorption has received relatively little attention compared to its non-absorbing counterpart. Generally speaking, the extinction length in a scattering and absorbing medium is simply given by $1/\ell_\text{e} = 1/\ell_\text{s} + 1/\ell_\text{a}$, with $\ell_\text{a}$ the absorption mean free path, that yet remains to be determined.

For particulate media, the absorption may come from the particles themselves, in which case it is incorporated in the transition operator $\mathbf{T}_0$ of the individual particle, and from the host medium, in which case the wavenumber $k_\text{h}$ should have a non-zero imaginary part due to absorption. Initial efforts to take correlations into account were made in the 1990s~\cite{kumar1990dependent, ma1990enhanced}, but were limited to very small (Rayleigh) particles. A multiple-scattering model for the average field in assemblies of spherical particles of arbitrary size has been developed by \citet{durant2007light}, leading to analytical expressions of the effective wavenumber (directly related to the extinction mean free path) as a function of the pair correlation function $g_2(r)$. The approach has later been generalized by \citet{mishchenko2008multiple} to arbitrary particulate media, including non-spherical particles and polydispersity. The theory unfortunately does not provide analytical expressions for the absorption mean free path as a function of correlations. This limitation can be lifted via a model for the average intensity, as shown, though for dipolar particles only, by~\citet{wang2018effect} under the QCA. The authors of the latter study find that short-range correlations have a weaker effect on the absorption mean free path than on the scattering mean free path.

The case of fluctuating permittivity media with absorption has been tackled only recently, to our knowledge, by~\citet{sheremet2020absorption}, who relied a diagrammatic description of multiple scattering for the average field and average intensity to reach an analytical expression for the absorbed power as a function of structural correlations. Although the theory is made specifically for scalar waves in fluctuating media with short-range correlations, it leads to the same conclusion as above on the impact of correlations on the scattering and absorption lengths.

\section{Structural properties of correlated disordered media} \label{sec:3}

Correlated disordered media can exhibit a rich variety of complex morphologies, which will impact light scattering and transport in many different ways. This section is concerned with the statistical description of the structural properties of these materials. For a more complete and thorough description, we recommend the textbook by \citet{torquato2013random}.

After comparing the quantities describing spatial correlations and derived in the previous section on realistic systems [Sec.~\ref{sec:3-continuous-vs-particulate}] and introducing a fundamental relation between fluctuations of particle number and spatial correlations in point patterns [Sec.~\ref{sec:3-fluctuation-correlation}], we define the main classes of correlated disordered media according to their pair correlation function $g_2(r)$ and structure factor $S(q)$ [Sec.~\ref{sec:3-classes}]. We then review the main techniques to numerically generate correlated disordered materials and simulate their optical properties [Sec.~\ref{sec:3-numerical}]. Experimental fabrication techniques for correlated disordered media are then presented [Sec.~\ref{sec:3-fabrication}]. The section is concluded with a brief summary of the experimental techniques to characterize structural correlations in real materials [Sec.~\ref{sec:3-measuring}].

\subsection{Continuous permittivity versus particulate models in practice} \label{sec:3-continuous-vs-particulate}
	
As shown in the previous section, light scattering in correlated disordered media can be described either by a continuous permittivity model that relies on a spatial permittivity correlation function or by a particulate model that relies on a two-point correlation function in the specific case of localized permittivity variations. To start this section, it is interesting to compare these two pictures in practical cases.
	
\begin{figure}[htbp]
    \centering
	\includegraphics[width=8cm]{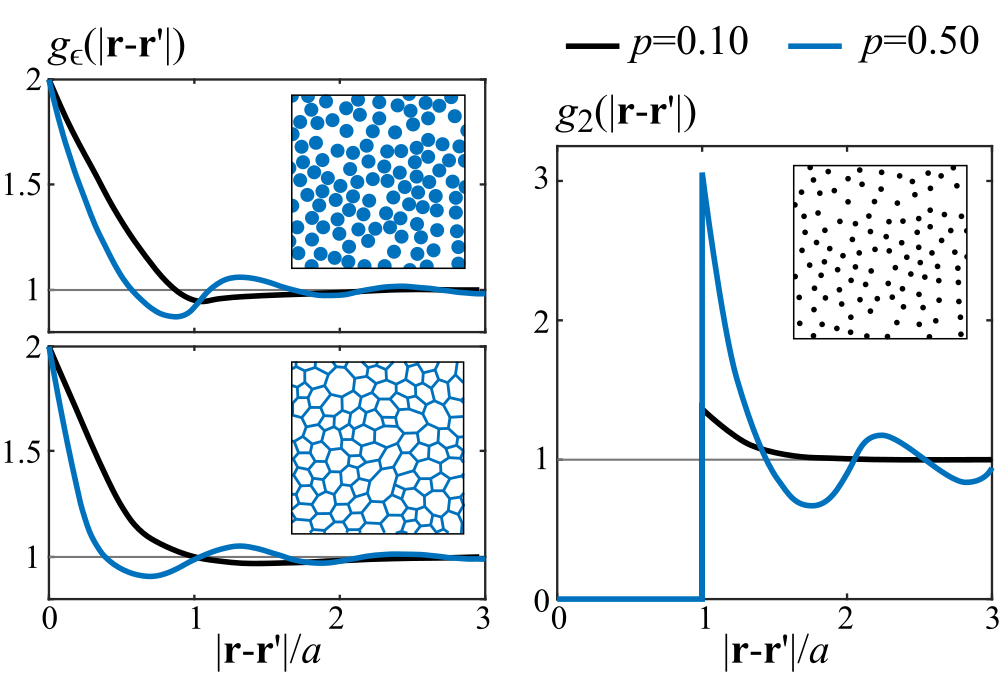}
	\caption{\label{fig:disk_packing} (Color online) Description of structural correlations for hard disk (diameter $a$) packings at packing fractions $p=0.10$ (black curves) and $p=0.50$ (blue (gray) curves). (Left) Correlation function $g_{\epsilon}(r) = h_\epsilon(r) + 1$ describing correlations due to the finite-size of the disks and to their positional correlation, for two different topologies: (top) a packing of disks and (bottom) a continuous network. (Right) Pair correlation function $g_2(r) = h_2(r) + 1 $, which describe only the positional correlation between the disk centers.}
\end{figure}

We consider a classical and very relevant example for photonics, that is a 2D assembly of impenetrable disks (diameter $a$) at two different packing fractions $p = 0.10$ and $0.50$, see Fig.~\ref{fig:disk_packing}. The disk packings were generated with a compression algorithm~\cite{skoge2006packing} that will be briefly described in Sec.~\ref{sec:3-numerical}. The top-left panel of Fig.~\ref{fig:disk_packing} shows the permittivity correlation function $g_\epsilon(|\mathbf{r}-\mathbf{r}'|) = h_\epsilon(|\mathbf{r}-\mathbf{r}'|)+1$ for the disk packings.
This correlation function was introduced to describe media with a fluctuating continuous permittivity, but can in fact be applied to any type of microstructure given its generality. The function first displays a rapid decrease of correlation followed by regular, vanishing oscillations. The very short-range correlation here is mostly associated to the finite size of the disks and the oscillations, which are stronger for higher packing fractions, are mostly a signature of correlations between neighboring particles. The difficulty to formally distinguish distinct types of correlations is a downside of the generality of $h_\epsilon$.
	
The generated disk positions can also serve as a basis to generate more complex structures. An example often encountered in photonics is based on Delaunay tesselations (described in Sec.~\ref{sec:3-numerical}). In the bottom-left panel of Fig.~\ref{fig:disk_packing}, we show the permittivity correlation function for these ``inverted'' structures, as generated from the disk packing above. Strikingly, the behavior of the correlation functions are very similar to those of the disk packing. The major difference is observed at small distances due to a very different morphology, but the curves become indistinguishable for $|\mathbf{r}-\mathbf{r}'| \gtrsim a$.
	
Finally, we can consider the pair-correlation function $g_2(|\mathbf{r}-\mathbf{r}'|)$, applicable specifically to ensembles of identical scatterers. The results are shown in the right panel of Fig.~\ref{fig:disk_packing}. The pair-correlation function for the disk packing is zero for $|\mathbf{r}_a-\mathbf{r}_b|$ between $0$ to $2R$ due to the impenetrability of the disks, and exhibits strong oscillations indicating structural correlations in the relative position between particles. Note that the amplitude of the oscillations is much larger than for $ g_{\epsilon}(|\mathbf{r}-\mathbf{r}'|)$.
	
This simple comparison shows that the use of the two-point permittivity correlation is hardly sufficient to distinguish different types of correlated disordered media. In fact, the permittivity correlation of the inverted disk structure at $p=0.50$ would be strictly identical to that of the direct structure by definition, while light scattering would evidently be markedly different. This is a strong indication that scattering is both dramatically affected by the local morphology of the system, which yields optical resonances, and by structural correlations in the relative position between scattering elements. Since a ``scattering element'' is not well defined for inverted structures such as connected networks, for clarity and simplicity, we will focus on the particulate description of scattering media in the remainder of this section.

\subsection{Fluctuation-correlation relation} \label{sec:3-fluctuation-correlation}
    
The description of point patterns underlying the structure of correlated disordered media is central, and many descriptors may be used in general. 
    

An important attribute of point patterns is the variance of the number $N$ of points contained within a window $\Omega$ with volume $V$. This quantity has a long history, several derivations are found for both continuous and discrete disorder models~\cite{ornstein1914accidental, de1949molecular, van1977v, landau1980statistical, martin1980charge, torquato2003local}. Considering a spherical window of radius $R$ for simplicity, probabilistic calculations eventually lead to a closed-form expression for the variance of $N$~\cite{torquato2003local}
\be \label{eq:fluctuation-correlation}
    \frac{\langle N^2 (R) \rangle - \langle N (R) \rangle^2}{\langle N (R) \rangle} = 1 + \rho \int h_2(\mathbf{r}) \Lambda(\r;R) d\mathbf{r},
\ee
where $\Lambda(\r,R) = v_2^\text{int}(\r;R)/V$ is the intersection volume $v_2^\text{int}(\r;R)$ of two windows separated by $\r$ normalized to the window volume $V=\frac{4}{3}\pi R^3$.
In the limit of large windows, one finds that
\be
    \lim_{R \rightarrow \infty} \frac{\langle N^2 (R) \rangle - \langle N (R) \rangle^2}{\langle N (R) \rangle} &=& \lim_{|\mathbf{q}| \rightarrow 0} S(\mathbf{q}), \label{eq:fluctuation-correlation-large} \\
    &=& 1 + \rho \int h_2(\mathbf{r}) d\mathbf{r}, \label{eq:fluctuation-correlation-large-real}
\ee
which corresponds to the simplified definition given in Appendix~\ref{subsec:fluctNumber}.

\begin{figure}[htbp]
    \centering
	\includegraphics[width=8cm]{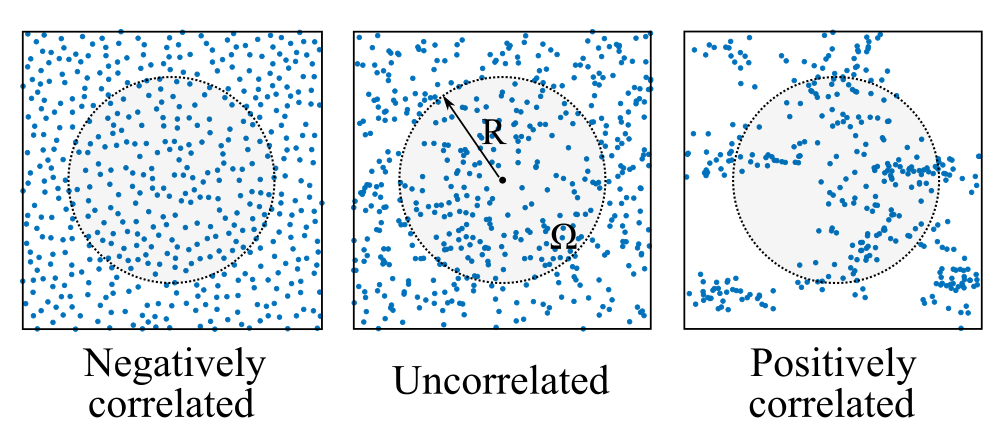}
	\caption{\label{fig:fluctuation-correlation} (Color online) Illustration of the fluctuation-correlation relation with three point patterns: (left) A negatively-correlated disordered medium, where points tend to repel themselves; (middle) A Poisson point pattern, that is uncorrelated; (right) A positively-correlated disordered medium, wherein clustering is present. The variance of the number of points in a spherical window $\Omega$ of radius $R$ is related to the total pair correlation function $h_2$ via Eq.~\eqref{eq:fluctuation-correlation}.}
\end{figure}

Equations~(\ref{eq:fluctuation-correlation})-(\ref{eq:fluctuation-correlation-large-real}) are remarkable in that they describe the spatial fluctuations in the number of points in the pattern from its pair correlation between points -- or equivalently, its structure factor near 0, which is a measurable quantity (e.g., by small-angle scattering, see Sec.~\ref{sec:3-measuring}). A Poisson point pattern $p_N = \langle N \rangle^N \exp \left[ - \langle N \rangle \right]/N!$ yields $\langle N^2 \rangle = \langle N \rangle + \langle N \rangle^2$, which, as expected, corresponds to a fully uncorrelated system with $h_2(\mathbf{r})=0$ or $\lim_{|\mathbf{q}| \rightarrow 0} S(\mathbf{q}) = 1$. Implementing structural correlations at constant density $\rho$ therefore results into weaker or stronger point density fluctuations. Negative correlations are obtained when $\rho \int h_2(\mathbf{r}) d\mathbf{r} <0$, leading to a sub-Poissonian fluctuations, while positive correlations are obtained when $\rho \int h_2(\mathbf{r}) d\mathbf{r} > 0$, leading to a super-Poissonian fluctuations. In the literature, such structures are sometimes denoted as negatively- and positively-correlated, respectively~\cite{davis2008radiation}. As illustrated in Fig.~\ref{fig:fluctuation-correlation}, they correspond to situations in which the points either repel or attract themselves. As we will see in Sec.~\ref{sec:5}, the impact of negative and positive correlations on optical transport are markedly different.

\subsection{Classes of correlated disordered media} \label{sec:3-classes}

\begin{figure*}[htbp]
\centering
\includegraphics[width=0.9\textwidth]{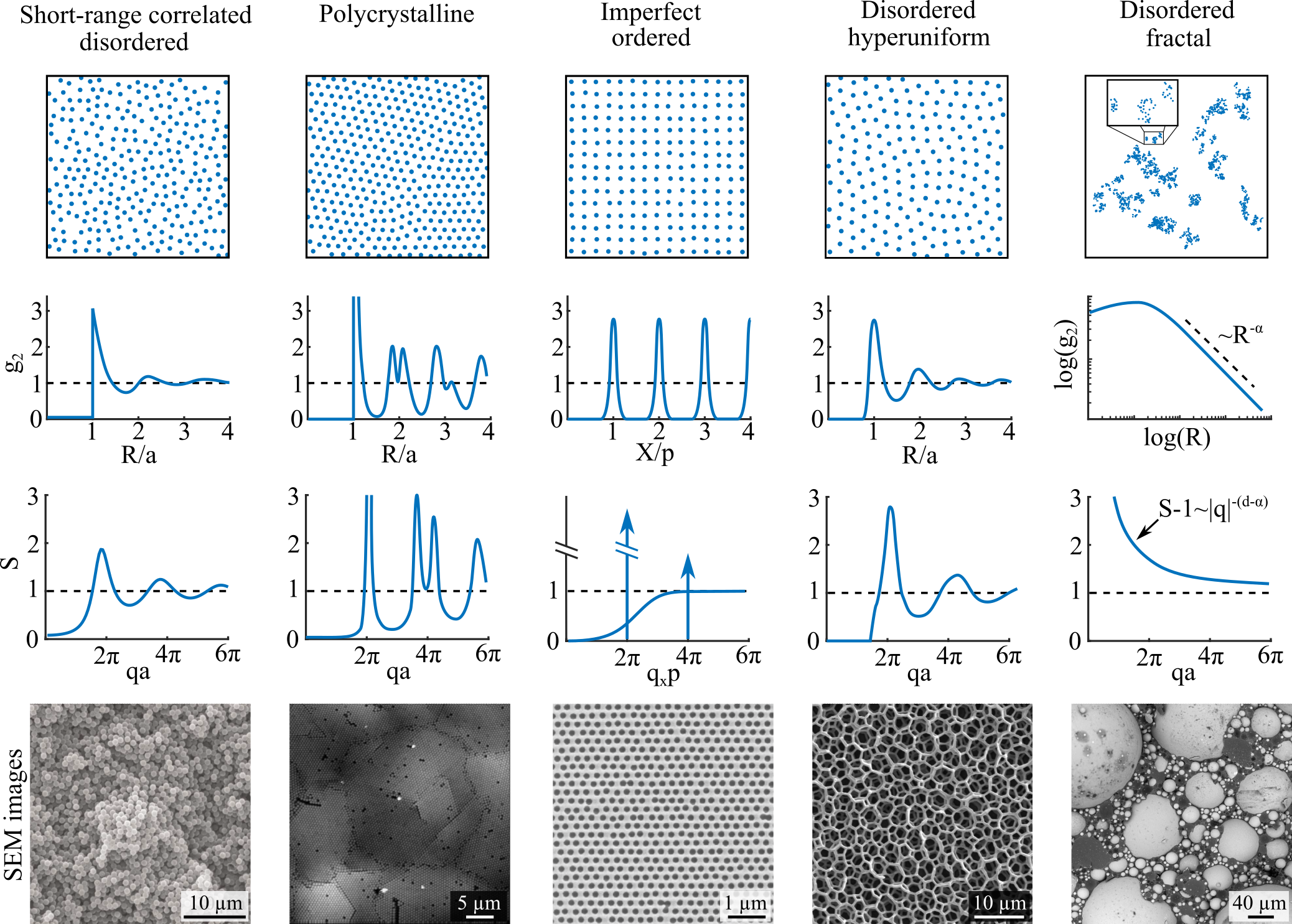}
\caption{\label{fig:classes_disordered_media} (Color online) Classes of correlated disordered media. (From top to bottom) Illustration of a correlated disordered medium; Pair correlation function; Structure factor; SEM image of a fabricated correlated disordered structure. (From left to right) Disordered short-range correlated structures, SEM image adapted with permission from \cite{garcia2008resonant}; Polycrystalline structure, SEM image adapted with permission from \cite{salvarezza1996edward}; Imperfect ordered structures, SEM image adapted with permission from \cite{garcia2012nonuniversal}; Disordered hyperuniform structures, SEM image adapted with permission from \cite{haberko2013fabrication}; Disordered hierarchical structures, SEM image adapted with permission from \cite{burresi2012weak}.}
\end{figure*}

	
Figure~\ref{fig:classes_disordered_media} summarizes the most important classes of correlated disordered media and their properties, that we will now describe specifically. Note that this panel is non-exhaustive -- other families of correlated disordered media exist, such as paracrystals~\cite{hosemann1963crystalline}, characterized by regular point patterns deformed on scales typically larger than the distance between neighboring points. Our focus here is on the classes that led to a substantial body of work in optics and photonics.
	
\subsubsection{Short-range correlated disordered structures}

Consider a volume containing a disordered ensemble of mobile, impenetrable particles (i.e., a fluid of hard particles) at a low density. With increasing particle density, the particles tend to organize themselves to fill space. In this regime of low to moderate densities, the system exhibits no structural correlation in the long range yet the impenetrability of the particles impose a short-range correlation that increases with the packing fraction~\cite{hansen1990theory}. As shown in Fig.~\ref{fig:classes_disordered_media} (1st column), short-range structural correlations give rise to decaying oscillations in the pair correlation function $g_2$. The most likely distance to find a neighboring particle is given by the position of the first peak and the decay of the higher-order peaks, which is generally rapid, allows defining a correlation length. In reciprocal space, such oscillations are also observed. When increasing short-range correlations, the structure factor goes from a flat response around 1 to sharper peaks whose amplitudes decrease with increasing $q$. Short-range structural correlations can be described, for instance, by analytical or semi-analytical solutions of the Ornstein-Zernike equation using the so-called Percus-Yevick approximation~\cite{percus1958analysis, wertheim1963exact} in three dimensions and the Baus-Colot approximation in two dimensions~\cite{baus1987thermodynamics}, respectively. At higher densities, these models become less accurate, although, for slightly polydisperse systems, errors appear to cancel, and predictions by Percus-Yevick approximation can describe experimental data up to random close packing or jamming~\cite{frenkel1986structure, scheffold2009scattering}.

Short-range correlated disordered systems constitute the primary class found in colloidal suspensions with isotropic interactions, since short-range correlations stem from the impenetrability of particles in suspension. The behavior of other repulsive particles, such as charge-stabilized particles, can often be mapped onto the isotropic hard-sphere case~\cite{pusey1986phase, gast1998simple}. The recent advent of colloids interacting via sticky patches could open a pathway towards more complex structures through self-assembly~\cite{he2020colloidal}. 

In general, short-range structural correlations are not limited to sphere assemblies but can also be encoded in connected networks \cite{florescu2009designer, liew2011photonic, muller2013silicon}, in which case the individual scattering centers are more difficult to identify. 
This form of correlated structure is widespread in natural photonic structures such as bird feathers~\cite{saranathan2012structure} and very popular in artificial photonic structures fabricated by top-down techniques~\cite{liew2011photonic, muller2013silicon, muller2017photonic}. Dry foams are promising candidates for correlated network structures that can be made by self-assembly~\cite{ricouvier2019foam, klatt2019phoamtonic, maimouni2020micrometric}. 

\subsubsection{Polycrystalline structures}

For a disordered ensemble of identical hard particles, one reaches a liquid-crystal coexistence at about 49\% and a purely crystalline phase for concentrations above 54.5\% \cite{pusey1986phase, pusey1991liquids, gast1998simple, zhu1997crystallization}. The equilibrium structure appears to be face-centered-cubic but hexagonal close packed structures are also observed and are found to be at least meta-stable~\cite{pusey1989structure}. This liquid to crystal phase transition, also known as the Kirkwood-Alder transition~\cite{gast1998simple}, is purely driven by the higher entropy of the crystalline phase compared to the liquid phase. The densest packing of monodisperse spheres in three dimensions is approximately 74\%, also referred to as the close-packing of equal spheres. Monodisperse particles usually assemble in finite-sized crystal clusters. These clusters are randomly arranged and form a polycrystalline material~\cite{astratov2002interplay, yang2010photonic}, see Fig.~\ref{fig:classes_disordered_media} (2nd column). 
In the bulk, crystallites are formed by homogeneous nucleation throughout the sample~\cite{pusey1989structure}. The size of the crystal clusters is then typically several tens of $\mu$m, much larger than the particle diameter $\sim \lambda$ but smaller than the usual sample size, which is typically in the millimeter to centimeter range. The radially-averaged pair correlation function exhibits peaks indicating the position of the $n$th-order neighboring particles, as well as minima approaching zero. Positional correlations vanish for distances exceeding the size of the crystal clusters. Similarly, the structure factor shows well-defined Debye-Scherrer rings due to Bragg scattering from randomly oriented crystal planes, that can be identified in light scattering~\cite{pusey1989structure}, similarly to powder diffraction in X-ray crystallography. 
	
The formation of clusters of regular arrays in fluids of hard particles is strongly influenced by the polydispersity of the particles, since particles of very different sizes do not naturally arrange in a crystal. Indeed, for hard sphere fluids with a polydispersity larger than 6-12\%, crystallization is avoided in three dimensions~\cite{pusey1987effect}. The spheres remain disordered and particles enter a solid glass phase at about 58\%. The glass can be further compressed until the spheres 'jam' forming what is known as a ``randomly closed packed'' or ``maximally jammed structure'' in the literature~\cite{torquato2000random}. 
The presence of some hidden structural order, crystalline precursors or locally favoured structures in the glass and jammed phase is still being discussed~\cite{zhang2016dynamical}.

Due to the unavoidable finite polydispersity, experimental realizations of crystalline photonic structures based on colloidal suspensions are the exception rather than the rule even at high packing fractions. 
By careful synthesis of colloidal particles made from polystyrene or silica (SiO$_2$), it is however possible to induce crystallization rather easily~\cite{salvarezza1996edward}. These materials are usually polycrystalline and display some defects and stacking faults to a varying degree. Polycrystalline structures are also observed in natural photonic structures, such as opals. Interestingly, relatively little is known about the comparison of scattering and light transport between random-close-packed assemblies of spheres and polycrystalline materials~\cite{yang2010photonic}, in particular when the size of crystallites is gradually reduced to smaller length scales.

\subsubsection{Imperfect ordered structures}

The two previous classes of correlated disorder were obtained by ``adding order'' into a fully-disordered (uncorrelated) system. Materials with correlated disorder can also be obtained starting from the other limit, that is a periodic system with random perturbations, see Fig.~\ref{fig:classes_disordered_media} (3rd column). In systems of infinite size, both the pair-correlation function $g_2$ and the structure factor $S$ are characterized by a series of Dirac peaks located at $\mathbf{r}-\mathbf{r}'=u_1 \mathbf{a}_1+u_2 \mathbf{a}_2+u_3 \mathbf{a}_3$ with $u_i \in \mathbb{Z}$ and $\mathbf{a}_i$ the lattice vectors, and $\mathbf{G}=v_1 \mathbf{b}_1+v_2 \mathbf{b}_2+v_1 \mathbf{b}_2$ with $v_i \in \mathbb{Z}$ and $\mathbf{b}_i$ the reciprocal lattice vectors, respectively~ \cite{kittel1976introduction, joannopoulos2011photonic}. If the position of a point of the lattice is randomly shifted (e.g., with normal distribution) around its nominal position, this results in a broadening of the Dirac peaks with a width that depends on the disorder amplitude. By contrast with the previous classes of disordered systems, disorder in such a periodic-on-average structure does not impact the correlation length, which remains infinite. A striking consequence of this is that the structure factor is characterized by Dirac peaks of vanishing width (for systems of infinite size) and decreasing amplitude with increasing wavenumber $q$ on top of a diffuse background. The latter, which is due to the random nature of the point pattern, equals 1 for large values of $q$ and quadratically goes to zero when $q$ goes to zero, as discussed, for instance, by~\citet{klatt2020cloaking}. Real systems are however never exactly periodic, due to fabrication imperfections and in practice this also leads to a finite correlation length~\cite{meseguer2002synthesis, lopez2003materials, koenderink2005optical, nelson2011epitaxial}.

A plethora of studies of ordered photonic crystal structures with imperfections, both numerical and experimental, can be found in the literature~\cite{soukoulis2012photonic}. Defects were also added intentionally, either at random or selected positions, to study the interplay between defect states, density of states, wave tunneling and percolation, random diffuse scattering, and directed Bragg scattering of light~\cite{florescu2010effects, garcia2009strong, aeby2020scattering}. Moreover, the interaction between the band structures and defect scattering is fascinating, since it might lead to other critical coherent transport phenomena such as Anderson localization of light~\cite{john1987strong}. Defect states can also be introduced in a photonic crystal to deliberately implement a particular function, such as optical sensing applications, lasing, or optical circuitry~\cite{joannopoulos2011photonic, soukoulis2012photonic, nelson2011epitaxial}. 

\subsubsection{Disordered hyperuniform structures}\label{HUsec}

One of the important characteristics of point patterns is how the number of points contained in a given volume fluctuate with various disorder realizations~\cite{torquato2013random}. This quantity is related to the notion of spatial uniformity. For a Poisson point process, one shows from Eq.~\eqref{eq:fluctuation-correlation} that the variance in the number of points $N$ contained in a $d$-dimensional sphere of radius $R$ grows as the sphere volume (i.e., $\langle N^2 \rangle - \langle N \rangle^2 = \langle N \rangle \sim R^d$). This result holds for many disordered point patterns. By contrast, the same analysis performed on a periodic pattern shows that the variance grows with the surface of the sphere, $\langle N^2 \rangle - \langle N \rangle^2 \sim R^{d-1}$. In a founding work, \citet{torquato2003local} proposed to define a general class of point patterns, dubbed ``hyperuniform'', the property of which is to exhibit point number fluctuations scaling as the surface of the window, that is slower than expected for usual disordered media. Hyperuniformity encompasses periodic, quasi-periodic but also -- very interestingly in the framework of this review -- a subclass of disordered systems, see Fig.~\ref{fig:classes_disordered_media} (4th column) for an illustration and \cite{torquato2018hyperuniform} for a recent review. It was observed numerically that maximally jammed packings of spheres and platonic solids tend to a hyperuniform structure~\cite{donev2005unexpected, zachary2011hyperuniform, jiao2011maximally}. While such long-range fluctuations can hardly be observed on the pair-correlation function, hyperuniform point patterns can be recognized from the behavior of the structure factor at low values
\begin{equation}
\lim_{\mathbf{q} \rightarrow 0} S(\mathbf{q}) = 0.
\end{equation}
Of particular interest in photonics are so-called ``stealthy'' hyperuniform structures, for which $S(\mathbf{q}) = 0$ for $0 < q \leqslant q_\text{max}$, where $q_\text{max}$ may be set to an arbitrary value. The region of zero structure factor is often followed by oscillations similar to those found in short-range disordered correlated media~\cite{froufe2016role}.
		
The concept of hyperuniformity in photonics has first been introduced in a numerical study by~\citet{florescu2009designer}.
Important efforts have been put since then on the fabrication of hyperuniform disordered systems, which could be achieved so far by lithography in 2D~\cite{man2013isotropic} and 3D~\cite{muller2013silicon}, block copolymer assembly~\cite{zito2015nanoscale}, emulsion routes~\cite{weijs2015emergent, ricouvier2017optimizing, piechulla2018fabrication, piechulla2022toward}, and spinodal solid-state dewetting~\cite{salvalaglio2020hyperuniform}.
		
\subsubsection{Disordered fractal structures}

In all classes of disordered point patterns discussed above, the average number of points $N$ contained in a $d$-dimensional sphere of radius $R$ is expected to grow as $\langle N \rangle \propto R^d$ - by doubling the observation radius for a 3D point pattern, the number of points increases by a factor $2^3=8$. This scaling is however not a general rule. Introduced by~\citet{mandelbrot1967long}, the concept of fractals encompasses systems for which the power-law scaling of the mass with the system size does not have the Euclidean dimension as an exponent. More specifically, for fractal point patterns, we have $\langle N \rangle \propto R^{d_\text{f}}$, where $d_\text{f}$ is a non-integer fractal dimension. Fractality has a dramatic impact on the structure, as illustrated in Fig.~\ref{fig:classes_disordered_media} (5th column). First, it is statistically self-similar, meaning that the structure is statistically identical whatever the scale on which it is looked at (though lower and upper bounds are always met in practice). Second, it exhibits enormous local density fluctuations and high lacunarity~\cite{allain1991characterizing}, meaning that both very dense and very empty regions are found. As a result, the pair correlation function can be shown to decay as a power-law as $g_2(r) \sim r^{-\alpha}$ and similarly for the structure factor, $S(q)-1 \sim |q|^{-(d-\alpha)}$ assuming that $0 < \alpha <d$. Depending on the process of structure formation, one can directly relate the exponent $\alpha$ with the fractal dimension $d_\text{f}$. For instance, clustering described by the Soneira-Peebles model gives $\alpha=d-d_\text{f}$, leading to $S(q)-1 \sim |q|^{-d_\text{f}}$~\cite{soneira1977there}. It is important to note that real systems generally exhibit lower and upper bounds in their fractal nature. This implies that the power-law decays are observed on finite range ($g_2(r)$ eventually goes to 1 at large $r$).
		
Fractal disordered optical materials are encountered in a wide variety of colloidal aggregates that form naturally for certain charged particles~\cite{meakin1987fractal} as well as in certain emulsions~\cite{bibette1993structure}. They can also be designed in a laboratory by inserting spacing particles with a size distribution that covers several orders of magnitude in a statistically-homogeneous disordered medium~\cite{barthelemy2008levy, bertolotti2010engineering}.

\subsection{Numerical simulation of correlated disordered media} \label{sec:3-numerical}

Numerical simulations of the complex heterogeneous morphologies play a key role in colloidal chemistry and soft matter physics. For light scattering studies, modelled structured materials are taken as input data for solving Maxwell's equations. Here, we present some standard numerical approaches that have been used in the literature to generate correlated disordered structures and simulate their optical properties.

\subsubsection{Structure generation}


Random packings of hard spheres in different dimensions is of great interest due to their structural and thermodynamic properties. The numerical generation of such ensembles plays a key role in research, especially in the case of random close packings (RCP)~\cite{song2008phase, torquato2010jammed, parisi2010mean}. When the packing fraction of the system is kept below a few tens of percent, a random sequential absorption (RSA) model~\cite{widom1966random} is suitable to generate large (non-equilibrium) ensembles in any dimensionality. In the RSA model, new points are randomly added to the system following a uniform distribution. The new point is rejected if closer than a given distance to any of the previous points in the pattern. A careful management of the coordinates storage in appropriate structures lead to very efficient algorithms but the computational time nevertheless diverges due to high rejection rate when approaching the maximum possible RSA filling fraction, being  about $54\%$ and $38\%$ for disks in 2D~\cite{wang2000series} and spheres in 3D~\cite{meakin1992random}, respectively. This limitation has been lifted by~\citet{zhang2013precise}, who proposed a precise algorithm to generate a saturated RSA configuration \textit{within a finite time}.

In order to achieve larger packing fractions up to the jamming packing (at a filling fraction $\phi\simeq64\%$ in 3D) several approaches leading to efficient algorithms have been developed. The Lubachevsky-Stillinger (or compression) algorithm \cite{lubachevsky1990geometric} generates random packings of any physically-realistic $\phi$ by placing a set of $N$ particles of vanishingly small size in a closed or periodic domain at random, and then letting the particles grow at a given rate, move and collide (elastically), until the desired $\phi$ is reached. This algorithm has been successfully used in the study of sphere packings in any dimensionality~\cite{skoge2006packing} and is largely used in photonics~\cite{conley2014light, froufe2016role}. An approach named ``ideal amorphous solids'' was also proposed by~\citet{lee2010geometry} to realize maximally random jammed packings of polydisperse particles. The method relies on building aggregates of touching spheres by placing spheres one by one around a center of mass.

Enlarging the kind of correlations encountered in sphere packings requires the use of interaction potentials beyond the hard sphere model. Simple two-body interaction potentials such as Lennard-Jones together with standard Monte-Carlo techniques have been used to generate assemblies of scatterers in different phases~\cite{de2016self}. Two and three-body interaction models like the Stillinger-Weber model~\cite{stillinger1985computer} can be used to generate fully connected dielectric networks showing the same statistical structural properties (coordination and angle statistics) as amorphous silicon or diamond. Stealthy hyperuniform point patterns have been generated numerically using a suitable pairwise, long-range potential in the real space~\cite{froufe2016role}. Quite often, the structures are generated using molecular dynamics. Being a very vast and mature field, various softwares are nowadays available, including NAMD~\cite{phillips2005scalable} and CHARMMS~\cite{brooks2009charmm}, which are both extensively used in the field of chemistry and biochemistry. Other software packages include HOOMD-blue~\cite{anderson2008general}, which is implemented for GPU, and LAMMPS~\cite{LAMMPSweb}, which exploits massive parallelization~\cite{Plimpton1995Fast}.

Instead of using constraints in real space as in the case of hard spheres, targeted interaction potentials can be obtained by imposing constraints on the structure factor in reciprocal space~\cite{uche2004constraints}, which allowed realizing, for instance, stealthy hyperuniform structures~\cite{batten2008classical, florescu2009designer}. A more general approach to the problem is obtained using constrained Fourier transforms as collective coordinates~\cite{kim2018inversion}.

Materials forming a continuous correlated disordered network are very relevant on different levels~\cite{wright2013eighty}. On the one hand, a network presents the necessary structural stability required by different fabrication methods~\cite{gaio2019nanophotonic}. On the other hand, the topology of the network apparently plays an important role in the emergence of different optical properties such as photonic gaps in disordered networks~\cite{weaire1971existence,florescu2009designer}. Besides the Stillinger-Weber model considered above, there are different protocols described in the literature to generate continuous random networks. The Wooten-Winer-Weaire (WWW) algorithm~\cite{wooten1985computer} considers a collection of points and bonds connecting pairs of points. The initial network, that can be ordered, is randomized after a number of bond reassignments followed by a relaxation of the structure (for instance following the Stillinger-Weber interaction potential). In this way, accurate predictions of the electronic structure, bond geometry statistics and atomic structure of amorphous semiconductors are obtained \cite{barkema2000high}. Replacing the chemical bonds by dielectric rods leads to amorphous dielectric materials~\cite{edagawa2014photonic}.

A protocol to generate strongly correlated continuous random networks was first proposed by~\citet{florescu2009designer} in two dimensions and used by~\citet{liew2011photonic} in three dimensions. The idea is to create a uniform topology network starting from an arbitrary point pattern. The Delaunay tessellation \cite{watson1981computing} is constructed from the seed point pattern. By definition each Delaunay cell is surrounded by 3 (in 2D) or 4 (in 3D) neighbors. The protocol indicates that the centroids of neighboring triangles (2D) or tetrahedrons (3D) are linked. This connected network shows a uniform connectivity since each node of the network is linked to the same number of neighbors. When the seed pattern is strongly correlated, for instance using random closed or stealthy hyperuniform packings, the resulting dielectric network presents interesting photonic properties, for instance complete gaps in its density of states \cite{florescu2009designer, liew2011photonic, froufe2016role} in 2D and 3D. The optical properties of these structures will further be discussed in Sec.~\ref{sec:6}.

\subsubsection{Electromagnetic simulations}\label{sec:EM-simulations}

No numerical method has been specifically developed to model the optical properties of correlated disordered structures. Generic electromagnetic methods are being used instead, the choice of a specific method depending on the type of structure to simulate, the quantity of interest and the computational load. We refer the reader to \citet{wriedt2009light, gallinet2015numerical} for an overview of the computational techniques used in photonics and light scattering, and to the internet portal \textit{Scattport} by~\citet{Scattport_web}, which conveniently provides a large collection of freely available software packages dedicated to light scattering problems.

The most widely used numerical methods for the quantitative analysis of 2D and 3D correlated disordered media are (i) the T-matrix method and its variants~\cite{mishchenko1996t}, (ii) the finite-difference time-domain (FDTD) method~\cite{sullivan2013electromagnetic}, and (iii) the planewave expansion (PWE) method~\cite{ho1990existence,johnson2001block}.

The T-matrix method, initially proposed by~\citet{waterman1965matrix} and further developed over the years \cite{mishchenko1996t}, is probably the most adapted to solve light scattering problems by particulate media, including in layered environments~\cite{kristensson1980electromagnetic, videen1991light, mackowski2008exact}. The method essentially relies on the possibility to decompose the incident and scattered fields around a particle as a superposition of vector spherical wave functions (VSWFs). Formally, the T-matrix relates the amplitude coefficients of the incident wave functions to those of the scattered wave functions, and the multiple-scattering problem is solved with a high degree of analyticity by making use of the translation addition theorem for VSWFs~\cite{stein1961addition,cruzan1962translational}. Several publicly available codes exist, among which the long-established MSTM~\cite{mackowski2011multiple, MSTM_web} and the more recent GPU-parallelized CELES \cite{egel2017celes} and SMUTHI \cite{egel2021smuthi}, which have been used to model large disordered clusters of particles, as in~\cite{aubry2017resonant, yazhgur2021light}.

In assemblies of very small scatterers, the T-matrix of the individual elements reduces to their electric polarizability and the electromagnetic interaction between scatterers can be described directly with the dyadic Green tensor, Eq.~\eqref{GFsing}. More straightforward to implement than the T-matrix method, the so-called coupled dipoles method (CDM)~\cite{foldy1945multiple,lax1952multiple} is a standard to test new concepts or theoretical models, for instance, on homogenization~\cite{schilder2017homogenization}, light emission statistics~\cite{pierrat2010spontaneous,sapienza2011long} or mesoscopic transport regimes~\cite{leseur2014probing}. For resonant scatterers with high quality factors (e.g., cold atoms), the coupled dipoles equations in absence of an incident field become a linear non-Hermitian eigenvalue problem~\cite{rusek1996localization}, whose solutions are the so-called quasinormal modes (QNMs) of the system with complex-valued frequencies~\cite{ching1998quasinormal}. The statistical properties of QNMs provide information on collective phenomena, like polaritonic modes~\cite{schilder2016polaritonic} and the Anderson transition~\cite{skipetrov2014absence, sgrignuoli2022subdiffusive, monsarrat2022pseudogap}.

The FDTD method is instead the most popular choice for non-particulate structures (e.g., connected networks). In essence, the method provides numerical solutions of the time-dependent Maxwell's equations with discretized space and time partial derivatives~\cite{yee1966numerical} over a (necessarily) finite volume and for a certain duration. Quantities related to light emission, scattering, transport and localization can be computed using appropriate boundary and initial conditions on the fields and sources, see \cite{scheffold2022transport, yamilov2022anderson} for very recent examples. The FDTD method is extremely versatile but the requirement to discretize the entire space for large disordered structures and perform simulations over long times implies a high computational load, which is yet mitigated by efficient parallelization. Many commercial and non-commercial software packages are nowadays available. Among those, MEEP is a powerful, maintained and open-source solution that is extensively used by the community~\cite{oskooi2010meep}.

Last but not least, the PWE method is the most common choice for identifying photonic gaps in non-absorbing (non-dispersive) dielectric structures. In short, the method solves the source-free wave propagation equation with periodic boundary conditions, expanding the fields and space-dependent permittivity in a Fourier series in reciprocal space, to solve the photonic band structure of the geometry~\cite{ho1990existence}. Primarily applied to photonic crystals, the use of the supercell approach (see Sec.~\ref{sec:6-photonic-gaps}) on large disordered structures generated with periodic boundary conditions, can provide quantitative information on the photonic density of states~\cite{florescu2009designer}. \textit{De facto}, the standard tool used by the community is the MIT Photonic Bands (MPB) open-source software package~\cite{johnson2001block}.

\subsection{Fabrication of correlated disordered media} \label{sec:3-fabrication}

Here, we present an overview of the different strategies and important design parameters for the experimental fabrication of strongly scattering correlated disordered media. We mainly focus on the fabrication of 3D materials but note that many of the concepts discussed here also hold for 2D materials. We will illustrate the fabrication concepts with a few examples but will not attempt to provide a comprehensive overview of this field of materials research, which is beyond the scope of our work. We note that the fabrication of disordered correlated photonic materials faces the same challenges than other optical metamaterials such as photonic crystal circuits or other 3D arrangements of structural units~\cite{soukoulis2011past}. The trade-offs one needs to consider are simplicity, freedom of design, speed or throughput, accuracy and resolution. Those parameters vary enormously and therefore no \emph{one-method-fits-all} fabrication route can be singled out.

The range of interest to observe strong scattering and coherent phenomena due to structural correlations is when typical length scales of the structure are on the order of the wavelength (typically half a wavelength) in the medium. This mandates, first of all, the sub-micron structuring of dielectric materials on length scales comparable to the wavelength of light with a refractive index contrast $(n/n_\text{h}) -1 \gg 0.1$. In practice, finding the optical material properties is also influenced by the fact that the effective refractive index $n_\text{eff}$ is often higher than the nominal background material index $n_\text{h}$ which tends to reduce the scattering and transport coefficients~\cite{reufer2007transport, naraghi2015near, schertel2019tunable}. 
As a general rule, the higher the space filling fraction of the scattering material the higher $n_\text{eff} \ge n_\text{h}$ and the stronger this effect.
The optimum space filling fraction is often found around $\phi \sim 0.3$ which is much lower than the space filling fraction obtained naturally by randomly packing spheres $\phi \sim 0.64$. In addition to these fundamental scattering parameters, structural correlations are key for the design and fabrication of optimally white materials. 

In general we can distinguish global structural properties, that is, properties that can be expressed by statistical averages and a corresponding structure factor $S(q)$, and local properties related the local topology, filling fraction and the scatterer morphology. In the following, we describe different approaches that have been used to fabricate disordered and strongly scattering media. Structural correlations then appear naturally or by design. 

\begin{figure*}[htbp]
\centering
\includegraphics[width=0.9\textwidth]{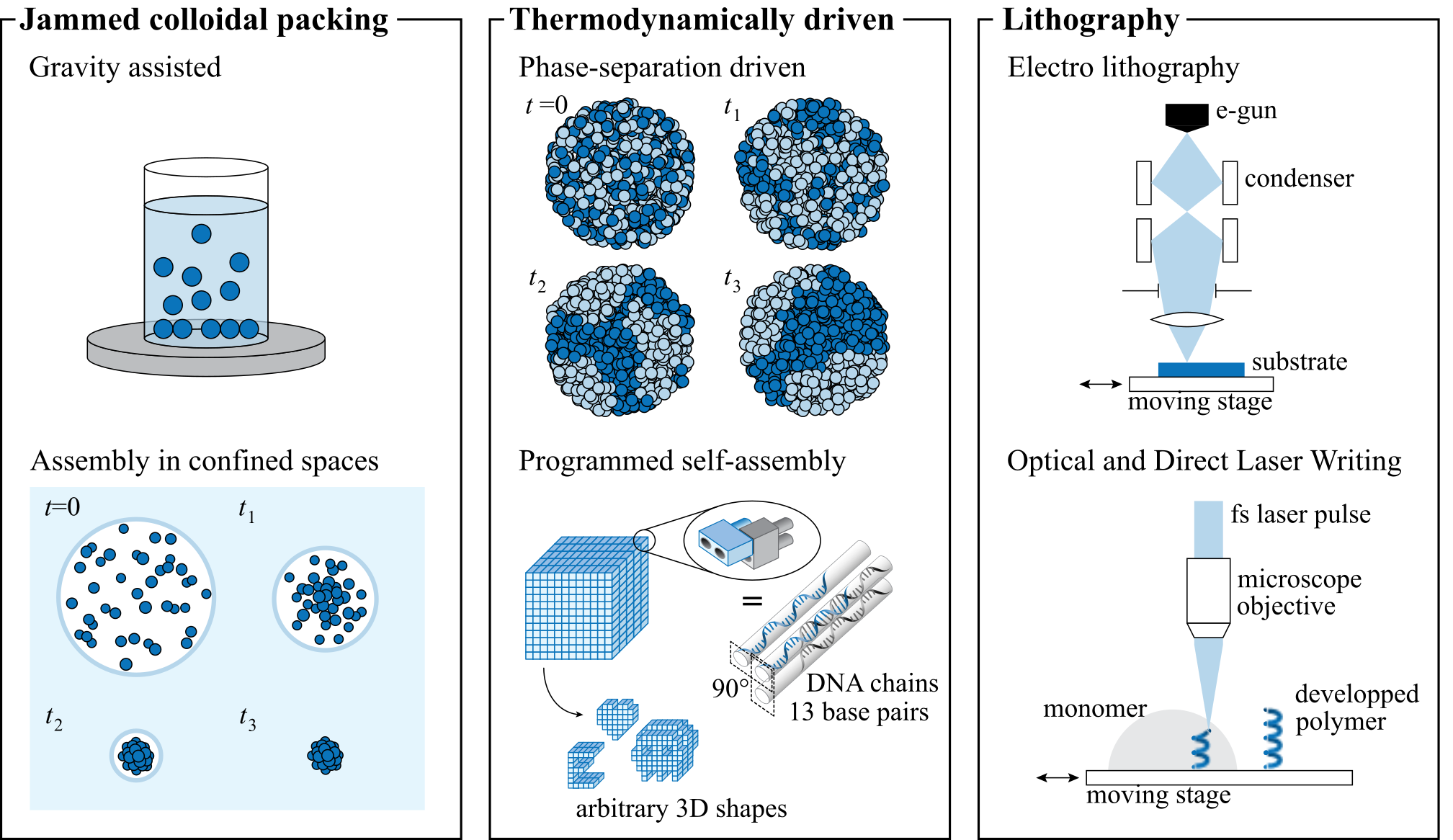}
\caption{\label{fig:classes_fabrication} (Color online) Overview of fabrication methods. (From left to right) \emph{Jammed colloidal packing}, where the colloids deposit via gravity assisted methods, or where they assemble in confined spaces due to local interactions. \emph{Thermodynamically driven assembly}, where the system goes through a phase-separation and rearrangement, driven by entropy or by pre-programmed interactions as for example using DNA strands. \emph{Optical and electron lithography}, where samples are fabricated by direct sculpturing of a material, using optical or electron beams to modify the local physical and chemical properties (subtractive) or where material is locally added to the structure via selective polymerisation or deposition (additive). Courtesy from M\'elanie M. Bay (University of Cambridge, UK).}
\end{figure*}

Figure~\ref{fig:classes_fabrication} summarizes the most important fabrication methods of correlated disordered media, that we will describe consecutively in detail below. Thermal or assisted self-assembly are bottom-up processes driven by a combination of entropy and external forces, such as gravity. Equilibrium and non-equilibrium self-assembly design routes following predefined pathways are frequently found in nature but are also increasingly considered as alternatives in the laboratory. Top-down approaches based on lithography come in many flavors and take advantage of powerful technology at hand. Lithography is very powerful for structuring two dimensional materials but only more recently significant progress has been made to fabricate 3D structured materials on submicron length scales. We will discuss some of the strengths and limitations of the different methods. 
	
\subsubsection{Jammed colloidal packing}

Dispersing submicron sized colloidal particles in a solvent phase with a lower index of refraction is the most common and most simple way to fabricate a strongly scattering, disordered medium. Ubiquitous examples are white paints, often based on a dispersion of submicron TiO$_2$ or polymer latex particles, or milk. For uniform suspensions of spherical particles, structural correlations appear naturally owing to the interactions between the particles which can be longer range DLVO-type double-layer repulsion or short-range excluded volume interactions 
The preparation is fairly simple and only requires some command over the stability of the suspension to avoid the formation of very large aggregates or flocks~\cite{galisteo2011self}. The degree of structural correlations can be controlled by the composition in particle volume fraction, electrolyte and type of solvent.

Colloidal particles can also be processed as powders which provides a higher refractive index contrast to air ($n_\text{h}^\text{air}=1$), as compared to solvent based dispersions (e.g., $n_\text{h}^\text{water}=1.33$), but offers less control over the microstructure. The statistically well defined structure in a liquid can be transferred to a solid film by film drying often preceded by sedimentation or centrifugation, see Fig.~\ref{fig:classes_fabrication} (left column, top)~\cite{reufer2007transport}. Such a colloidal film has the structural properties of a frozen colloidal liquid at random close packing conditions. For identical spheres, this results in pronounced short-range correlations. It is however often difficult to avoid crystallization. To avoid the formation of crystallites one can employ size polydispersity, the pre-formation of aggregates or the use of non-spherical particles, but this usually leads to a reduction of structural correlations. Photonic crystals with controlled disorder can also be fabricated by combining spherical colloids of two different polymers, and to selectively etch one after the crystal deposition~\cite{peng2007light}. In this way, controlled defects in an otherwise periodic lattice are formed~\cite{garcia2009strong}. Optimizing this fabrication process has been subject of active research~\cite{garcia2007photonic}.

Finally, densely-packed colloidal aggregates, typically of micron-sized spherical shapes, can be realized by selective solvent evaporation or spray-drying~\cite{manoharan2003dense, yi2003generation, moon2004electrospray, vogel2015color, yazhgur2021light}, see Fig.~\ref{fig:classes_fabrication} (left column, bottom). These so-called ``photonic balls'', which may be composed of dielectric or metallic particles and be suspended in air or in a solvent, have been used to realize angle-independent structural colors~\cite{park2014full}, artificial (meta) materials~\cite{dintinger2012bottom} or micron-sized random lasers~\cite{ta2021biocompatible}. The finite size of the photonic balls breaks the translational invariance and, for small photonic balls, this leads to additional metaball-scattering contribution.  Structural correlations within the balls are likely to depend on the size of the aggregate and and the quenching rate. Crystallisation of the surface layer is often observed. 

\subsubsection{Thermodynamically-driven self-assembly}

Colloidal self-assembly proceeds via a random process that arranges prefabricated scatterer of a given size $a$ in space. The fabrication process is stochastic and driven by thermal motion or external fields such as gravity and the resulting structures are relatively simple. In contrast, biology and recent DNA based nano-fabrication processes rely on well defined fabrication pathways or cascades that can be programmed which leads to beautiful and complex optical materials in nature \cite{prum1998coherent, vignolini2012pointillist}. In biology, it is known that many species are able to produce optical materials that show color or whiteness with optimal morphology and structural correlations, short or long ranged \cite{prum2009development, luke2010structural, burresi2014bright}. The physical mechanisms that underlies the assembly of photonic structures in living organism is still not understood~\cite{onelli2017development, dufresne2009self, wilts2019nature, prum2009development}. Even without a complete understanding of the biological processes, it is possible to use such architectures as materials for biotemplating. To this end, the biological material is used as a template or cast for a synthetic material with a high refractive index such as TiO$_2$ \cite{galusha2010diamond}. Analogous three-dimensional architectures have been produced on the tens of nanometer-scale via block-copolymers self-assembly~\cite{stefik2015block}, and the interplay between order and disorder on a slightly larger-scale (few hundreds of nanometers) have been shown to be controllable via block-copolymers brush systems~\cite{song2018photonic}, see Fig.~\ref{fig:classes_fabrication} (middle column, top).

Another very promising route is based on DNA-nanotechnology \cite{he2020colloidal}. DNA-origami techniques, invented a decade ago~\cite{rothemund2006folding}, are considered one of the breakthroughs in nanotechnology. Recently, methods for making micrometre-scale DNA-Origami objects have been developed~\cite{zhang2017dna}, see Fig.~\ref{fig:classes_fabrication} (middle column, bottom). The use of DNA-origami or bioinspired assembly techniques is still in its infancy. It is however the only fabrication route that may possibly allow the design of complex, correlated disordered three-dimensional optical materials in the visible range owing to the nanoscale control over the fabrication process. 



\subsubsection{Optical and e-beam lithography}

Despite the rapid advances in nano-assembly, such as DNA origami, it is still difficult and often impossible to fabricate tailored disordered optical materials at will. In particular optimized structures designed \emph{in silico} cannot be readily transferred into real materials yet using such approaches. Lithography is an established and powerful alternative to self-assembly. Its leading performance is unchallenged in the fabrication of two dimensional materials, such as silicon, owing to the decades of optimization in the semiconductor industry. The resolution of deep UV based optical lithography is now at $10-20$~nm \cite{sanders2010advances}. The use of a predefined photographic mask means that this is a highly parallelized method and the resolution can be reached over a large area, such as entire 30 cm silicon wavers. High resolution optical lithography however has a very high start-up cost for instrumentation and for the fabrication of individual photo masks. E-beam lithography is a serial fabrication tool with similar resolution capacity. It is versatile and can fabricate any 2D structure but it is much slower and thus not suited for high-output volumes~\cite{altissimo2010beam}. Early attempts in the late 1990s focused on the fabrication of structured photonic materials in 2D for visible and near-infrared wavelengths using lithographic patterning followed by reactive ion etching to produce long air holes in high index materials~\cite{krauss1996two, zoorob2000complete}. It is very challenging to generalize the use of these powerful 2D methods for the fabrication of 3D materials. Small sized three-dimensional infrared photonic crystal on a silicon wafer were reported based on stacking several layers of 2D structures, fabricated with fairly standard microelectronics fabrication technology~\cite{lin1998three}. In principle, this approach can also be applied to correlated disordered materials but owing to its extreme cost and complexity as well as limitations in size it has not been widely used. More recently, the etching of air rods has been applied to fabricate 3D hole-arrays using 3D masks~\cite{grishina2015method}. This method is in an early stage of development and the evaluation of the optical performance of the materials obtained is still in progress, nonetheless, it offers potential also for the template-free, direct fabrication of correlated disordered 3D photonic materials with a very high refractive index contrast. 

The inherent limitations of conventional colloidal self-assembly strategies have led to the development of a class of 3D high resolution lithography tools in the late 1990s and the early 2000s known as direct laser writing (DLW)~\cite{sun1999three, deubel2004direct}. The most popular implementation of direct laser writing is based upon the development of the two-photon microscope by~\citet{denk1990two}. Using a focused femtosecond pulsed laser two photons are absorbed simultaneously in the focal spot, but not elsewhere, owing to the highly nonlinear absorption cross section. In microscopy, the re-emission of a photon is used for imaging, in direct laser writing the absorbed energy is used to initiate a chemical reaction in the photoresist. By scanning a near infrared fs-pulsed laser beam in 3D, a polymeric structure can be written with a resolution of approximately $200$nm laterally and $500$nm axially. The resolution is limited by the point spread function of the microscope objective and the two photon cross section as well as the photoresist. Recently, it was shown that the resolution can be further enhanced using a stimulated-emission-depletion (STED) microscopy inspired approach \cite{fischer2011three, klar2014sub} or by controlled heat-induced shrinkage of polymeric network structures~\cite{aeby2022fabrication}.. DLW has been used to fabricate polymer templates for a variety of optical metamaterials such as woodpile photonic crystals, quasicrystals and polarizers~\cite{deubel2004direct, ledermann2006three, soukoulis2011past, gansel2009gold}. It has also been instrumental for the experimental realization of 3D correlated disordered network materials, based on hyperuniform point patterns or other types of disordered correlated photonic materials~\cite{renner2015spatial}. Despite its power and versatily, the DLW method also suffers from imperfections due to shrinkage of the polymer structure during development and deformations~\cite{deubel2004direct, haberko2013direct, renner2015spatial}. Moreover, due to the relatively low refractive index of the polymer photoresist ($n\simeq 1.5$), it is usually necessary to transfer the cast or template into another, higher index, material such as TiO$_2$ or silicon. This can be done using single or double inversion protocols which can be parallelized~\cite{tetreault2006new, staude2010fabrication, muller2013silicon, marichy2016high, muller2017photonic}. Therefore, such single or double inversion of the template is, in principle, not a time limiting step in the fabrication protocol. However, the complex chemical and etching procedures needed often lead to an incomplete infiltration (and thus a lower refractive index)~\cite{staude2010fabrication, marichy2016high}, a general deterioration of the quality of the structure and additional surface roughness~\cite{muller2017photonic}.

\subsection{Measuring structural correlations} \label{sec:3-measuring}

Structural correlations in disordered photonic media can be measured using microscopy, tomography and scattering. Scattering can only be employed if the materials structure is translationally invariant and isotropic or aligned in a well-defined direction. This is the case for colloidal photonic liquids and glasses or randomly close packed particles or rods which are correlated on short length scale but are statistically uncorrelated (at least asymptotically) on large length scales~\cite{rojas2004photonic, garcia2007photonic, reufer2007transport}. A challenge is the fact that the material has to be fairly transparent to the used radiation and therefore light is not a suitable probe, unless some form of refractive index matching or clearing is possible. In the latter case, confocal microscopy has also been applied successfully~\cite{haberko2013direct}. Optical materials are usually fairly transparent to neutrons or X-rays. Ultra Small Angle Neutron and X-Ray scattering instruments are in principle suitable for this task and available at large scale facilities~\cite{bahadur2015small} but these experiments are difficult and time-consuming and they are thus not routinely carried out to measure structural correlations in complex photonic media. Small Angle Neutron Scattering (SANS) has been used successfully to measure the structure factor of photonic liquids composed or relatively small colloids in suspension~\cite{rojas2004photonic}. The direct visualization of the materials local and global structure is often more useful or, for many novel systems, even required. To this end, electron microscopy is routinely applied often in tandem with focused ion beam milling and cutting. More recently, X-ray imaging and X-ray tomography have been developed as non-invasive tools for the real space characterization of correlated photonic materials \cite{grishina2019xray, wilts2018evolutionary}. Another promising route to study the internal structure of 3D photonic materials is destructive tomography using ion beam milling or etching techniques in conjuction with electron or atomic force microscopy~\cite{magerle2000nanotomography, burresi2014bright}.

\section{Modified transport parameters} \label{sec:5}

The primary effect of structural correlations is to modify the light scattering and transport parameters. This section offers a survey of the theoretical predictions and experimental observations of modified transport properties due to structural correlations. We first focus on colloidal systems and photonic materials, typically characterized by short-range correlations (i.e., negatively-correlated) [Sec.~\ref{sec:5-glasses}]. We discuss optical transparency and enhanced single backscattering phenomena on the basis of the theory developed in Sec.~\ref{sec:2}, and survey progress on resonant and Bloch-mediated scattering. In a second part, we describe the markedly different transport properties of materials with large-scale heterogeneities (i.e., positively-correlated) [Sec.~\ref{sec:5-fractal}]. Transport in such systems requires a generalization of the radiative transfer equation and can become anomalous in presence of a fractal heterogeneity.


	
\subsection{Light scattering and transport in colloids and photonic materials} \label{sec:5-glasses}

\subsubsection{Impact of short-range correlations: first insights} \label{sec:5-glasses-insight}

\begin{figure}[htbp]
\centering
\includegraphics[width=8cm]{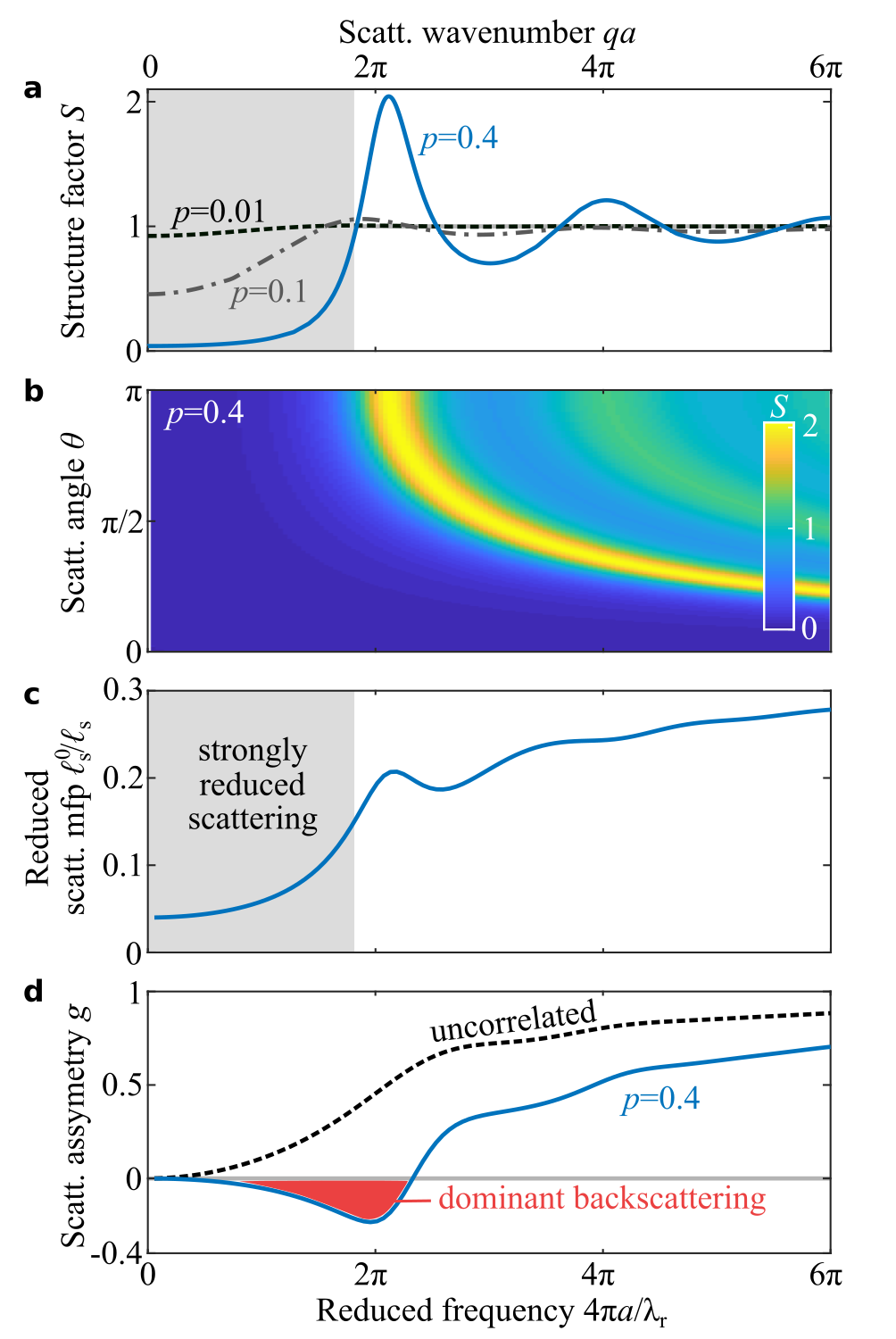}
\caption{\label{fig:short-range-correlated} (Color online) Impact of structural correlations on light scattering and transport in colloids. (a) Structure factor $S$ of a hard-sphere liquid for three different packing fractions $p= \pi/6 a^3 \rho$, with $a$ the particle diameter and $\rho$ the particle number density. The gray-shaded area indicates the low scattering wavenumber range where the structure factor is strongly diminished. (b) Angular and spectral response of the structure factor taking $q=4\pi/\lambda_\text{r} \sin{\theta/2}$ and $\theta$ the scattering angle for $p=0.4$. Exact backscattering ($\theta \rightarrow \pi$) is particularly pronounced when $\lambda_\text{r} \approx 2 a$. (c) Ratio of the scattering mean free paths neglecting structural correlations ($\ell_\text{s}^0$) and considering structural correlations ($\ell_\text{s}$), and (d) scattering anisotropy parameter $g$ without and with structural correlations, obtained in the realistic case of particles of diameter $a=100$ nm and refractive index $n_\text{p}=1.6$ (e.g., polystyrene) in a host medium with index $n_\text{h}=1.33$ (e.g., water). The observed features are directly linked to the structure factor. The red-shaded (gray-shaded) area in panel (d) highlights the range where single scattering is dominantly backward, implying $\ell_\text{t}<\ell_\text{s}$.}
\end{figure}

Let us start this section by examining the expressions derived for the scattering and transport mean free paths for assemblies of spherical particles, Eqs.~\eqref{eq:th_scattering_length_structure_factor} and \eqref{eq:th_transport_length_structure_factor}. In deriving these expressions, we have assumed that an effective permittivity $\epsilon_\text{eff}$ for the system can be defined, leading to an effective wavenumber $k_\text{r} = k_0 \text{Re}[n_\text{eff}]$ and a scattering wavevector $q = k_\text{r} |\u - \u'|$, where $\u$ and $\u'$ are the scattered and incident directions. The form factor is given by $F(q)= k_\text{r}^2 \frac{\ud \sigma}{\ud \Omega}(q)$ [Eq.~\eqref{eq:form_factor_intensity}], where $\frac{\ud \sigma}{\ud \Omega}$ is the differential scattering cross-section of the individual particle in the host medium evaluated at the wavenumber $k_\text{r}$. The structure factor $S(q)$ is the key quantity describing structural correlations for particulate media. Figure~\ref{fig:short-range-correlated}(a) shows the structure factor predicted within the Percus-Yevick approximation for hard spherical particles~\cite{wertheim1963exact} at different filling or packing fractions $p = (\pi/6) a^3 \rho$, where $a$ is the particle diameter and $\rho$ is the particle density. Increasing the density and/or the particle diameter leads to short-range correlations characterized by a reduction of $S$ at small values of $q$ (gray-shaded area), the emergence of a peak slightly above $qa = 2\pi$~\cite{liu2000improved}, and oscillations with a decaying amplitude at larger values of $qa$.

The effect of these short-range correlations on light scattering can be apprehended by rewriting the scattering wavevector as $q=2 k_\text{r} \sin{\theta/2}$ with $\theta$ the scattering angle. Insightful expressions for the scattering and transport mean free paths can in fact be derived from this change of variables, leading to
\be \label{eq:ls_theta}
	\frac{1}{\ell_\text{s}} = \rho \int_{4\pi} \frac{\ud \sigma}{\ud \Omega}(\theta) S(\theta) \ud \Omega,
\ee
and
\be \label{eq:lt_theta}
	\frac{1}{\ell_\text{t}} = \rho \int_{4\pi} \frac{\ud \sigma}{\ud \Omega}(\theta) S(\theta) (1- \cos \theta) \ud \Omega.
\ee
respectively, where $\Omega$ is the solid angle. Similarly, the scattering anisotropy parameter is given by
\be \label{eq:g_theta}
	g = \frac{\int_{4\pi} \frac{\ud \sigma}{\ud \Omega}(\theta) S(\theta) \cos \theta \ud \Omega}{\int_{4\pi} \frac{\ud \sigma}{\ud \Omega}(\theta) S(\theta) \ud \Omega}.
\ee
The structure factor can thus be seen as a quantity that modifies the scattering diagram $\frac{\ud \sigma}{\ud \Omega}(\theta)$ of the individual particle due to far-field interference. In absence of correlations ($S=1$), the scattering mean free path is simply given by $\ell_\text{s}^0 = (\rho \sigma_\text{s})^{-1}$ with $\sigma_\text{s} = \int_{4\pi} \frac{\ud \sigma}{\ud \Omega}(\theta) \ud \Omega$ the scattering cross-section of an individual particle.

Figure~\ref{fig:short-range-correlated}(b) shows the structure factor for $p=0.4$ expressed as a function of $a/\lambda_\text{r}$ with $\lambda_\text{r} = \lambda/\text{Re}[n_\text{eff}]$ and $\theta$. The most remarkable features here are the systematic reduction of scattering around the forward direction $\theta \approx 0$ and strong increases in the backward direction $\theta \approx \pi$ at specific frequencies, especially near $qa = 2\pi$, similarly to the Bragg condition in crystals.

We apply Eqs.~\eqref{eq:ls_theta}-\eqref{eq:g_theta} to a practical situation, namely spherical polystyrene particles ($n_\text{p}=1.6$) with diameter 100 nm dispersed in water ($n_\text{h}=1.33$). We use the CPA to get the effective refractive index~\cite{soukoulis1994propagation}, resulting in $n_\text{eff} \approx 1.44$ on the entire wavelength range considered here, although the actual choice of the effective medium theory is of little importance for such low-index contrast systems. Figures~\ref{fig:short-range-correlated}(c)-(d) show the variation of scattering efficiency $\ell_\text{s}^0 / \ell_\text{s}$, and the scattering asymmetry parameter $g$ in the limit of an uncorrelated medium (asymmetric scattering is then entirely due to the particle alone) and for a strongly correlated system, $p=0.4$. Two main conclusions can be drawn here. First, structural correlations lead to a reduction of the scattering efficiency, that is particularly pronounced in the low frequency range, where the wavelength is much larger than the characteristic length of the system. Thus, an incident wave propagates ballistically on longer distances (on average). Second, the angular dependence up to the first peak in the structure leads to a negative scattering anisotropy parameter $g$, meaning that light is predominantly scattered backward, leading to $\ell_\text{t} < \ell_\text{s}$. Both these effects have been observed experimentally, as reported below.

We considered here particles with a fairly low-index contrast to emphasize the role of short-range structural correlations on light scattering and transport. The range of optical properties is significantly enriched when considering the possibility of having spectrally sharp Mie resonances in high refractive index contrast materials or longer-range structural correlations, as will be discussed hereafter (Sec.~\ref{sec:photonic-glasses} and \ref{sec:imperfect-PhC}).

\subsubsection{Enhanced optical transparency}

The impact of structural correlations on light scattering in colloids emerged in the 1950s when it was noticed that the light intensity scattered either by protein solutions~\cite{doty1952macro} or by collagen fibrils in the cornea stroma~\cite{maurice1957structure} was not following the behavior expected for small scattering elements uncorrelated in position. In the celebrated article by~\citet{maurice1957structure}, it was supposed that a periodic organization of the fibrils was at the origin of a surprising optical transparency. Later works shown theoretically that this transparency could be explained by short-range correlated disorder~\cite{hart1969light, benedek1971theory, twersky1975transparency}. More recently, dense nanoemulsions, a kind of synthetic mayonnaise made from smaller than usual oil droplets with a diameter around 50 nm, have been shown to be much more transparent than more dilute suspensions $\phi\sim 0.1$ of the same droplets~\cite{graves2008transmission}. A transparency window has been observed in scattering fibrillar collagen matrices as a function of collagen concentration~\cite{salameh2020origin}. All these observations are explained by the strongly reduced scattering efficiency observed in the long-wavelength regime and shown in Fig.~\ref{fig:short-range-correlated}(c).

The notion of transparency relies on the proportion of ballistic light after a sample and therefore depends on the ratio between the extinction mean free path $\ell_\text{e}$, or scattering mean free path $\ell_\text{s}$ in absence of absorption, and the sample thickness $L$. Thus, materials exhibiting short-range correlated disorder unavoidably become opaque for very large thicknesses. The question of whether this conclusion holds for stealthy hyperuniform media, for which the structure factor strictly equals 0 on a range of scattering wavevectors $q$, naturally follows. The perturbative expansion of the intensity vertex (or equivalently of the phase function) up to the second order [Eq.~\eqref{eq:expansion-gammaop}] predicts that scattering is completely suppressed for such media [Eq.~\eqref{eq:th_scattering_length_structure_factor}]. Scattering may however occur due to the higher-order terms. Taking these into account leads to the definition of a criterion for optical transparency that reads~\cite{leseur2016high}
\be
    \frac{L}{\ell_\text{s}^0} \ll k_\text{r} \ell_\text{s}^0,\label{eq:transparency-criterion}
\ee
derived here for point scatterers ($\ell_\text{s} = \ell_\text{t}$). This shows that stealth hyperuniformity does \textit{not} completely suppress scattering. Optical transparency can be achieved in situations in which an uncorrelated disordered medium would be opaque ($L/\ell_\text{s}^0 \gg 1$) but only provided that the ratio $\ell_\text{s}^0/\lambda_\text{r}$ is sufficiently large.

\subsubsection{Tunable light transport in photonic liquids}


Spherical colloids are often considered as big atoms in soft matter physics~\cite{poon2004colloids}. From this viewpoint, each colloidal particle takes the place of an atom that is interacting with its peers via specific colloidal interactions. For colloids in suspensions these interactions are often tunable both in strength and sign, such as the well-known double layer repulsion between charged microspheres suspended in salty water. Thus, depending on the volume fraction occupied by the particles and the interaction strength, different colloidal phases can be found such as correlated liquids, entropic glasses, jammed packings, or a crystal~\cite{pusey1986phase}.

Early experiments of light scattering by charged particles, typically made of polystyrene or PMMA, were initiated in the mid 1970s, notably by~\citet{brown1975light}, who could measure the structure factor of colloidal suspensions of subwavelength particles beyond the first peak by conventional light scattering. The impact of structural correlations on light transport in the multiple-scattering regime was later studied by~\citet{fraden1990multiple} and~\citet{saulnier1990scatterer}, who reported transmission and coherent backscattering measurements of the transport mean free path in optically thick materials composed of resonant (wavelength-scale) particles at various packing fractions, see also~\cite{kaplan1994diffuse, rojas2002diffusing, yazhgur2021light, sbalbi2022effect}. Both works observed an increase of the transport mean free path due to structural correlations. A further step forward was made by~\citet{rojas2004photonic}, where it was shown that a fine control over structural correlations via Coulomb repulsion could induce a strong wavelength dependence of the optical properties of colloidal liquids and even negative values of the scattering anisotropy parameter, $g<0$ (i.e., $\ell_\text{t}<\ell_\text{s}$). In such ``photonic liquids'', the strong spectral variations of transport parameters make that samples of intermediate optical thicknesses and/or partly absorbing become structurally colored in reflection. Note the overall excellent agreement between experiments and theoretical predictions based on direct measurements of $S(q)$ with small angle neutron scattering (SANS)~\cite{rojas2004photonic}. 

Initial works~\cite{fraden1990multiple, saulnier1990scatterer, rojas2004photonic} have not considered an effective index to correct the scattering wave number $q$. Fortunately, the outcome of doing so does not significantly impact the results due to the low index contrast.



\subsubsection{Resonant effects in photonic glasses}\label{sec:photonic-glasses}

Photonic glasses are solid materials composed of closed-packed dielectric spheres, with size comparable to the wavelength of light, arranged in a disordered way~\cite{garcia2007photonic}. This is usually achieved by intentional colloidal flocculation and subsequent deposition. The mono-dispersity of the building blocks that compose them induces Mie resonances, which remain observable in the closely-packed systems~\cite{aubry2017resonant}. The resonances are all the stronger as a higher index contrast is achieved by evaporation of the host liquid. Besides, when the scattering material is solid, material stability is ensured by physical contacts between neighboring particles, thereby resulting in stronger short-range correlations compared to photonic liquids, and strong near-field interaction between particles. The latter impacts both the magnitude and frequency of the Mie resonance of the individual sphere~\cite{sapienza2007observation} and can transmit more light than expected from classical scattering theory (as developed in Sec.~\ref{sec:2})~\cite{naraghi2015near}.

The strong short-range correlation and near-field interactions makes the modelling of realistic photonic glasses very challenging. Recent works~\cite{aubry2017resonant, schertel2019tunable} have argued that the effect of the near-field coupling on transport in photonic glasses could be captured by defining an effective wave number $k_\text{r}$ with an index obtained from the energy-density coherent potential approximation (ECPA)~\cite{busch1995transport}. This appears in contradiction with our rigorous derivation of Eqs.~\eqref{eq:th_scattering_length_structure_factor} and \eqref{eq:th_transport_length_structure_factor}, which required neglecting near-field interaction between particles. Besides, it is surprising that an approach based on the evaluation of the energy density would correctly predict the average field phase velocity. The most advanced formalism to date to describe scattering and transport by dense, particulate media possibly with high-index materials is the quasicrystalline approximation (QCA), exploited recently by~\citet{wang2018achieving} to study the interplay between Mie resonances and structural correlations.

Numerical simulations can be useful to validate theoretical models and provide physical insight. For example, the strong-contrast formulas derived by~\citet{torquato2021nonlocal} for two-phase composites have been tested with FDTD simulations in the case of 2D and 3D dense packings of spheres with hard-sphere (equilibrium) and stealthy hyperuniform correlated disorder, showing in passing the existence of a transparency window up to a finite wave number in the latter. The complexity of the relation between structural correlations and light transport was evidenced in a recent work by~\citet{pattelli2018role}, where a graphics processing unit (GPU) implementation of the T-matrix method~\cite{egel2017celes} was used to investigate scattering by large assemblies of particles on a wide range of parameters. Simulations reveal that, given the wavelength and the particles size and refractive index, the shortest transport mean free path is obtained at intermediate degrees of correlations and particle densities.

Although much remains to be understood, photonic glasses and all resonant dense scattering media have demonstrated their great versatility and efficiency to harness light scattering and transport, with interesting applications in, for instance, structural colors and random lasing, as will be discussed in Sec.~\ref{sec:7}.

\subsubsection{Modified diffusion in imperfect photonic crystals}\label{sec:imperfect-PhC}

An extreme case of correlated disordered media is that of a disordered photonic crystals, in which long range order is established by the almost-periodic structure and scattering can be induced by imperfections, defects or (intentional) contamination with additional scattering elements. In a crystal, light propagation is dictated by the photonic band diagram which maps the frequency-wavevector relation of propagating Bloch modes~\cite{joannopoulos2011photonic}. Perfectly periodic structures are typically characterized by strong variations of the group velocity and the formation of partial (or even complete) photonic gaps corresponding to a lack of propagating states. The scattering cross-section of a defect typically increases with the reduction of the group velocity and light will scatter only where propagating states exist, therefore very anisotropically.

Multiple scattering and transport of light are expected to be strongly affected, while a more quantitative prediction requires a precise modelling of the kind of scattering and the crystal topology. Pioneering experiments on coherent backscattering~\cite{koenderink2000enhanced, huang2001anomalous} and diffuse light transport~\cite{astratov1995optical, vlasov1999different} in photonic crystals searched for signatures of Bloch-mode mediated scattering but have merely shown standard light diffusion~\cite{koenderink2005optical,rengarajan2005effect,aeby2020scattering}. Single light scattering in a disordered photonic crystal have been measured, with clear modification of the scattering mean free path around the bandgap~\cite{garcia2009strong}, reflection studies have shown anisotropic scattering~\cite{haines2012anisotropic}, while dynamical studies have shown exceptionally reduced diffusion constants~\cite{toninelli2008exceptional}.

Instead of relying on natural imperfections in otherwise ordered photonic crystal, correlated disordered media can be made by creating lattice vacancies in photonic crystals~\cite{garcia2011photonic}. In these structures transport and scattering mean free path and the diffusion constant have been measured to present strong dispersion~\cite{garcia2011photonic}. The transition from order to disorder in the structure and its impact on the  transport parameters is still an active field of research~\cite{schops2018inhibited}, which has also motivated the development of hyperuniform materials, where such a transition can be driven by a single parameter, as will be discussed in Sec.~\ref{sec:6}.

\subsection{Anomalous transport in media with large-scale heterogeneity} \label{sec:5-fractal}

We have been concerned so far with systems for which the distribution $p_N$ for the number of scatterers in a window of volume $ V \gg 1/\rho $ has a small variance, thereby making the system appear quite homogeneous on the scale of tens or hundreds of scatterers. Here, we will be concerned with disordered systems exhibiting large-scale heterogeneities leading to a large variance, also known as positively-correlated systems, as described in Sec.~\ref{sec:3}. Such systems are ubiquitous in nature, a well-known example being cloudy atmospheres~\cite{marshak20053d}. The density of droplets in suspension in clouds can indeed fluctuate over orders of magnitude. As illustrated in the right panel of Fig.~\ref{fig:fluctuation-correlation}, one may find very sparse as well as denser regions, implying a strongly fluctuating scattering efficiency.
Research on the topic has experienced numerous developments, most notably in the framework of transport theory in so-called non-Markovian stochastic mixtures, also known as non-classical transport theory~\cite{pomraning1991linear}. For a recent and thorough review of the literature on non-classical transport, we refer the reader to~\citet{deon2022hitchhiker}.
As we shall now see, despite the absence of coherent interference effects between neighboring scatterers, such long-range correlations have a dramatic impact on transport.

\begin{figure}[htbp]
\centering
\includegraphics[width=\columnwidth]{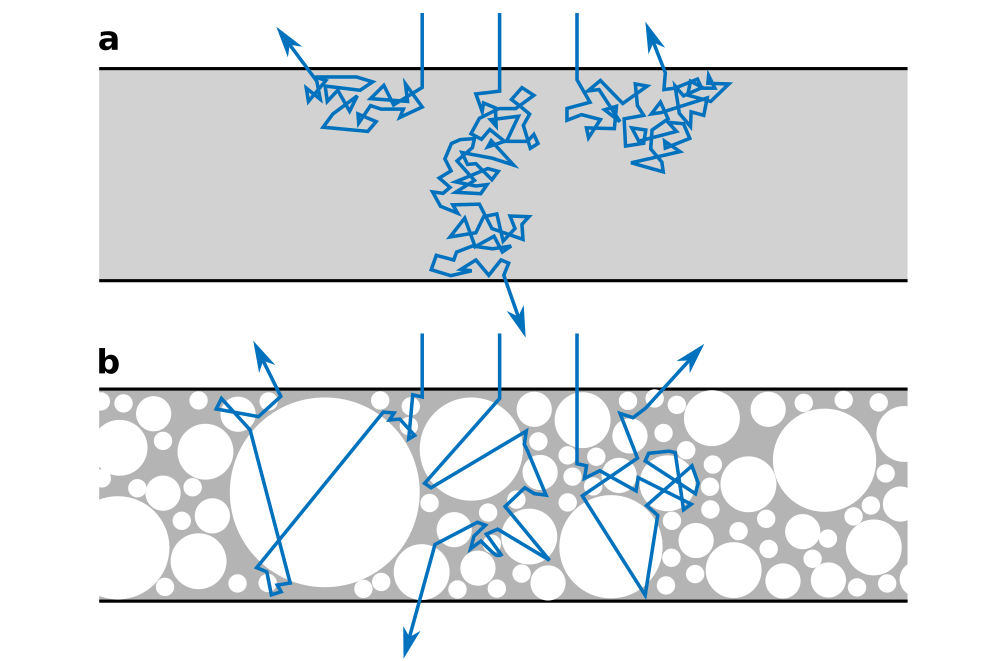}
\caption{\label{fig:Levy} (Color online) Impact of large-scale heterogeneity on transport in multiple-scattering media. (a) Sketch of a transport process in a statistically homogeneous medium (gray shaded area). Within radiative transfer, transport can be described as a random walk process with exponentially-decaying step length distribution. For thick media, transport is well described by the diffusion equation. (b) Sketch of transport in a scattering medium containing large non-scattering regions (white disks). Transport is driven by long steps, making the step length distribution no longer exponential. For certain systems with fractal heterogeneity, such as L\'evy glasses~\cite{bertolotti2010engineering}, transport can experience a transient superdiffusive behavior.}
\end{figure}

\subsubsection{Radiative transfer with non-exponential extinction}

The first element to describe radiative transfer is extinction. Equations~\eqref{pwt} and \eqref{keff} in Sec.~\ref{sec:2} impose that the coherent intensity $|\bra \E \ket|^2$ should decay exponentially on an average distance given by the extinction mean free path $\ell_\text{e}$. In strongly heterogeneous media, however, one may anticipate that the decay will be slower than exponential. An intuitive explanation is that spatially-extended non-scattering or weakly scattering regions promote trajectories much longer than the average decay length (i.e., the extinction mean free path).
Discussions on non-exponential extinction in strongly heterogeneous media date back to the mid-twentieth century with studies on neutron propagation in pebble bed reactors~\cite{behrens1949effect, randall1962stochastic} and light absorption in suspensions of photosynthesizing cells~\cite{rabinowitch1951photosynthesis, duyens1956flattering}. The topic gained further attention with later studies on radiative transfer in cloudy atmospheres~\cite{natta1984extinction, davis2004photon, varosi1999analytical} and, more recently, in the framework of computer graphics~\cite{jarabo2018radiative, bitterli2018radiative}.

A common approach to describe the non-exponential decay of the coherent intensity, proposed by several authors about two decades ago~\cite{kostinski2001extinction, borovoi2002extinction, kostinski2002extinction, marshak1998radiative}, consists in describing the heterogeneous medium as local ``patches'' or clusters of particles exhibiting a varying average extinction rate (or equivalently average particle densities). To describe this heuristic model, let us define a position-dependent particle density $\rho(\mathbf{r})=\langle N(\mathbf{r}) \rangle /V$, with $\langle N \rangle$ the average number of scatterers in volume $V$. We consider a system that is dilute at all points of space ($\rho(\mathbf{r}) \lambda^3 \gg 1$) such that radiative transfer applies. The key point of the approach is to assume that the distribution of number $N$ of particles in the volume $V$, and consequently the distribution of number of extinction events, follows a Poisson distribution, $p_{N|\langle N(\mathbf{r}) \rangle} = \langle N \rangle^N \exp \left[ - \langle N \rangle \right]/N!$. The patchiness leads to variations of $\langle N(\mathbf{r}) \rangle$ via a distribution $p_{\langle N \rangle}$. The distribution of extinction counts in a volume $V$ should then be
\begin{eqnarray}
    p_N &=& \int_0^\infty p_{N|\langle N(\mathbf{r}) \rangle)} p_{\langle N \rangle}  d\langle N \rangle \nonumber \\
    &=&\int_0^\infty \frac{\langle N \rangle^N \exp \left[ - \langle N \rangle \right]}{N!} p_{\langle N \rangle} d\langle N \rangle,
\end{eqnarray}
The relation with the classical extinction (Beer-Lambert) law is established by noting that the probability to cross the volume with no extinction event over a depth $z$ is given by $p_0$, and invoking the law of large numbers with $\langle N \rangle = z/\ell_\text{e}$. Taking $p_{\langle N \rangle}=\delta(\langle N \rangle - \beta)$ and the ballistic transmission $T_\text{b}=|\bra \E \ket /E_0|^2$ of a planewave along the $z$-direction through a medium leads to
\be
    T_\text{b}(z) \equiv p_0(z) = \exp[-\langle \rho \rangle \sigma_\text{e} z],
\ee
where we have set $\beta = \langle \rho \rangle \sigma_\text{e} z$ and $\sigma_\text{e}$ is the extinction cross-section. By contrast, the use of $\Gamma$ or fractional Poisson distributions lead to asymptotic power-law decays with varying exponents $m$~\cite{kostinski2001extinction, casasanta2018towards}
\begin{equation}
    T_\text{b}(z) \sim (1+\beta)^{-m}.
\end{equation}
One key aspect of course is the determination of an actual function in realistic systems. Important efforts have notably been dedicated to the determination of particle density distribution in clouds~\cite{kostinski2000spatial}. Slower-than-exponential decays of the coherent intensity have also been observed in photosynthetic cultures~\cite{knyazikhin1998influence}.


Very importantly, the radiative transfer equation [Eq.~\eqref{eq:th_rte}] has been generalized to account for arbitrary non-exponential extinction. Defining a probability density function $f_s$ of the random step length $s$ as $f_s(s) \equiv T_\text{b}(s)/\int_0^\infty T_\text{b}(s) ds$, one reaches a generalized (scalar) radiative transfer equation ~\cite{larsen2011generalized}
\begin{eqnarray}\label{eq:th_rte_nonexp}
    &&\left[\frac{\partial}{\partial s} + \bm{u}\cdot\bm{\nabla}_{\bm{r}}+\Sigma_\text{e}(s)\right]I(\bm{r},\bm{u},s) \nonumber \\
    && = \delta(s) \gamma \int p(\bm{u},\bm{u}')\Sigma_\text{e}(s')I(\bm{r},\bm{u}',s')\ud s'\ud\bm{u}',
\end{eqnarray}
where $\gamma = \ell_\text{e}/\ell_\text{s}$ is the single-scattering albedo (the probability to be scattered upon an extinction event) and $I(\bm{r},\bm{u},s)$ now depends on the step length $s$ via
\begin{equation}
   \Sigma_\text{e}(s) = \frac{f_s(s)}{1-\int_0^s f_s(s') \ud s'}.
\end{equation}
Equation~\eqref{eq:th_rte} is recovered by taking $f_s(s) = \exp[-s/\ell_\text{e}]/\ell_\text{e}$. A few remarks are in order. First, the distribution $f_s(s)$ should have a finite mean as to allow the definition of a mean free path $\ell_\text{e}$. Second, one of the key features of transport with non-exponential step length distributions is the fact that it is a non-Markovian process (i.e., implying memory in the construction of individual steps), contrary to the classical Beer-Lambert law which is a Markovian (memoryless) process ($\exp[x+y]=\exp[x]\exp[y]$). Third, Eq.~\eqref{eq:th_rte_nonexp} describes transport in the volume of a medium. Care should be taken on its applicability to bounded domains, since an incorrect treatment of the initial steps (light entering the medium) can result in a breaking of reciprocity. An extension of the formalism to bounded domains has been proposed by~\citet{deon2018reciprocal}. Finally, -- and this aspect has not been significantly emphasized previously -- the formalism assumes an ``annealed'' disorder, meaning that the medium is randomized after each scattering event. As we will see below, correlations between successive scattering events due to a ``quenched'' disordered potential can have a significant impact on transport. 

Along similar lines, radiation transport can be efficiently modelled numerically in arbitrary geometries via random-walk (Monte Carlo) simulations either in heterogeneous media with spatially-varying scattering parameters~\cite{glazov1977integral, boisse1990radiative, audic1993monte} or in statistically homogeneous media using arbitrary step length distributions, including those with diverging second moment~\cite{nolan2003stable}. The latter are known as L\'evy walks~\cite{zaburdaev2015levy} and have been proposed as a tool to describe radiation transport in clouds~\cite{davis1997levy}, leading to enhanced ballistic transmission and transmitted intensity fluctuations.

\subsubsection{From normal to super-diffusion}

In classical radiative transfer, the average (incoherent) intensity is expected to follow the laws of diffusion after many scattering events, Eq.~\eqref{eq:th_diffusion_approximation}. Physically, diffusion is related to the Brownian motion of many independent moving elements (i.e., random walkers). As long as the second moment of the step length distribution $f_s(s)$ is finite, the central limit theorem shows that the average step length converges towards a normal distribution, eventually leading to a diffusive process, characterized by a mean square displacement $\langle r^2(t) \rangle \sim 2dDt$. Under these circumstances, the presence of large heterogeneities does not prevent the diffusion limit but leads to a modified diffusion constant. From a simple isotropic random walk consideration~\cite{ben2000diffusion}, it is possible to show that the diffusion constant for a step length distribution $f_s(s)$ is given by~\cite{svensson2013holey}
\begin{equation}
    D = \frac{v}{2d} \frac{E[s^2]}{E[s]},
\end{equation}
with $E[X]$ is the expectation value of the random variable $X$. This expression is apparently not well known and yet very interesting. It shows that the fluctuations in the step length are as important as the mean step length. In practice, any correlated system exhibiting slower-than-exponential decay will experience an increased diffusion constant. The known expression $D = \frac{v\ell_\text{t}}{d}$ is only recovered for an exponentially-decaying function of $f_s(s)$ and using the similarity relation $\ell_\text{t} = \ell_\text{s}/(1-g)$.

A fundamentally different behavior is observed when heterogeneities are so strong that they make the second moment of $f_s(s)$ diverge. This is the case for power-law decays $f_s(s) \sim s^{-(\alpha+1)}$ with $\alpha < 2$, defining the so-called L\'evy walks. By virtue of the generalized central limit theorem~\cite{gnedenko1954limit}, one shows that the average step length should follow an $\alpha$-stable L\'evy distribution, which is identically heavy-tailed. L\'evy walks lead to superdiffusive transport, characterized by a mean-square displacement growing faster than linear with time~\cite{zaburdaev2015levy}
\begin{equation}
    \langle r^2(t) \rangle \sim t^\gamma, \text{ with } 1 < \gamma \leq 2.
\end{equation}

L\'evy statistics and anomalous diffusion are widespread in science, from the random displacement of molecules in flows~\cite{solomon1993observation} to the foraging strategy of animals~\cite{bartumeus2005animal}. While early studies had already evidenced modified path length distributions of light in fractal aggregates of particles~\cite{dogariu1992enhancement, dogariu1996enhancement, ishii1998optical}, the first experiments aiming to control the anomalous diffusion of light in disordered systems have been initiated by~\citet{barthelemy2008levy}. So-called L\'evy glasses are fabricated by incorporating in a disordered medium containing small scattering particles, a set of transparent, non-scattering spheres with sizes ranging over orders of magnitude acting as spacers (see the last panel in Fig.~\ref{fig:classes_disordered_media}). By controlling the distribution of sphere diameters and assuming single scattering in the interstices between the spheres and annealed disorder, one can control the step length distribution $p(s)$ in the medium~\cite{bertolotti2010engineering}. Latest time-resolved experiments on L\'evy glasses showed indeed a (transient) superdiffusive light transport~\cite{savo2014walk}. L\'evy statistics in light transport has also been observed in hot atomic clouds~\cite{mercadier2009levy, baudouin2014signatures, araujo2021levy}, as a result from Doppler broadening~\cite{pereira2004photon, baudouin2014signatures}, not from structural correlations.

An important aspect of transport in L\'evy glasses is the fact that the disorder is frozen or quenched. Classical transport models assume annealed disorder, in the sense that there is no correlation between successive scattering events -- a photon ``sees'' a new structure after each scattering event. In real samples, successive steps are however not independent. There exists correlations due to the large empty regions. As shown by theory and experiments on scattering powders containing large monodisperse voids~\cite{svensson2014light}, quenched disorder leads to an effective reduction of the diffusion constant compared to annealed disorder. The impact of quenched disorder in L\'evy-like systems has been subject to several numerical and theoretical investigations~\cite{beenakker2009nonalgebraic, barthelemy2010role, burioni2010levy, buonsante2011transport, groth2012transmission, burioni2012scattering, burioni2014superdiffusion}, eventually showing that the actual observation of superdiffusive transport in finite-size system (hence with truncated step-length distribution) requires a proper finite-size scaling analysis and packing strategy~\cite{burioni2014superdiffusion}.

\section{Mesoscopic and near-field effects} \label{sec:6}

The interplay of order and disorder in photonic structures not only impacts light transport but also promotes strong coherent effects, resulting in the emergence of sometimes unexpected phenomena for disordered systems. This section is devoted to the main mesoscopic and near-field phenomena that have attracted attention in the past decades, namely the opening of photonic gaps in disordered systems [Sec.~\ref{sec:6-photonic-gaps}], transitions between various mesoscopic transport regimes [Sec.~\ref{sec:6-mesoscopic2D}], non-universal speckle correlations [Sec.~\ref{sec:6-near-field-speckles}] and large local density of states fluctuations [Sec.~\ref{sec:6-LDOS-fluctuations}]. We attempt to provide a clear picture of the current understanding in the field.

\subsection{Photonic gaps in disordered media} \label{sec:6-photonic-gaps}

Photonic gaps are one of the most striking manifestations of structural parameters on optical transport. Similarly to electronic gaps in semiconductors, a photonic gap corresponds to a spectral range in which no propagating modes exist. The concept of photonic gap is known in optics since the early works on (one-dimensional) thin-film optical stacks~\cite{yeh1988optical}, emerging as a consequence of the periodic modulation of the refractive index on the wavelength scale. The idea was generalized in the late 1980s to two and three-dimensional periodic structures~\cite{yablonovitch1987inhibited, john1987strong} and has been at the heart of research in optics and photonics for about two decades. The interest in photonic gaps largely comes from the possibility to engineer defects states with high-quality factors and wavelength-scale confinement, opening unprecedented opportunities to control spontaneous light emission and light propagation for applications in all-optical integrated circuits~\cite{joannopoulos2011photonic}.

Probably because of the convenience of Bloch's theorem and the development of numerical methods exploiting periodicity to solve Maxwell's equations, it is widely believed in the optics and photonics community that the opening of photonic gaps requires the refractive index variation to be periodic in space (i.e., the structure to exhibit long-range periodic correlations). Early works investigating the impact of structural imperfections on optical properties however realized, by drawing a parallel with semiconductor physics were similar questions have been tackled~\cite{weaire1971existence, phillips1971electronic, thorpe1973note}, that certain gaps could persist even in absence of periodicity~\cite{chan1998photonic, jin2001photonic}, thanks to local (Mie or short-range correlated) resonances~\cite{lidorikis2000gap}. Later reports on photonic gaps in 3D disordered structures exhibiting short-range correlations~\cite{edagawa2008photonic, imagawa2010photonic, liew2011photonic} and the proposition that hyperuniformity was a requirement for photonic gaps~\cite{florescu2009designer} greatly stimulated the community to unveil the relation between local morphology and structure, and the opening of spectrally wide gaps~\cite{froufe2016role, sellers2017local, ricouvier2019foam, klatt2019phoamtonic}.

\subsubsection{Definition and identification of photonic gaps in disordered media}

\begin{figure}[htbp]
\centering
\includegraphics[width=\columnwidth]{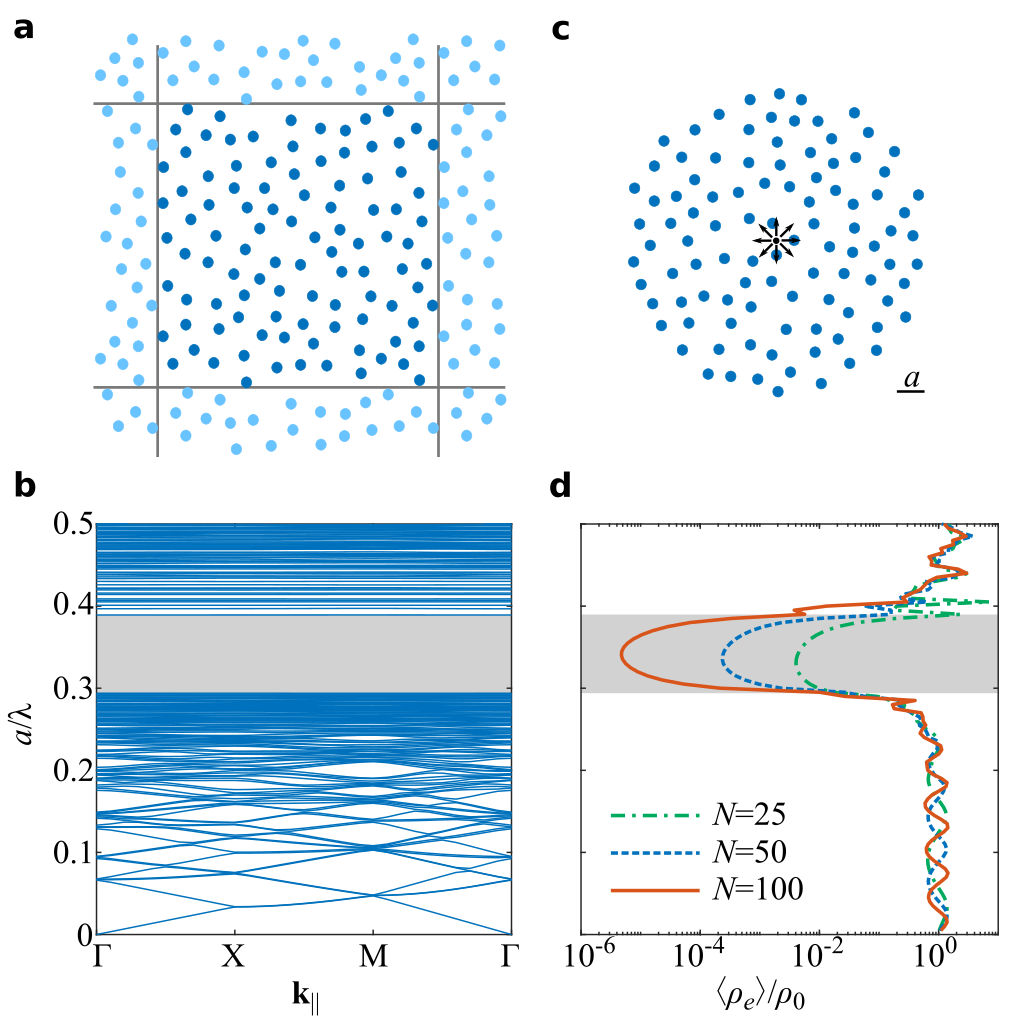}
\caption{(Color online) Signatures of photonic gaps in short-range correlated ensembles of dielectric rods in TM polarization. The rods have a permittivity $\epsilon=11.6$, a radius $r=0.189a$, are placed in air and are packed by RSA at a surface filling fraction $f=11.2\%$ (number density $n=a^{-2}$). (a) A disordered ensemble of rods is generated in a square region with periodic boundary conditions. (b) Photonic band structure  with a gap that correspond to an absence of eigenmodes in a finite spectral range $a/\lambda$. Numerically, the eigenmodes can be computed using the planewave expansion method with the supercell approach. The band structure was calculated here for a supercell containing $N=100$ rods. (c) A photonic gap can be identified by monitoring the spontaneous emission rate of a dipole source in the center of the system for varying system sizes. The emitter here is always placed in air. (d) Finite-size scaling of the spontaneous emission rate (or LDOS) for systems containing 25, 50 and 100 rods. $\langle \rho_\text{e} \rangle$ is the projected LDOS averaged over disorder configurations and $\rho_0$ is the projected LDOS in air. A gap leads to a strong damping of spontaneous emission with system size.}\label{fig:gap-rods}
\end{figure}

The notion of photonic gap is intimately linked with that of density of states (DOS). Formally, the DOS describes the spectral density of eigenmodes in the medium (i.e., the solutions of the source-free Maxwell's equations) around frequency $\omega$. For instance, the DOS of a closed and non-absorbing system with volume $V$ is simply
\begin{equation}\label{eq:DOS}
\rho(\omega) = \frac{1}{V} \sum_m \delta(\omega-\omega_m),
\end{equation}
where $\omega_m$ is the frequency, that is the eigenvalue, associated to resonant mode $m$. For non-dissipative systems, this frequency is real. In this framework, a photonic gap thus corresponds to a spectral region wherein $\rho(\omega)=0$, and can therefore easily be found from an eigenmode  analysis. In the case of disordered media, a classical strategy, illustrated in Fig.~\ref{fig:gap-rods}a in the case of parallel dielectric cylinders in TM polarization, is to employ the PWE method~\cite{ho1990existence, johnson2001block}, briefly presented in Sec.~\ref{sec:EM-simulations}, on a large periodic supercell. The system being conservative, a photonic gap is easily recognized as a spectral region containing no propagating modes [Fig.~\ref{fig:gap-rods}b]. A photonic gap can also be identified by time-domain simulations in real space (FDTD) using the order-$N$ spectral method~\cite{chan1995order}. Special attention should be given to whether the gap persists in the thermodynamic limit (i.e., when the system size increases) -- an aspect that has been overlooked until recently~\cite{klatt2022wave}. This approach can only be numerical as the DOS is not directly accessible experimentally and real systems are always of finite size and open. The latter makes that the DOS can in fact never be strictly equal to zero.

A second approach, which may now be performed experimentally~\cite{lodahl2004controlling, leistikow2011inhibited, sapienza2011long, aubry2020experimental}, consists in performing a finite-size scaling analysis of the emitted power (or spontaneous emission rate) of a quantum emitter embedded in a finite-size systems, see Fig.~\ref{fig:gap-rods}c for an illustration. The power $P_\text{em}$ emitted by a dipole source with moment $\mathbf{p}=p\mathbf{u}$, rigorously derived from Maxwell's equations~\cite{novotny2012principles, carminati2015electromagnetic}, reads
\begin{equation}\label{eq:power-emitted-dipole}
P_\text{em} = \frac{\pi \omega^2}{4 \epsilon_0} |\mathbf{p}|^2 \rho_\text{e} (\mathbf{r},\mathbf{u},\omega),
\end{equation}
where $\rho_\text{e}$ is the projected local density of states (LDOS) (in units of s.m$^{-3}$) defined as
\begin{equation}\label{eq:projected-LDOS}
\rho_\text{e}(\mathbf{r},\mathbf{u},\omega) = \frac{2 \omega}{\pi c^2} \text{Im}\left[ \mathbf{u} \cdot \mathbf{G}(\mathbf{r},\mathbf{r},\omega) \mathbf{u}\right].
\end{equation}
$\mathbf{G}(\mathbf{r},\mathbf{r}',\omega)$ is the total Green tensor in the structured environment. Decomposing it as a sum of the Green tensor in the homogeneous background and the Green tensor due to the fluctuating permittivity, one readily understands that the suppression (resp., enhancement) of the LDOS results from destructive (resp., constructive) interference at the dipole position between the field radiated in the homogeneous background and the field scattered by the heterogeneities. 

Equation~\eqref{eq:projected-LDOS} is general, yet it does not explicitly depend on the actual states of the system. To gain some physical insight, it is possible to express $\mathbf{G}$ in terms of the eigenmodes of the system, see Appendix~\ref{App:QNM} for more details. The eigenmodes of open (non-Hermitian) systems, also known as quasinormal modes (QNMs)~\cite{ching1998quasinormal, lalanne2018light}, are described by complex frequencies $\tilde{\omega}_m=\omega_m - i \gamma_m/2$ and normalized fields $\tilde{\mathbf{E}}_m (\mathbf{r})$, where the non-zero imaginary part stems from leakage. Physically, the existence of a photonic gap translates into the absence of resonant modes in the volume of the medium: the resonant modes may only be confined to the boundaries of the medium (i.e., on a length scale of the order of the extinction (scattering) mean free path). Thus, their excitation by a source deep inside the system, the LDOS and the resulting emitted power are all expected to tend towards zero with increasing size in the photonic gap, while it should remain quite unchanged in presence of propagating modes [Fig.~\ref{fig:gap-rods}d].

Despite the conceptual simplicity of this second strategy, care should be taken with the interpretation of emitted power (or spontaneous emission decay rate) spectra measurements. Indeed, as we will see in Section~\ref{sec:6-LDOS-fluctuations}, LDOS fluctuations can be enormous in complex media, depending considerably on the local environment around the emitter position as well as on the emitter orientation. The interplay between near-field interaction and the far-field radiation has been studied by FDTD simulations in finite-size photonic crystals by~\citet{mavidis2020local}. Additionally, in disordered systems, only average quantities acquired over a large set of disorder realizations are statistically relevant. This raises a second difficulty related to the fact that quantum emitters like quantum dots have a finite size, thereby inducing a local spatial correlation, and they are usually not distributed uniformly in all materials composing the complex medium (e.g., a semiconductor and air). Thus, the configurational average of the LDOS for a real emitter will often not be strictly equal to the average LDOS, as may be computed numerically for instance. Although it seems reasonable to assume that the average LDOS should converge towards the DOS in the limit of infinite system size, it appears that the link between photon emission statistics and the existence of photonic gaps has only been established phenomenologically to date.

\subsubsection{Competing viewpoints on the origin of photonic gaps}

Discussions on the origin on the photonic gaps in disordered media started to emerge in the late 1990s, inspired by earlier works on electronic gaps in (periodic and amorphous) semiconductors. Two main mechanisms have been identified.

The first, generally accepted, mechanism is that photonic gaps build up from interference between counterpropagating waves on a periodic lattice, thereby placing long-range structural correlations at the core of the picture. It is the photonic analog of the nearly free electron model in solid-state physics~\cite{kittel1976introduction}. Formally, the Bloch modes -- the eigenmodes of periodic systems -- result from a coupling between forward and backward propagating planewaves on the periodic lattice~\cite{yeh1988optical}. In spectral gaps, they form stationary patterns that do not carry energy (in the lossless case) due to a backscattering phenomenon with precise phase-matching condition. Their propagation constant is complex, leading to a damping of an incident wave in the specular direction, without scattering. Photonic (band) gaps in 1D media, or in one particular direction in a 2D or 3D photonic crystal \cite{spry1986theoretical}, can exist even for vanishingly small refractive index contrasts. Omnidirectional gaps in higher dimensions requires higher contrasts and a finely-optimized structure and morphology~\cite{joannopoulos2011photonic}. Because such spectral gaps are created by long-range periodicity, they are expected to be very sensitive to lattice deformations.
 
The second proposed mechanism is that photonic gaps are formed by coupled resonances between short-range correlated neighboring scatterers. It is the photonic analog of the tight-binding model in solid-state physics, developed in particular to explain the origin of the electronic density of states of amorphous semiconductors~\cite{weaire1971existence}. Intuitively, similarly to the level repulsion observed in a pair of coupled resonances, interaction between nearest neighbors in ensembles of identical resonators may ``push'' the states of the coupled system away from the resonant frequency. This is typically obtained with high refractive index Mie scatterers at moderate densities in low refractive index media. In this picture, the interaction between distant resonators, and thus long-range structural correlations, are irrelevant. As a consequence, one expects photonic gaps to exist in both periodic and disordered systems, provided that the resonances of the individual scatterers and the coupling coefficient between scatterers remain nearly constant throughout the entire structure. Care should however be taken regarding the analogy with the electronic tight-binding model due to the polarized nature of light waves~\cite{monsarrat2022pseudogap}.

A different, complementary viewpoint on this second mechanism is provided by considering the effective material parameters of assemblies of resonant objects~\cite{lagendijk1996resonant}. In particular, the effective permittivity $\epsilon_\text{eff}$ is predicted to exhibit a polaritonic response that possibly becomes negative in its real part for sufficiently strong resonances and high density. Having $\text{Re}\left[\epsilon_\text{eff} \right]<0$ implies that $\text{Im}\left[ n_\text{eff} \right] > \text{Re}\left[ n_\text{eff} \right]$, corresponding to a coherent field propagating in the effective medium that is overdamped, as in a metal. Stealthy hyperuniform structures (be they ordered or disordered) suppress scattering in the long-wavelength regime (up to the second order in the expansion of the intensity vertex). This implies that $\text{Im}\left[\epsilon_\text{eff} \right] \simeq 0$. Thus, one arrives to a situation where propagation is damped by coupled resonances and scattering is suppressed by structural correlations. This describes a system behaving as a homogeneous medium with no propagating states, that is a system exhibiting a photonic gap.

\subsubsection{Reports of photonic gaps in the litterature}

\begin{figure}[htbp]
\centering
\includegraphics[width=\columnwidth]{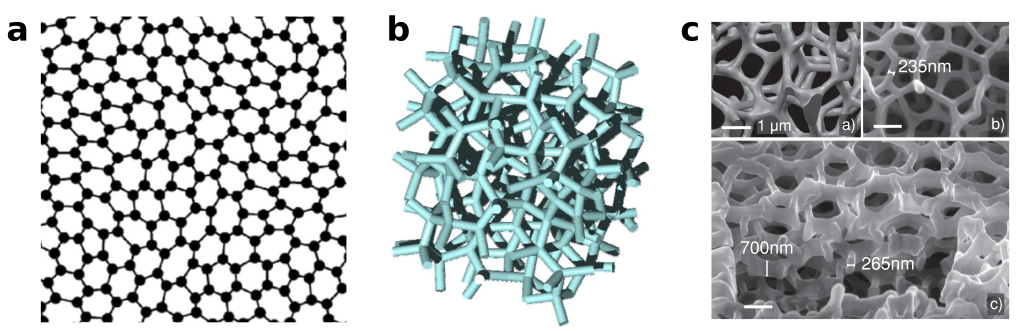}
\caption{\label{fig:photonic-gap} (Color online) Photonic structures lacking long-range order that were shown to exhibit large photonic gaps. (a) 2D stealthy hyperuniform structure exhibiting a gap for both TM and TE polarizations. The TM gap is due to the resonances of the dielectric rods and the TE gap to the air pores surrounded by dielectric holes. Adapted with permission from \cite{florescu2009designer}. (b) 3D amorphous diamond structure exhibiting an omnidirectional photonic gap. The structure consists in a network of dielectric rods forming air pores of comparable sizes. Adapted with permission from \cite{edagawa2008photonic}. (c) First experimental realizations of 3D disordered structures potentially exhibiting a photonic gap at optical frequencies. The silicon photonic medium was realized by direct laser writing followed by a double inversion process. Adapted with permission from \cite{muller2013silicon}.}
\end{figure}

The formation of photonic gaps very much depends on the dimensionality of the system and on the light polarization, for both periodic and disordered media. For instance, early works using numerical simulations with scalar waves have suggested that 3D face-centered cubic lattices of dielectric spheres could exhibit an omnidirectional gap, but vector wave calculations disproved this prediction~\cite{ho1990existence}. Figure~\ref{fig:photonic-gap} shows various examples of disordered photonic structures exhibiting photonic gaps.

It was suggested and demonstrated numerically already about two decades ago that photonic gaps in 2D ensembles of dielectric (e.g., silicon) rods in TM polarization (electric field normal to the propagation plane) are created by the strong electric dipole resonance of individual rods~\cite{jin2001photonic}. Those gaps were actually observed previously in an experiment on light localization~\cite{dalichaouch1991microwave} and interpreted as ``vestiges'' of the photonic band diagram of the periodic system. The structures do not need to be hyperuniform to exhibit gaps, but require a reasonable amount of short-range correlations. It is interesting to note that both the first and second gaps (in periodic arrays) are actually due to the same electric dipole resonance, while the intermediate conduction band is associated to the magnetic dipole resonance~\cite{vynck2009all}. In TE polarization (electric field in the propagation plane), a similar resonant behavior leading to a gap was pointed out by~\citet{o2002photonic} for very high index materials, but the gap closes for typical dielectric materials in the optical regime.

2D inverted structures made of circular air holes in dielectric exhibit photonic gaps that, by comparison, are much more sensitive to lattice deformations~\cite{yang2010photonic}, suggesting that periodicity, at least on a few periods, is required. It was shown however that connected networks made of thin dielectric walls on a stealthy hyperuniform pattern are favorable to exhibit a photonic gap in TE polarization~\cite{florescu2009designer}. First reports in non-periodic arrays were made on quasiperiodic structures~\cite{chan1998photonic}. This result is more unexpected than for the direct structures since one cannot define a unique scattering element in this case. Nevertheless, short-range correlations tend to homogenize locally the size distribution and shape of air pores, which, surrounded by dielectric walls in TE polarization, could be seen as nearly identical resonant scatterers. Though short-range correlations appeared being sufficient to open a photonic gap~\cite{froufe2016role}, a recent numerical investigation by \citet{klatt2022wave} showed that the apparent gap of many non-stealthy-hyperuniform structures actually closes at sufficiently large system sizes. This supports the conjecture that three attributes -- hyperuniformity, high degree of stealthiness ($\chi$-parameter) and bounded holes -- are necessary for a photonic gap to exist in the thermodynamic limit.

A greater challenge is to form photonic gaps in 3D disordered media. Studies nowadays tend to agree that the best solution for the purpose are 3D connected networks, basically consisting of air pores of nearly identical size surrounded by an array of dielectric rods~\cite{edagawa2008photonic, imagawa2010photonic, liew2011photonic, yin2012amorphous}. Recently, \citet{sellers2017local} put forward the idea of ``local self-uniformity'' to explain the formation of wide gaps. Such structures strongly ressemble foams~\cite{ricouvier2019foam, klatt2019phoamtonic}, which suggests the possibility to fabricate them with bottom-up techniques~\cite{maimouni2020micrometric, bergman2022macroporous}. Important efforts are underway to demonstrate experimentally 3D photonic gaps in the optical regime. First results on samples realized by direct laser writing and double inversion to increase the refractive index contrast indicate a depletion of transmission~\cite{muller2017photonic}. This feature could recently be pushed down close to the telecom wavelengths at 1.5~$\mu$m by heat-induced shrinkage of the network polymer template prior to silicon coating~\cite{aeby2022fabrication}.

\subsection{Mesoscopic transport and light localization} \label{sec:6-mesoscopic2D}

Mesoscopic transport in disordered systems refers to a regime wherein interferences between multiply-scattered waves lead to significant transport parameter deviations compared to classical approaches such as radiative transfer. Coherent effects, at the mesoscopic length scales, often lead to statistical distributions that are much broader and more complex than those expected from thermodynamic considerations. Signatures of mesoscopic effects, such as large sample-to-sample transmittance fluctuations, non-self-averaging transport parameters or long-range speckle intensity correlations, may still be visible on macroscopic scales provided that the signal has a sufficiently long coherence length compared to the characteristic lengths of the system. If many concepts in mesoscopic physics have been developed in the context of electronic transport~\cite{mello2004quantum, altshuler2012mesoscopic, sheng2006introduction, akkermans2007mesoscopic}, research on classical waves brought a great deal of new ideas and challenges to the topic~\cite{rotter2017light}, stimulated by the unique possibility to engineer the scattering materials at the subwavelength scale.

One of the most fascinating phenomena in mesoscopic physics of classical waves is the Anderson localization~\cite{anderson1958absence}, see~\cite{lagendijk2009fifty} for an historical overview of the topic and \cite{abrahams201050} for more technical details. The phenomenon takes its roots in the so-called weak localization effect, which describes a small reduction of the diffusion constant (compared to that predicted from radiative transfer) due to interference between counter-propagating waves. This effect requires reciprocity to hold~\cite{vantiggelen1998reciprocity}, which is generally the case in non-magnetic optical materials. Strong (Anderson) localization is obtained by a progressive renormalization of the diffusion constant that eventually leads to a complete halt of transport, as described by the self-consistent diagrammatic theory due to~\citet{vollhardt1980diagrammatic, vollhardt1982scaling}. In open finite-size systems, the localized regime is characterized by exponentially-decaying transmittance~\cite{vantiggelen2000reflection}, anomalous time-dependent response~\cite{skipetrov2006dynamics}, large transmitted speckle intensity fluctuations~\cite{chabanov2000statistical} and multifractality of the field~\cite{mirlin2006exact}.

A transition between extended and localized regimes is expected in three-dimensional (3D) systems when the scattering mean free path becomes comparable with the effective wavelength in the medium, $k_\text{r} \ell_\text{s} \approx 1$, also known as the Ioffe-Regel criterion~\cite{ioffe1960non}. Experiments on high-index semiconductor powders and photonic glasses, which offer amongst the smallest $\ell_\text{s}$ in optics, have failed to provide evidence of light localization~\cite{wiersma1997localization, scheffold1999localization, scheffold2013inelastic, sperling2013direct, skipetrov2016red}, contrary to studies on ultrasounds in elastic networks~\cite{hu2008localization} and matter waves in optical potentials~\cite{kondov2011three, jendrzejewski2012three}. It turned out that the key role of polarization for electromagnetic waves and near-field effects had been largely underestimated~\cite{skipetrov2014absence, bellando2014cooperative, naraghi2015near, cobus2022crossover}, thereby placing a finer engineering of the local morphology -- and of structural correlations -- at the heart of the problem.

In the literature, the challenge of reaching a localized regime in 3D in optics appears closely related to that of creating a photonic gap. In a founding work, \citet{john1987strong} proposed that a slight disorder in a periodic medium exhibiting a  photonic gap would promote Anderson localization near the gap edge, where some (but not all) propagation directions are inhibited. Anderson localization occurs in the band and differs in that sense from classical light confinement, where defect (cavity) modes -- or bound states -- are formed in the gap. This distinction has remained somewhat fuzzy in the literature in optics. Localized modes have been observed via numerical simulations in randomly-perturbed periodic inverse opals~\cite{conti2008dynamic}, where it was found that the strongest light localization was obtained at an optimal degree of disorder [Fig.~\ref{fig:Introduction}(d)], as well as in amorphous diamond structures~\cite{imagawa2010photonic} [Fig.~\ref{fig:Introduction}(e)], but their precise nature is unclear. The transition between extended and Anderson-localized regimes (outside the photonic gap) has been evidenced only recently in a numerical study on disordered hyperuniform structures thanks to a statistical analysis based on the self-consistent theory of localization~\cite{haberko2020transition, scheffold2022transport}. Although the effect of the kind of structural correlation on mesoscopic transport remains to be clarified, disorder engineering has clearly given a new hope for the experimental observation of 3D Anderson localization of light.

Light localization in two-dimensional (2D) disordered systems has experienced much less difficulties in comparison. Theoretical arguments developed for electronic transport~\cite{abrahams1979scaling} let us expect that all waves be localized on some length scale $\xi$ in two dimensions independently of the scattering strength of the medium. Despite the absence of a ``true'' transition, 2D systems have been very appealing because they can be fabricated, characterized (structurally and optically) and modelled much more easily than their 3D counterpart. The first report of localization of classical waves date back to~\citet{dalichaouch1991microwave} with a study of microwave propagation in high-index dielectric cylinders in TM polarization, where a link with photonic gaps was already made. The first experimental demonstration of Anderson localization in the optical regime was obtained by~\citet{schwartz2007transport} in photonic lattices consisting of evanescently coupled parallel waveguides wherein localization occurs in the transverse direction~\cite{deraedt1989transverse}. In this configuration, the electromagnetic problem is mapped onto the time-dependent Schr\"odinger equation where the propagation direction plays the role of time, enabling the exploration of many interesting problems in condensed matter physics~\cite{segev2013anderson, rechtsman2013photonic, weimann2017topologically}. 2D Anderson localization has later been reported for in-plane propagation of near-infrared light in suspended high-index dielectric membranes perforated by disordered patterns of holes~\cite{riboli2011anderson}. Quantum dots incorporated in the membrane are excited locally by a near-field probe and their photoluminescence is collected by the same probe at the same position. A post-treatment allows recovering spatial and spectral information on the resonant modes of the system~\cite{riboli2014engineering}.

Structural correlations have not been considered specifically in these early works, probably because they were not necessary to observe localized modes in 2D. Nevertheless, they can impact mesoscopic transport in mainly two ways: First, by creating a photonic gap in the vicinity of which localized modes appear, as shown experimentally in photonic lattices with short-range correlations~\cite{rechtsman2011amorphous} and randomly-perturbed periodic hole arrays in dielectric membranes~\cite{garcia2012nonuniversal}. The latter study shows that the stronger confinement in periodic systems is obtained with an optimal level of disorder, similarly to~\cite{conti2008dynamic}. Second, by modifying the scattering and transport parameters of disordered systems, which in turn modify the localization length $\xi$, as suggested by~\citet{conley2014light}. In 2D, small changes of structural correlations may induce variations of $\xi$ over orders of magnitude, since this quantity is expected to grow exponentially with the mean free path~\cite{abrahams1979scaling, sheng2006introduction}. This allows moving very easily from a quasi-extended regime to a localized regime in finite-size systems. Numerical simulations performed for TE-polarized waves in 2D disordered patterns of holes in dielectric confirm this possibility, although only a qualitative agreement with theoretical predictions is obtained. Note also that the study covers a short-range correlation up to the onset of polycrystallinity, which might affect localization.

\begin{figure}[htbp]
\centering
\includegraphics[width=\columnwidth]{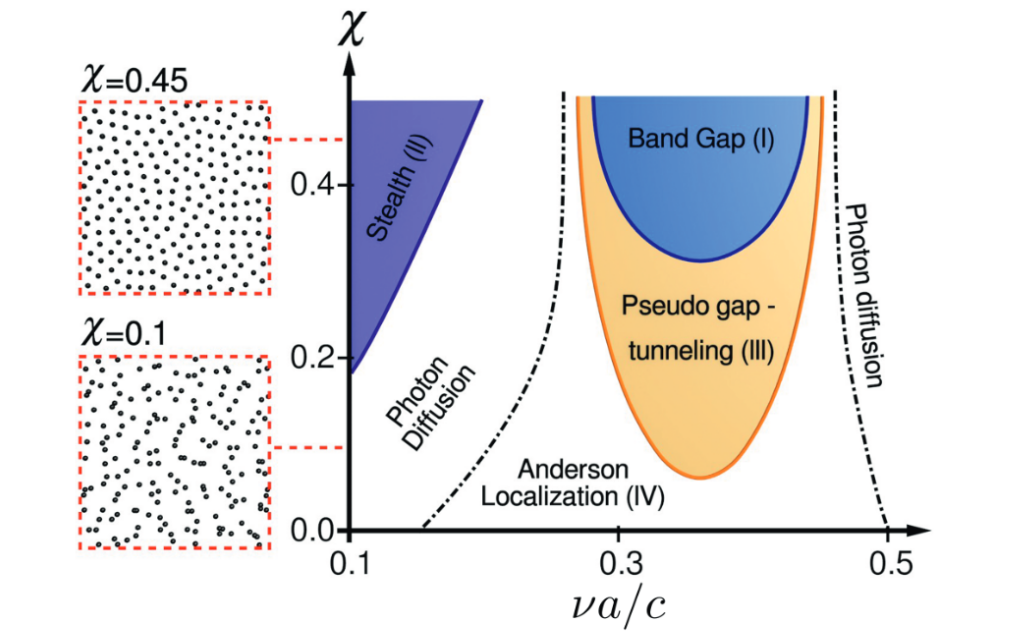}
\caption{\label{fig:TPD} (Color online) Correlation-frequency ($\chi-\nu$) transport phase diagram for 2D disordered hyperuniform media. The system is stealthy hyperuniform array of high-index dielectric rods and the wave is TM-polarized. $\chi$ is the degree of stealthiness~\cite{batten2008classical}, $\nu a/c \equiv a/\lambda$ is a reduced frequency with $a$ the mean distance between scatterers (related to the cylinder density). Five transport regimes may be identified, as discussed in the main text. Note that the transition between photon diffusion (quasi-extended regime) and Anderson localization depends on system size. Adapted with permission from~\cite{froufe2017band}.}
\end{figure}

The variety of mesoscopic transport regimes in 2D disordered media was investigated recently by~\citet{froufe2017band}, who proposed a transport phase diagram, shown in Fig.~\ref{fig:TPD}, for 2D stealthy hyperuniform structures made of high-index cylinders in TM polarization. Uncorrelated media (small values of $\chi$) experience the standard behavior with quasi-extended and localized regimes depending on the scattering strength of the cylinders and the system size. At the opposite, strongly correlated media (high values of $\chi$) are very transparent at low frequencies due to the suppressed single scattering over a finite range of scattering wavenumbers and exhibit a photonic gap (i.e., zero DOS in infinite media) at intermediate frequencies near the resonant frequency of the cylinder. The gap is surrounded by a low-DOS region containing weakly-coupled resonant states (defect modes) and the Anderson-localized regime. The phase diagram has been validated numerically~\cite{froufe2017band} and experimentally in the microwave regime~\cite{aubry2020experimental}. This diagram is specific to the considered system (including system size) and polarization. Nevertheless, it is quite representative of the different transport regimes that may be observed in correlated disordered media.

Localization of light in 2D ensembles of resonant scatterers in TE polarization encounters similar difficulties as in the 3D case due to the vectorial nature of light~\cite{maximo2015spatial}. A recent theoretical study by~\citet{monsarrat2022pseudogap} on hyperuniform patterns of high-quality-factor resonant dipole scatterers shows that localization of TE-polarized waves can occur at moderate scatterer densities concomitantly with the opening of a pseudogap, provided that a sufficiently high degree of short-range correlation is implemented. In essence, imposing a typical distance between resonant scatterers enables efficient destructive interference of vector waves, which leads to a depletion of the density of states and promotes the formation of localized states. Analytical expressions for the density of states and the localization length are rigorously established and found to agree well with numerical simulations. Clearly, the generalization of this theoretical framework to 3D resonant systems could contribute to unveiling the microscopic mechanisms behind the 3D Anderson localization of light.

\subsection{Near-field speckles on correlated materials} \label{sec:6-near-field-speckles}

Upon scattering by one specific realization of a disordered medium, a speckle pattern is formed~\cite{goodman2007speckle}. Universal intensity statistics are found in far-field speckle patterns, that are independent on the microscopic features of disorder. When speckle patterns are observed in the near field (i.e., at a distance from the output surface smaller than the wavelength of the incident light), the statistical properties of the speckle become dependent on the statistical features of the medium itself. In particular, as we will see, near-field speckles may exhibit direct signatures of the presence of spatial correlations in the scattering medium~\cite{carminati2010subwavelength, naraghi2016disorder, parigi2016near}. 

\subsubsection{Intensity and field correlations in bulk speckle patterns}

A standard observable in the study of speckles is the correlation function of the intensity fluctuations $\delta I$ at two different points $\r$ and $\r'$, defined as
\begin{equation}
\langle \delta I(\rg) \, \delta I(\rg^\prime) \rangle = \langle I(\rg) \,  I(\rg^\prime) \rangle
-\langle  I(\rg) \rangle \langle I(\rg^\prime) \rangle,
\end{equation}
with the intensity $I(\r) = |\E(\r)|^2$. As a measure of the degree of correlation of the intensity, one uses the normalized correlation function
\begin{equation}
C^I(\rg,\rg^\prime) = \frac{ \langle \delta I(\rg)\, \delta I(\rg^\prime) \rangle }
{\langle  I(\rg) \rangle \langle I(\rg^\prime) \rangle } 
\end{equation}
which, in terms of the field amplitude, is a fourth-order correlation function. In the weak-scattering regime $k_\text{r} \ell_\text{s} \gg 1$, the field is a Gaussian random variable. Indeed, the field at any point in the speckle results from the summation of a large number of independent scattering sequences, leading to Gaussian statistics by virtue of the central-limit theorem~\cite{goodman2015statistical}. Moreover, in a statistically homogeneous and isotropic medium, and far from sources, the speckle pattern can be considered to be unpolarized. In these conditions, the intensity correlation function factorizes in the form~\cite{carminati2021principles}
\be
    C^I(\r,\r')= \sum_i \left| C_{ii}^E(\r,\r') \right|^2, 
\label{eq:factor_CI_CU}
\ee
where $C_{ij}^E$  is the $(i,j)$ component of the normalized correlation function between two vector components of the field, or normalized coherence tensor, defined as
\begin{equation}
\mathbf{C}^E(\r,\r') = \frac{ \langle \E(\r) \otimes \E^*(\r') \rangle }
{\sqrt{\langle  I(\r) \rangle} \sqrt{\langle I(\r') \rangle } }.
\end{equation}
Recall that the spatial correlation of the field in the numerator is described by the Bethe-Salpeter equation [Eq.~\eqref{eq:th_bethe_salpeter_full_vector}].

Let us first consider the simplest model of an infinite medium illuminated by a point source at position $\rg_0$. For large observation distances ($|\rg- \rg_0| \gg \ell_\text{s}$ and $|\rg^\prime- \rg_0| \gg \ell_\text{s}$), one can derive the following general result~\cite{dogariu2015electromagnetic, vynck2014polarization, carminati2021principles}
\begin{equation}
\mathbf{C}^E(\r,\r') =  \frac{2 \pi}{k_\text{r}} \, \mathrm{Im} \langle \mathbf{G}(\r,\r') \rangle,
\label{eq:correlImG_scalar}
\end{equation}
where $\langle \mathbf{G} \rangle$ is the averaged Green function in the medium. This form of the field correlation function is always found under general conditions of statistical homogeneity and isotropy of the field~\cite{setala2003universality}.

In order to characterize the field spatial correlation averaged over the polarization degrees of freedom, one often introduces the degree of spatial coherence
\begin{equation}
\gamma^E(\r,\r') = \text{Tr} \left[ \mathbf{C}^E (\r,\r') \right].
\end{equation}
In an infinite medium, and for short-range correlation with $k_\text{r} \lcor \ll 1$, with $\lcor$ the correlation length of disorder, it is known that~\cite{carminati2015electromagnetic,carminati2021principles}
\begin{equation}
\gamma^E(\r,\r')  =\mathrm{sinc}\left(k_\text{r} R \right)\, \exp[-R/(2\ell_\text{s})] \, ,
\label{eq:correl_field}
\end{equation}
where $R=|\rg-\rg^\prime|$.
This expression takes the same form as that initially derived for scalar waves in~\cite{shapiro1986large}. In an infinite medium, for a Gaussian and unpolarized speckle pattern, the field and intensity correlation functions have a range limited by the wavelength $\lambda_\text{r}=2\pi/k_\text{r}$ and by the scattering mean free path $\ell_\text{s}$.

The impact of structural correlations in the medium on speckle correlations (of the field or intensity) can occur on different levels. First, the value of $\ell_\text{s}$ is directly dependent on the degree of correlation of disorder. Second, the general shape of the field correlation function can also be substantially modified when near fields cannot be ignored in either the illumination process (e.g., under excitation by a localized source inside the medium or close to its surface), or the detection process (e.g., detection at subwavelength distance from the surface). An example is discussed in the next subsection.

\subsubsection{Near-field speckles on dielectrics}

A speckle pattern observed at subwavelength distance from the surface of a disordered medium (near-field speckle) exhibits statistical properties that may strongly differ from the universal properties of far-field speckles. In the case of near-field speckles produced by rough surface scattering, it is known that in the single-scattering  regime the spatial correlation function of the near-field intensity is linearly related to the spatial autocorrelation function of the surface profile~\cite{greffet1995relationship}. In the case of speckles produced by volume multiple scattering, the degree of spatial coherence can be evaluated in a plane at a distance $z$ from the sample surface, in regimes ranging from the far field to the extreme near field~\cite{carminati2010subwavelength}. For $z\gg \lambda$, we obtain
\begin{equation}
    \gamma^E(\rg,\rg^\prime)  = \sinc(k_0 \rho) \, ,
\end{equation}
where $\rho$ is the distance separating the two observation points $\rg$ and $\rg^\prime$ in a plane at a constant $z$ (parallel to the sample surface). The width $\delta$ of the correlation function, that measures the average size of a speckle spot, is limited by diffraction and scales as $\delta \sim \lambda/2$. At subwavelength distance from the medium surface, near fields are dominated by quasi-static interactions. The scale of variation of the field is driven by geometrical length scales, and no more by the wavelength~\cite{greffet1997image, novotny2012principles}. Characterizing the structure by the correlation length $\lcor$, and assuming $\lcor \ll \lambda$, we can distinguish two regimes. For $\lcor \ll z \ll \lambda$, we have
\begin{equation}
\gamma^E(\rg,\rg^\prime)  = 
\frac{1-\rho^2/(8z^2)}{[1+\rho^2/(4z^2)]^{5/2}} \, ,
\label{eq:QSz}
\end{equation}
showing that $\delta \sim z$ due to quasi-static (evanescent) near fields. Finally, in the regime $\lcor \simeq z \ll \lambda$ (extreme near field), we obtain
\begin{equation}
\gamma^E(\rg,\rg^\prime) = M\left(\frac{3}{2},1,\frac{-\rho^2}{\lcor^2}\right)
\label{eq:QSlcor}
\end{equation}
where $M(a,b,x)$ is the confluent hypergeometric function, which here takes the form of a function decaying from 1 to 0 over a width $\delta \simeq \lcor$. In summary, according to the theory in \cite{carminati2010subwavelength}, we expect the speckle spot size to decrease in the near field as the distance $z$ to the surface, and to saturate at a size on the order of the correlation length of the medium. 

\begin{figure}[htbp]
\includegraphics[width=\columnwidth]{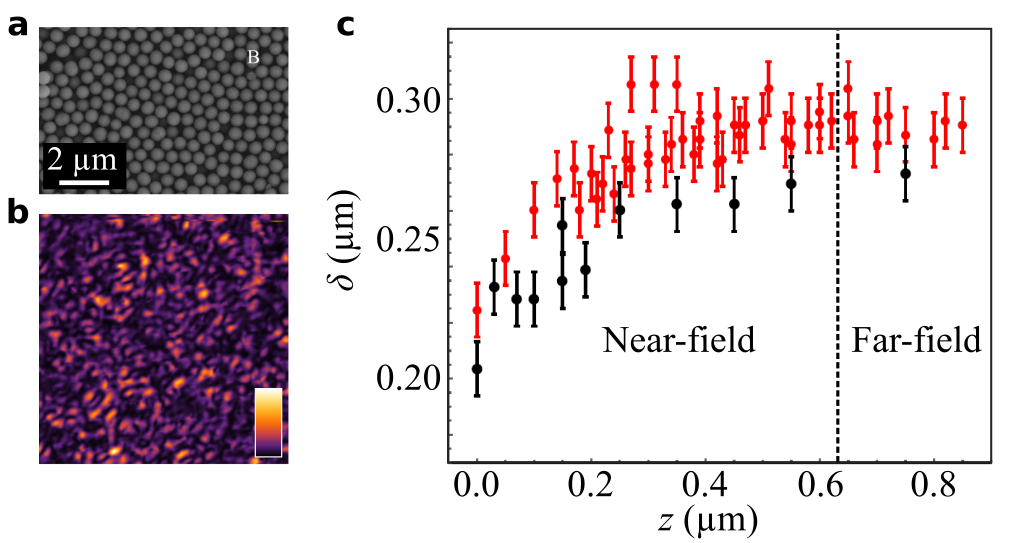}
\caption{(Color online) Signatures of structural correlations on near-field speckles. (a) Scanning electron microscope image of the surface of a typical sample, consisting of several layers of silica spheres in a partially ordered arrangement. (b) Example of speckle image recorded with a scanning near-field optical microscope at a wavelength $\lambda = 633$ nm. (c) Measured correlation length $\delta$ in the speckle pattern versus the distance $z$ to the sample surface, in the distance range for which the far-field to near-field transition is observed. The vertical dashed line corresponds to $z=\lambda$. The black and red (gray) markers correspond to two samples with an average diameter of the silica spheres $d =276$ nm and $d = 430$ nm. The very short-distance behavior is expected to depend on the level of short-range order in the sample (size and local organization of the spheres in space). Adapted with permission from \cite{parigi2016near}.\label{fig:NFspeckle_dielectrics}}
\end{figure}

The dependence of the speckle spot size at short distance can be probed experimentally using scanning near-field microscopy (SNOM). Studies have been reported in~\cite{emiliani2003near, apostol2003spatial, apostol2004first}. The behavior described above has been confirmed recently by~\citet{parigi2016near}, and the main result is summarized in Fig.~\ref{fig:NFspeckle_dielectrics}. The measurement provides the intensity correlation function $C^I(\rg,\rg^\prime)$, the width of which, according to Eq.~\eqref{eq:factor_CI_CU}, can be qualitatively compared to that of the degree of spatial coherence $\gamma^E$. By recording near-field speckle images at different distances $z$ from the surface of sample with correlated disorder, the dependence of the speckle spot size $\delta$ on the distance to the surface can be extracted. The result is displayed in Fig.~\ref{fig:NFspeckle_dielectrics}c. The decrease $\delta$ in the near field regime is clearly visible, as well as the non-universal dependence at very short distance (the two curves correspond to two samples with different structural correlations).

Retrieving information on structural correlations of disordered media from optical measurements is also possible via a stochastic polarimetry analysis of the scattered light~\cite{haefner2008stochastic}. As shown by~\citet{haefner2010scale}, the local anisotropic polarizabilities of a complex material generally depend on the volume of excitation (which may be controlled, for instance, via a near-field probe). It turns out that one can define a length scale corresponding to a maximum degree of local anisotropy, that is characteristic of the material morphology. This length scale has been evidenced in numerical simulations~\cite{haefner2010scale} but not yet experimentally to our knowledge.

\subsection{Local density of states fluctuations} \label{sec:6-LDOS-fluctuations}

The modification of the spontaneous emission rate from quantum emitters due to electromagnetic interaction with a structured environment is one of the major achievements in optics and photonics in the past decades~\cite{pelton2015modified}. As briefly discussed in Sec.~\ref{sec:6-photonic-gaps}, this effect is formally described by the local density of states (LDOS) that is expressed as a function of the Green tensor $\mathbf{G}(\mathbf{r},\mathbf{r})$ at the origin [Eq.~\eqref{eq:projected-LDOS}]. Expectedly, the LDOS should be highly sensitive to the local environment with which it interacts, especially in the near field.

The near-field interaction regime has been initially described using numerical simulations of LDOS distributions inside disordered media and a single-scattering theory~\cite{froufe2007fluorescence}. The model system is a spherical domain with radius $R$, filled with subwavelength dipole scatterers. The LDOS is calculated at the center of the domain and surrounded by a spherical exclusion volume of radius $R_0$. The length scale $R_0$ is a microscopic length scale that characterizes the local environment ($R_0$ can be understood as the minimum distance to the nearest scatterer). It was shown that the statistical distribution of the LDOS is strongly influenced by the proximity of scatterers in the near field, and by the local correlations in the disorder~\cite{caze2010near, leseur2017spatial}. As in the case of near-field speckle, this is a consequence of quasi-static near-field  interactions that make the LDOS sensitive to the local geometry.

\begin{figure}[htbp]
\includegraphics[width=\columnwidth]{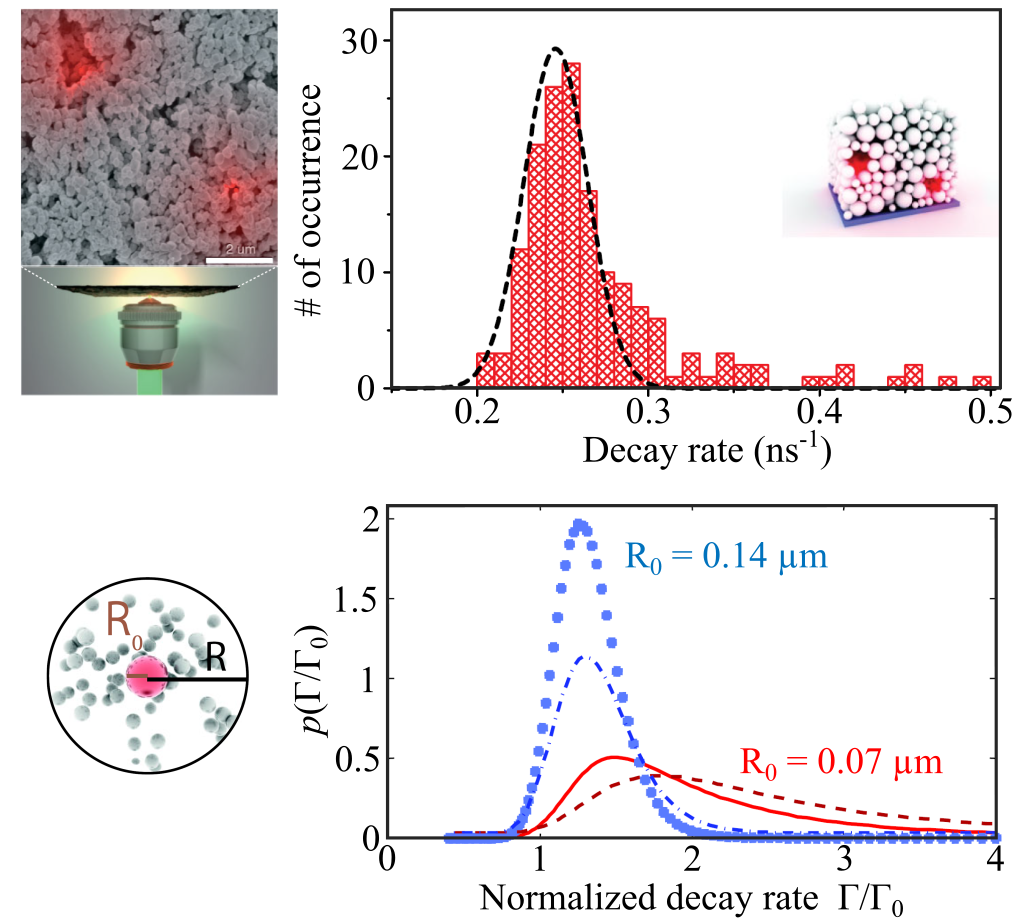}
\caption{(Color online) Impact of structural correlations on LDOS fluctuations. Top: Measured statistical distributions of the spontaneous decay rate $\Gamma \propto \rho$ (LDOS) of fluorescent beads (nanosources with 20~nm diameter) in a ZnO powder with transport mean free path $\ell_\text{t} = 0.9$~$\mu$m. A scanning electron microscope image of the sample is shown on the left, together with a schematic view of the illumination/detection geometry. The asymmetric shape of the statistical distribution of LDOS and the long tail is a signature of near-field interactions occuring inside the sample. Bottom: Numerical simulations of the statistical distribution of the normalized decay rate $\Gamma/\Gamma_0=\rho/\rho_0$ of a dipole emitter placed at the center of a disordered cluster mimicking the ZnO powder. The emitter is surrounded by an exclusion volume with radius $R_0$. This length scale describes local correlations in the positions of the scatterers in the sample. Blue (light gray) curves correspond to an exclusion radius $R_0=0.14$~$\mu$m while red (dark gray) curves correspond to an exclusion radius $R_0=0.07$~$\mu$m (the two curves in the same color correspond to two different densities of scatterers). The simulation demonstrates the substantial influence of $R_0$ (near-field interactions and local correlations in the disorder) on the shape of the distribution. Adapted with permission from~\cite{sapienza2011long}.\label{fig:C0_R0}}
\end{figure}

Statistical distributions of LDOS in strongly scattering dielectric samples have been measured experimentally at optical wavelength. The approach consists in dispersing fluorescent nanosources inside a scattering material~\cite{birowosuto2010observation, sapienza2011long}. Experiments mimicking the model systems studied theoretically use powders made of polydisperse spheres of high-index material (such a ZnO at wavelength $\lambda \sim 600-700$ nm). An example of measured LDOS distributions is shown in Fig.~\ref{fig:C0_R0}. The LDOS distribution (top panel), inferred from the distribution of the decay rate $\Gamma$ of nanoscale fluorescent beads, exhibits a high asymmetric shape with a long tail that is a feature of near-field interactions. This experiment confirms the sensitivity of LDOS fluctuations to the local environment in a volume scattering material in the multiple scattering regime. Comparison with numerical simulations (bottom panel) demonstrates the substantial role of the microscopic length scale $R_0$ on the shape of the distributions.

Disordered metallic films made by deposition of noble metals (silver or gold) on an insulating substrate (glass) are also known to produce large near-field intensity fluctuations close to the percolation threshold. On the surface of such materials, the near field intensity localizes in subwavelength domains (hot spots)~\cite{shalaev2007nonlinear,seal2005near,laverdant2008polarization}. The near-field LDOS exhibits enhanced spatial fluctuations in this regime, that reveal the existence of spatially localized modes~\cite{krachmalnicoff2010fluctuations,caze2013spatial,carminati2015electromagnetic}. Disordered metallic films close to percolation are an example of nanoscale disordered materials in which correlations in the disorder substantially influence the optical properties.


Being the LDOS very sensitive to small changes in the local environment of the emitter, the study of the statistics of LDOS, accessible through the decay rate $\Gamma$, can be related to the structural properties of a dynamical system of interacting, and hence correlated, scatterers. As an example, it has been numerically demonstrated that the statistical distributions of single emitter lifetimes in a scattering medium can evolve from a unimodal distribution to a different one when the system undergoes a phase transition. The regions of phase coexistence in small systems often turn out to be dynamical phase switching regions, where the entire systems switches between the two phases~\cite{briant1975molecular, berry1984melting, honeycutt1987molecular, labastie1990statistical, wales1994coexistence}. The signature of the phase switching regime in the $\Gamma$ statistics can be dramatic, since the distribution can be bimodal in the phase switching regime regions while unimodal in the pure phases. This striking behavior can be related to the statistics of neighboring scatterers surrounding the emitter and is not signaled by other light transport properties such as scattering cross section statistics for instance~\cite{de2016fluctuations}. Bimodal distributions of LDOS have been also described for emitters embedded in single layers of disordered but correlated lattices~\cite{de2014effect}.

\begin{figure}[htbp]
\includegraphics[width=\columnwidth]{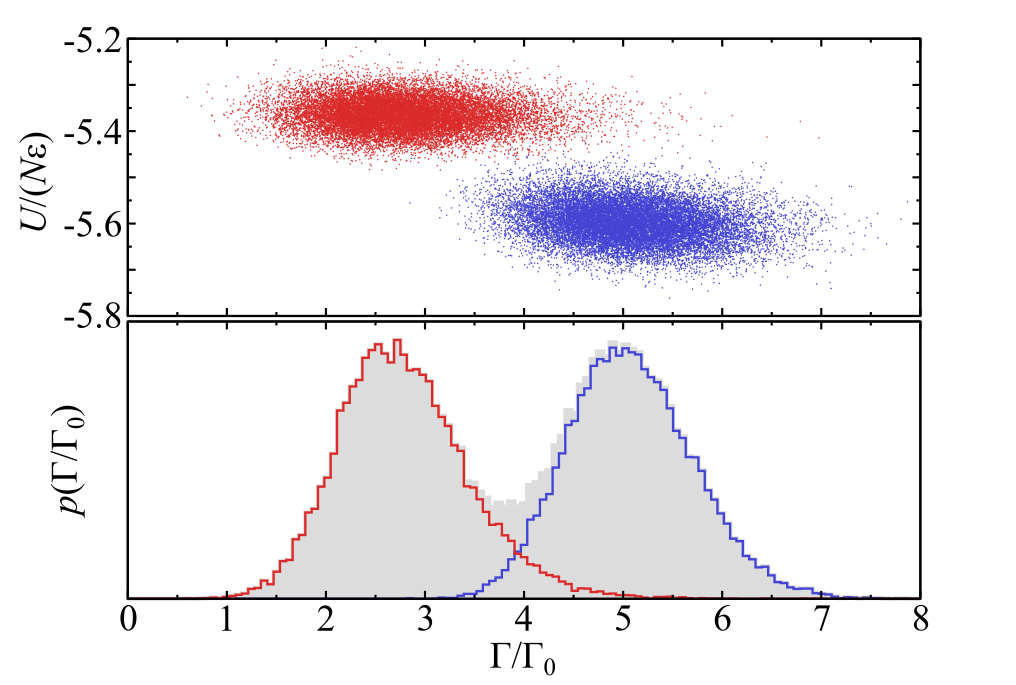}
\caption{\label{fig:LDOS_distrib_Sousa2016} (Color online) Signature of structural phase transition in LDOS statistics. Monte Carlo sampling of energy per particle normalized to the energy minimum $\varepsilon$ (top) and decay rate normalized to the vacuum one $\Gamma_{0}$ (bottom) for a single emitter placed at the center of a tightly confined system of resonant point scatterers interacting via Lennard-Jones potential. The temperature is such that the systems entirely switches between two phases randomly. On the bottom, the corresponding decay rates distributions are represented for the low and high energy branches. The gray area shows the sum of both distributions. Adapted with permission from \cite{de2016fluctuations}.}
\end{figure}

Figure~\ref{fig:LDOS_distrib_Sousa2016} shows numerical predictions for a system of $\sim 1000$ resonant point dipoles interacting through a Lennard-Jones potential and tightly confined within a spherical volume~\cite{de2016fluctuations}. The system is kept at a temperature corresponding to the liquid-gas transition. Due to strong finite size effects, the system is not in phase coexistence but rather switches randomly between the two phases. We see that the emitter decay rates are strongly correlated to the energetic state of the system, leading to two clearly-distinguishable modes. Thus, slight differences in structural correlations can clearly be identified by a statistical analysis of decay rate measurements.

Recent experiments have shown strongly inhibited spontaneous emission in systems undergoing an order-disorder phase transition~\cite{schops2018inhibited}. The formation of clusters exhibiting short-range correlations leads to a strong suppression of emission that is apparently comparable to that of an ordered structure. 

\section{Photonics applications} \label{sec:7}

The considerable advances in nanofabrication in the past decades have opened new opportunities in the engineering of disordered materials at the subwavelength scale. In this section, we describe the main applications of correlated disordered media in optics and photonics, namely in light management [Sec.~\ref{sec:7-thin-films}], random lasing [Sec.~\ref{sec:7-random-lasing}] and visual appearance design [Sec.~\ref{sec:7-visual-appearance}]. The interested reader will find more examples of photonic applications of correlated disorder in the recent review article by~\citet{cao2022harnessing}.

\subsection{Light trapping for enhanced absorption} \label{sec:7-thin-films}

Enhancing the interaction of light with matter is of paramount importance for various applications, including photovoltaics, white light emission and gas spectroscopy. The enhanced light-matter interaction generally translates into a stronger light absorption, be it exploited for photocurrent generation, converted into emission by fluorescence, or simply monitored.

The most popular light trapping strategy for thick ($L \gg \lambda$) bulk materials rely on randomly textured surfaces, acting as Lambertian diffusers to efficiently spread light along all directions within the medium for an arbitrary incoming wave~\cite{yablonovitch1982statistical, green2002lambertian}. Structural correlations on random rough surfaces provide angular and spectral control over scattering~\cite{martins2013deterministic}, described via the so-called Bidirectional Scattering Distribution Function (BSDF)~\cite{stover1995optical}. Volume scattering constitutes an interesting alternative to surface scattering, as multiple scattering tends to increase the interaction between light and matter~\cite{rothenberger1999contribution, muskens2008design, mupparapu2015path, benzaouia2019solar}. Quite counterintuitively, one should note that the average path length for a Lambertian illumination in non-absorbing media is independent of the scattering strength of the material~\cite{pierrat2014invariance} and equivalent to the surface scattering light trapping, as predicted from the equipartition theorem~\cite{yablonovitch1982statistical} and as verified experimentally recently~\cite{savo2017observation}. The absorption efficiency therefore depends strongly on the ratio between scattering and absorption. The benefit of structural correlations on light absorption in disordered media has only been considered recently by~\citet{bigourdan2019enhanced} and \citet{sheremet2020absorption}, who showed by theory and numerical simulations that stealthy hyperuniform patterns of absorbing dipolar particles enhance the overall absorption of the medium (compared to the uncorrelated system) close to an upper bound.

Stimulated by technological development in next-generation photovoltaic panels, considerable efforts have been dedicated to light trapping in thin films ($L \approx \lambda$) in the past two decades, exploiting coherent phenomena as a new means for enhancing light-matter interaction~\cite{fahr2008engineering, mokkapati2012nanophotonic, gomard2013photonic}. Fundamentally, coupling between an incident planewave and a resonant mode in the layered medium is enabled by fulfilling a matching condition between the projected wavevectors parallel to the interface $\mathbf{k}_{||}$ in the two media~\cite{yu2010fundamental}. For periodic photonic crystals, this condition is found for leaky Bloch modes having wavevectors in the light cone $k_{\text{B},||} < k_0 n_\text{str}$, where $n_\text{str}$ is the refractive index of the superstrate or substrate. In periodic structures, the absorption peaks are spectrally narrow and strongly depend on $k_{\text{B},||}$. A further improvement can be obtained by creating imperfections that broaden the spectral and angular response, leading to an overall improved optical efficiency~\cite{oskooi2012partially, peretti2013absorption}. 

\begin{figure}[htbp]
\centering
\includegraphics[width=\columnwidth]{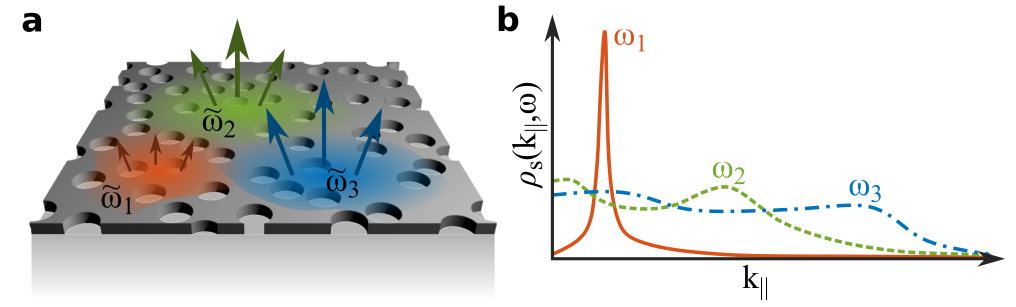}
\caption{\label{fig:light_trapping} (Color online) Process of light coupling and decoupling between a thin dielectric membrane and free-space modes. (a) Sketch of a photonic structure with correlated disorder containing several leaky resonant modes (QNMs). The QNMs are described by different complex frequencies and are coupled to free-space modes. (b) Spectral functions of a short-range correlated disordered photonic structure at different frequencies. At low frequencies, the spectral function is a narrow peak. The value at $k_{||}=0$ provides information on the coupling efficiency at normal incidence. At higher frequencies, the peaks broaden as a result from stronger scattering and reach higher values for small wavevectors, indicating more efficient coupling.}
\end{figure}

For disordered media, the quantity of interest is the so-called spectral function, defined as~\cite{sheng2006introduction},
\begin{equation}\label{eq:spectral-function}
\rho_\text{s}(\mathbf{k},\omega) = \frac{2 \omega}{\pi c^2} \text{Im}\left[ \text{Tr} \langle \mathbf{G}(\mathbf{k},\omega) \rangle \right],
\end{equation}
which is the average density of states resolved in spatial frequencies and is obtained from the Fourier transform of the average Green tensor $\langle \mathbf{G}(\r,\r',\omega) \rangle$. At a given frequency, the spectral function typically exhibits a peak centered on the effective wavevector in the disordered medium, $k_\text{r}$, and a width that is inversely proportional to the extinction mean free path, see Fig.~\ref{fig:light_trapping}. As shown by \citet{vynck2012photon}, short-range correlations allow a fine tuning of the spectral function, including in the radiative zone, eventually leading to a spectrally and angularly optimal light absorption~\cite{pratesi2013disordered, bozzola2014broadband}. It has been suggested that stealthy hyperuniform structures can lead to even higher overall absorption efficiency (integrated over a spectrum of interest) compared to short-range correlated and periodic media~\cite{liu2018role}. Experiments on correlated disordered hole patterns~\cite{trompoukis2016disordered}, nanowire arrays exhibiting fractality on some scale~\cite{fazio2016strongly}, complex nanostructured patterns~\cite{lee2017concurrent} and hyperuniform structures~\cite{piechulla2021antireflective, tavakoli2022over} have clearly demonstrated the benefit of disorder engineering for light trapping.

Finally, let us point out that the coupling process between free space and thin-film layers is very relevant also in the optimization of light extraction from light-emitting devices like organic LEDs~\cite{gomard2016photon}, where correlated disordered photonic structures could be realized on large scales, for instance, by inkjet-printing of polymer blends~\cite{donie2021phase}.

\subsection{Random lasing} \label{sec:7-random-lasing}

\begin{figure}[htbp]
\includegraphics[width=\columnwidth]{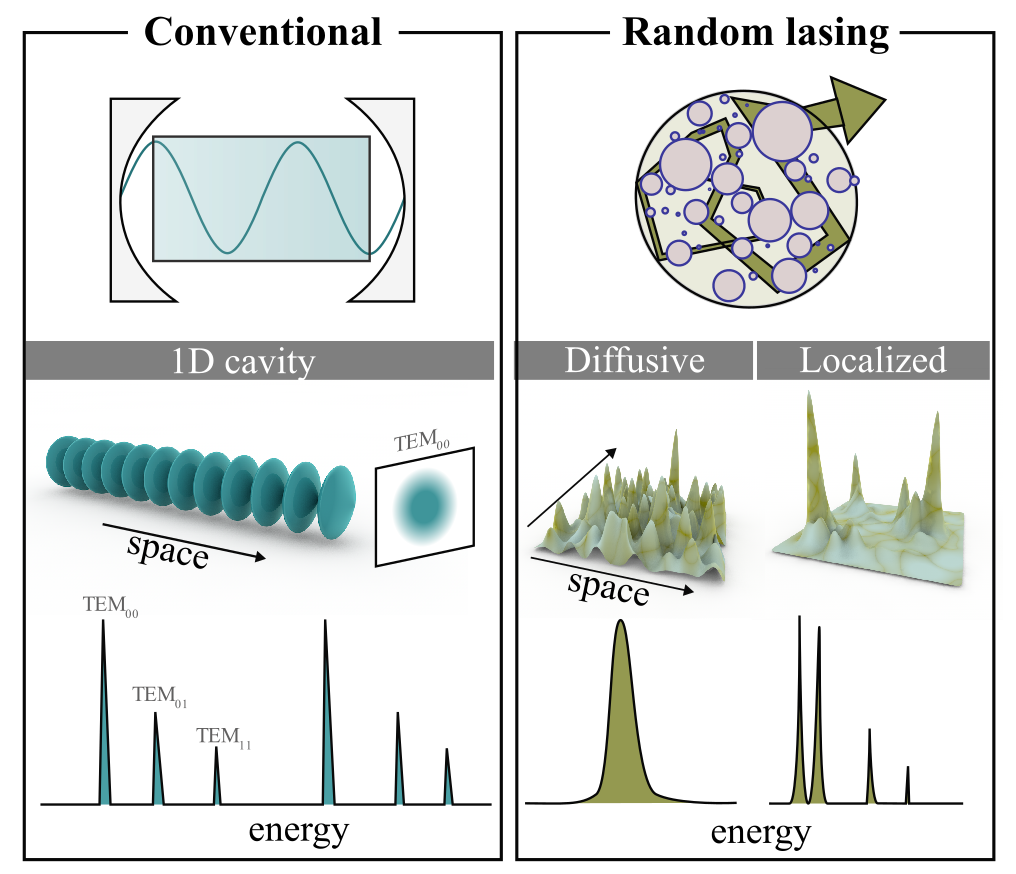}
\caption{\label{fig:randomlasingintro} (Color online) Conventional vs Random lasing. While a conventional laser (left) is usually composed of a two-mirrors cavity which defines the optical modes, a random laser (right) exploits the confinement by multiple scattering to enhance the probability of stimulated emission. It lases on the ``speckle'' modes of the disordered medium, either delocalized (bottom left) or localized (bottom right). In both lasers lasing occurs when the gain is larger than the losses, above a certain pumping threshold energy, when stimulated emission becomes the dominant emission process. Lasing peaks can appear in both diffusive or localized regimes, but are easily washed out by temporal or spatial averaging in diffusive media. Adapted with permission from~\cite{sapienza2019determining}.}
\end{figure}

Random lasers, where light is trapped in the gain medium by multiple scattering, offer new possibilities for efficient lasing architectures. The disordered matrix folds the optical paths inside the medium by multiple scattering, effectively increasing the probability of stimulated emission, which in turns provides optical gain and the amplification that triggers lasing~\cite{wiersma2008physics, cao2005review}, see Fig.~\ref{fig:randomlasingintro}.
	
Its functioning principle is the same as in conventional lasing but without the need for carefully aligned optical elements. The emission of a random laser is also surprisingly coherent, with photon statistics close to that of normal laser emission~\cite{florescu2004photon}, with strong mode coupling~\cite{tureci2008strong} non-trivial modes organization (symmetry replica breaking)~\cite{ghofraniha2015experimental} and unbounded (L\'evy distributed) intensity fluctuations~\cite{uppu2015exponentially}. Due to the volumetric and (on average) isotropic nature of its lasing patterns a random laser is expected to feature $\beta$-factors close to one (i.e., a threshold-less behaviour~\cite{van2002beta}). The result (Fig.~\ref{fig:randomlasingintro}) is an opaque medium in which laser light is generated along random paths in all directions, and over a broad spectral range, with complex temporal profiles~\cite{leonetti2011mode}. 

In the multiple scattering regime, the random lasing threshold can be related to a critical volume or size, such that lasing action can only be achieved for sample sizes larger than this critical dimension. Similar to the critical volume in a neutron bomb, the critical size ensures that the photons sustain net amplification and therefore that the light emerging from the sample is mostly due to spontaneous emission.
For a three-dimensional scattering medium, with isotropic scattering and no correlation, embedded in a slab geometry, the critical thickness $L_\text{cr}$ has been calculated by using the radiative transfer equation~\cite{pierrat2007threshold} and is solution of
\begin{equation}
    \frac{1}{\ell_\text{s}}-\frac{1}{\ell_\text{g}}
        =\frac{\pi}{(L_\text{cr}+2z_0)\tan\left(
            \pi\ell_\text{s}/(L_\text{cr}+2z_0)
        \right)}
    \label{eq:Dcr}
\end{equation}
where $z_0=0.7104\ell_\text{s}$ is the extrapolation length and $\ell_\text{g}$ the net-gain length. Eq.~\eqref{eq:Dcr} reduces to $L_\text{cr}= \pi\sqrt{\ell_\text{s}\ell_\text{g}/3}$ in the diffusive limit with $z_0=0$. For anisotropic scattering, we get $L_\text{cr}= \pi\sqrt{\ell_\text{t}\ell_\text{g}/3}$. In typical samples, the critical length is of the order of 1 to 100~$\mu$m (e.g., $L_\text{cr}\sim \ell_\text{t}$ with $\ell_\text{t} \sim 4~\mu$m in~\cite{caixeiro2016silk} and $L_\text{cr}\sim 300~\ell_\text{s}$ in~\cite{froufe2009threshold}).
	
The initial scattering architectures for lasing have been 3D disordered semiconductors powders or randomly fluctuating colloids in solution, which can be well thought of and described as a random cloud of dipoles (i.e., without any correlation). Pure randomness is the assumption which simplifies the complexity of the problem to make it treatable with theoretical models. Despite the many successes of this type of uncorrelated disorder, a new generation of disordered lasing architectures, with more robust and collective light-trapping schemes~\cite{gottardo2008resonance} and new topologies~\cite{gaio2019nanophotonic} has emerged.
	
In particular, spatial correlations between scatterers is a very effective approach for tuning the spectral properties, the number of lasing modes and their threshold by designing photonic band-edge states at the position of the gain. For example, localized modes near the edge of a (2D) photonic gap have been exploited for random lasing~\cite{liu2014random} and single-mode operation has been achieved in compositionally disordered photonic crystals~\cite{lee2019taming}. The role of the gap edge has been highlighted in semiconductor membranes with pseudo-random patterning~\cite{yang2010demonstration}, randomly mixed photonic crystals~\cite{kim2010band} and amorphous network structures~\cite{wan2011time}, while in photonic amorphous structures, the short-range order improves optical confinement and enhances the quality factor of lasing modes~\cite{yang2011lasing}.
	
Modelling lasing action in correlated disordered media is often a challenge. In particular lasing occurs for the modes with highest net gain, often escaping the transport models which instead deal with the average intensity. While a full-wave solution of the Maxwell’s equations, coupled to the dynamics of the gain, as for example Maxwell-Bloch models~\cite{conti2008dynamic} would contain all the relevant phenomena, it is very hard to implement in realistic samples due to its computational requirements. Advanced \textit{ab-initio} theoretical models have been developed. Let us mention the self-consistent laser  theory~\cite{tureci2008strong, ge2010steady}, which relies on a decomposition of the lasing field on a basis of resonant modes, and the Euclidean matrix theory by~\citet{goetschy2011euclidean}, which relies on analytical predictions for the random Green's matrix of a system. Alternatively, more simplified models that neglect the coherence of the modes and describes transport with the radiative transfer equation (or within the diffusion approximation) can be used. The diffusion approximation stems from the initial proposal by~\citet{letokhov1968generation} and simplifies the calculations significantly~\cite{gaio2015tuning, wiersma1996light}. These models can be extended to include scattering correlations, to modify the scattering and gain parameters, following the theory described in Sec.~\ref{sec:2}.

\subsection{Visual appearance} \label{sec:7-visual-appearance}

\subsubsection{Photonic structures in nature}

Living organisms produce a vast variety of photonic mechanisms to modulate their visual appearance by exploiting a wide range of biopolymers and architectures. Colors produced by these organisms are referred to as structural colors as they are mainly influenced by the nanostructural features of the materials, rather than pigments.
The lack of consistent methods and tools of analysis, as well as the incredibly large number of species showing different architectures however makes it difficult to categorize natural photonic structures.
Distinct species use different materials, structures and strategies for as many biological functions (to attract mates, hide from predators or act as a defence mechanism)~\cite{whitney2009floral, vignolini2012pointillist, seago2008gold, wilts2014sparkling}. Another degree of difficulty arises from the fact that such natural architectures are often hierarchical and their visual appearance depends on several factors, including geometrical features, the addition of absorbing pigments and finally, the visual system for which such structures are build (different animals and insects have different perceptions).

Only a limited selection of explanative examples will be discussed and analyzed hereafter. The reader should keep in mind that this represents a minimal fraction of the efforts that have been done in this field to systematically characterize and categorize natural photonic structures.
For consistency with the scope of this review, we only mention in passing the case of one-dimensional disordered multilayered structures, which are found in certain beetles~\cite{hunt2007comprehensive, del2016polarizing, onelli2017development}, butterflies~\cite{bossard2016evolving}, leaves~\cite{vignolini2016structural} and algae \cite{chandler2017structural}. Imperfect one-dimensional grating structures, which play an important role for flowers to  enhance signalling to pollinators~\cite{moyroud2017disorder}, for instance, will also not be discussed further.

\begin{figure}[htbp]
\includegraphics[width=8cm]{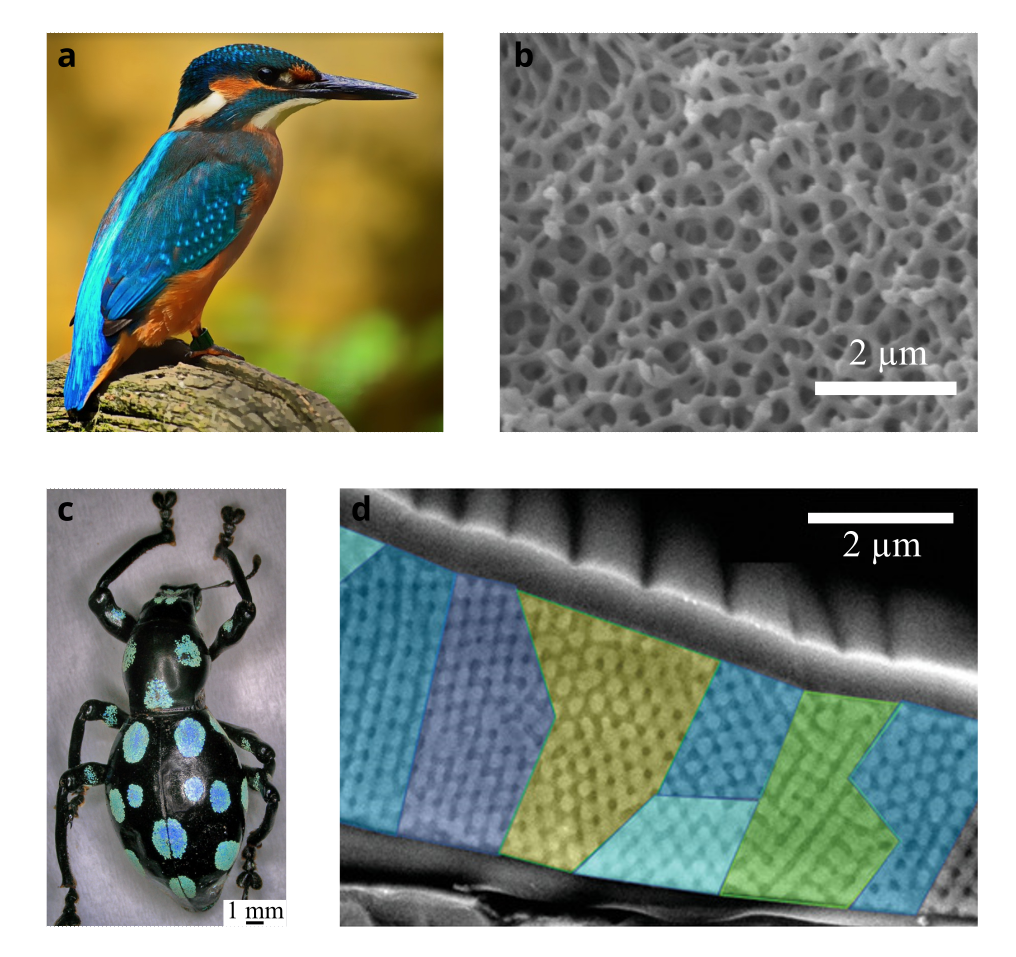}
\caption{\label{fig:colours_nature} (Color online)
(a) Kingfisher, the angular independent blue (light gray) coloration of the bird feather is the result of a correlated 3D structure, which is shown in the SEM image in (b). Image courtesy from Pixabay and Bodo Wilts (University of Salzburg, Austria), respectively. See \cite{stavenga2011kingfisher} for more information.
(c) Pachyrhynchus sarcitis Weevils, the blue (light gray) colored spot in the weevil exoskeleton are the results of a polydomain diamond photonic structures, shown in the SEM image on (d). Adapted with permission from \cite{chang2020hereditary}.}
\end{figure}

The most widespread family of (2D or 3D) correlated disorder in nature is the short-range correlation, which generally aims at producing angular-independent colorations. Short-range correlated structures are found in butterflies~\cite{prum2006anatomically}, bacteria~\cite{schertel2020complex}, and many animals. Probably the most famous examples are the structures found in the feathers of the eastern bluebird \emph{Cotinga maynana}~\cite{prum1998coherent} and of the Kingfisher~\cite{stavenga2011kingfisher}, which present a short-range correlation of keratin fibrillary network but also several other one such as the \emph{I. puella} \cite{noh2010noniridescent}. Correlated ensembles of collagen spheres producing an angle-independent color are found in avian skin~\cite{prum2003structural} and mammalian skin (primates)~\cite{prum2004structural}. Interestingly, these types of structures are always producing blue and green colorations in nature~\cite{magkiriadou2014absence, jacucci2020limitations, hwang2021designing, jeon2022eurasian}. 

Many examples of polycrystalline structures are exploited for coloration in insect wings using three-dimensional photonic crystal structures such as diamond and gyroids. In these cases, polydomains provide a more angular independent response which might again be functional for signalling and camouflaging~\cite{michielsen2007gyroid}. Examples are the diamond-like structures observed inside the scales of the weevils \textit{Lamprocyphus augustus} \cite{galusha2008discovery}, \textit{Entimus imperialis} \cite{wilts2012brilliant} and \textit{Pachyrhynchus weevils} \cite{chang2020hereditary}. Similarly, three-dimensional gyroid structures are found in many lycaenid and papilionid butterflies by \cite{michielsen2007gyroid} and in various species such as \textit{C. rubi} \cite{michielsen2009reflectivity, schroder2011chiral}, \textit{C. remus}, \textit{P. sesostris }\cite{wilts2011iridescence}, and \textit{T. opisena}~\cite{wilts2017butterfly}.

Distinct from the above cases, anisotropic network-like structures can be optimized to enhance whiteness. Several natural examples exist, see \cite{jacucci2021light} for a recent review. A notable example of optimized whiteness is found on the scales of the beetle genus \textit{Cyphochilus}, which shows a brilliant white coloration while being only 5-7~$\mu$m thick~\cite{vukusic2007brilliant, burg2019liquid, luke2010structural, burresi2014bright, wilts2018evolutionary}. Optical and anatomical studies confirm that anisotropy of the random polymeric network structure in such beetles scales are crucial for scattering optimisation at low refractive indices~\cite{cortese2015anisotropic, jacucci2019role, utel2019optimized, lee2020anisotropic}.

\subsubsection{Synthetic structural colors}
		
The ability to control visual appearance with correlated photonic structures, both in terms of color and scattering response, is critical in photonic pigments. With the improvements in the fabrication techniques, it is now possible to assemble such photonic materials cheaply and on a large scale and therefore, their use to replace traditional pigments is becoming a reality~\cite{goerlitzer2018bioinspired, lan2018unexpected, saito2018replication}. Of particular interest here are short-range correlated structures~\cite{shi2013amorphous}. The simplest way to achieve such materials in films consists in a rapid drying of colloidal suspensions to form photonic glasses~\cite{garcia2007photonic, forster2010biomimetic, schertel2019structural}. Combined with additive manufacturing techniques, one can fabricate complex-shaped objects exhibiting diffuse colors~\cite{demirors2022three}. Several types of colloidal particles, functionalisation and matrices have been proposed to enlarge the color palette with these structures \cite{forster2010biomimetic, schertel2019structural, ge2015robust, hantsch2019ysz, kim2021direct} and several tricks have been proposed to improve color contrast and appearance \cite{hwang2020effects, hantsch2021tailoring} also exploiting absorbing species, such as carbon black~\cite{takeoka2013production}. However, all these approaches have so far only been capable of providing only faint blue and green colors. In order to expand the visible palette toward red hues, core-shell photonic glasses~\cite{kim2017inverse, shang2018photonic} and inverse photonic structures have been exploited~\cite{zhao2020angular} -- however their color purity and the reflected intensity remain limited~\cite{jacucci2020limitations}. Short-range crystalline systems with carefully tuned domain orientations or the geometry of the system might allow to overcome this issue~\cite{song2019hierarchical}.


The design of structural colors with artificial materials also raises questions about their predictability with theoretical models or numerical methods, and the inverse design of artificial materials.

The starting point to predict a color is the computation of the reflectance or transmittance spectra of the disordered material. These spectra are then weighted by the spectral power distribution of the illuminant and by color matching functions for the chromatic response of the observer, to be finally projected onto a specific color space~\cite{ohta2006colorimetry}, such as CIE 1931 XYZ. The computation of the reflectance and transmittance spectra is evidently the most tedious step. Most studies have relied on finite-difference time-domain (FDTD) simulations~\cite{taflove2005computational}, for instance for 3D particulate media~\cite{dong2010structural} and porous dielectric networks~\cite{galinski2017scalable}, yet at the cost of heavy computational loads (although this is mitigated by efficient parallelization). Analytical expressions based on diffusion theory have also been used~\cite{schertel2019structural}, but care should be taken on the validity of diffusion approximation ($L/\ell_\text{t} \gg 1$). A good alternative is to rely on Monte Carlo light transport simulations~\cite{wang1992monte, alerstam2008parallel} -- the numerical counterpart of radiative transfer -- wherein positional correlations can be taken into account analytically via Eqs.~\eqref{eq:th_scattering_length_structure_factor} and \eqref{eq:th_transport_length_structure_factor} and assuming that an effective index can be defined. This numerical approach was used to unveil the importance of the packing strategy of photonic glasses on their color saturation and angle-dependence~\cite{xiao2021investigating}, explore efficiently the parameter space~\cite{hwang2021designing}, and investigate the potential of random dispersions of photonic balls~\cite{yazhgur2022inkjet} for coloring applications~\cite{stephenson2022predicting}.

The inverse design of structural colors is, by comparison, still in its infancy. The aforementioned Monte-Carlo approach by~\citet{hwang2021designing} has been combined with Bayesian optimization to determine the experimental parameters required to reach a target color. Powerful topology optimization techniques~\cite{jensen2011topology}, also known as adjoint methods, have been used recently to design complex dielectric network materials creating targeted colors in reflection~\cite{andkjaer2014inverse, auzinger2018computational}. Although the role of structural correlations on coloration is implicit in this case, a subsequent structural analysis of the optimal designs could lead to the definition of recipes for materials creating vivid colors.

\section{Summary and perspectives} \label{sec:8}

Research on disorder engineering in optics and photonics has considerably grown in the past decade stimulated by the advent of new concepts and applications. In this last section, we attempt to identify some of the most promising developments for future research along with the theoretical and experimental challenges that will need to be tackled.

\subsection{Near-field-mediated mesoscopic transport in 3D high-index correlated media}

Multiple light scattering in disordered media has been treated for many years as a process wherein the vector nature of light could be simplified either by keeping its transverse component only, as we have done in Sec.~\ref{sec:2}, or by treating light just as a scalar wave~\cite{akkermans2007mesoscopic}. Whereas these approximations may be well justified in dilute media (for both) and far from any polarized source in an opaque medium (for the latter), it recently turned out that the importance of the longitudinal component, which appears in the near-field regime, on mesoscopic transport in dense systems has been largely underestimated so far~\cite{skipetrov2014absence, naraghi2015near,  naraghi2016phase, escalante2017longitudinal, cobus2022crossover, monsarrat2022pseudogap}. We believe that this aspect deserves full attention from the community. A first attempt to incorporate the longitudinal component in the theory has been proposed by~\citet{vantiggelen2021longitudinal} for random ensembles of resonant point scatterers, giving physical ground to the existence of near-field channels in light transport. Near-field interaction processes are dramatically impacted by subwavelength-scale structural correlations, as we have seen in Sec.~\ref{sec:6-near-field-speckles}, and developing a theoretical framework to describe radiative transfer in arbitrary correlated media including the near-field contribution would certainly be an important step forward.

Related to this are the determination of effective material parameters for dense, resonant disordered media and their use to describe light scattering and transport, which are still matter of investigation~\cite{aubry2017resonant, yazhgur2021light, yazhgur2022scattering}, as quantitative agreement with experiments and numerics has remained difficult to reach with classical models. On this aspect, let us point out the recent works by~\citet{gower2019multiple, gower2019proof}, demonstrating that multiple coherent waves with different wavenumbers (at fixed frequency) should actually contribute to the average field. This may have important consequences for scattering by finite-size systems~\cite{gower2021effective} and raises the question of whether these multiple waves are affected in a similar way by structural correlations.


The prominent role played by the precise morphology of 3D disordered media on the emergence of photonic gap and Anderson-localized regimes also deserves clarification. 3D high-index connected (foam-like) structures appear as the best candidates for this purpose according to numerical simulations~\cite{imagawa2010photonic, liew2011photonic, sellers2017local, klatt2019phoamtonic, haberko2020transition} but the underlying physical mechanisms have remained difficult to grasp, thereby calling for further theoretical advances~\cite{scheffold2022transport}. In addition to near-field effects, future works may need to consider high-order $n$-point correlation functions (with $n>2$) in the description of structural characteristics~\cite{torquato2013random, torquato2021nonlocal} as well as high-order diagrams in the multiple-scattering expansion~\cite{vollhardt1980diagrammatic}, which can contribute significantly in strongly correlated media as shown recently~\cite{leseur2016high}. Numerical investigations will continue in parallel, and we believe that progress would greatly benefit from the development of numerical methods to solve electromagnetic problems on large systems more efficiently~\cite{egel2017celes, bertrand2020global, egel2021smuthi, valantinas2022computing, lin2022fast}.

Experimental demonstrations of photonic gaps and 3D Anderson localization of light in the optical regime have remained out of reach until now and would be great scientific milestones. The main challenge to overcome at this stage is the fabrication of 3D connected structures with finely-tuned correlated morphologies with sufficiently high refractive indices (ideally offering an index contrast above 3) and sufficiently large thicknesses ($L \gg  \ell_\text{t}$). The steady progress on bottom-up approaches such as bio-templating~\cite{galusha2010diamond}, DNA-origami~\cite{zhang2017dna} and microfluidic-based foam processing~\cite{maimouni2020micrometric} gives hope for first successful realizations in the next few years. As a longer-term objective, the design and fabrication of 3D \textit{stealthy} hyperuniform media would be an outstanding result. Ultimately, the availability of such high-index nanostructured materials will unlock the possibility to explore experimentally the physics of mesoscopic phase transitions~\cite{evers2008anderson} for (polarized) electromagnetic waves.

\subsection{Mesoscopic optics in fractal and long-range correlated media}

Light propagation in positively-correlated media is characterized by a non-exponential decay of the coherent intensity. As discussed in Sec.~\ref{sec:5-fractal}, materials exhibiting fractal heterogeneities in the form of non-scattering regions of varying sizes can lead in certain conditions to a superdiffusive behavior~\cite{burioni2014superdiffusion, savo2014walk}. Anomalous transport processes~\cite{klages2008anomalous} and dynamics on fractal networks~\cite{nakayama1994dynamical} have a long history, but optical studies on fractal media have so far been mostly concerned with structure factor measurements in the single scattering regime~\cite{lin1989universality} (note that the optical properties of semicontinuous metal films near percolation, on which there exists a vast litterature~\cite{shalaev2007nonlinear}, strongly rely on near-field plasmonic effects and not on light transport). Coherent optical phenomena in ``L\'evy-like'' media have been sparsely addressed until now~\cite{burresi2012weak}. Multiple scattering formalisms have been extended to media described by fractal dimensions~\cite{akkermans1988theoretical, wang1994scattering} or exhibiting superdiffusion~\cite{bertolotti2010multiple}, disregarding however several difficulties related to the definition of the self-energy and the effective index~\cite{tarasov2015fractal}, and perhaps more importantly to the quenched nature of disorder~\cite{burioni2014superdiffusion, barthelemy2010role}. All in all, the development of a rigorous \textit{ab-initio} theory for multiple light scattering in strongly-heterogeneous materials would be a formidable achievement.

Numerical and experimental studies on 1D and quasi-1D L\'evy-like systems have revealed anomalous conductance fluctuations and scaling~\cite{fernandez2014beyond, ardakani2015controlling, lima2019dirac}. Higher-dimensional systems are likely to exhibit a similarly rich physics, as illustrated recently~\cite{chen2022enhanced}, which yet remains to be explored. One example is the critical dimension of 2 above which the Anderson transition exists~\cite{abrahams1979scaling} that may be lowered, depending on the fractality or lacunarity of the system. Optical experiments and numerical simulations could be performed in this regard on high-index planar photonic structures similarly to~\citet{riboli2014engineering}, giving access to LDOS statistics, or to~\citet{yamilov2014position} for transmittance and internal light intensity measurements.

An alternative route for the experimental study of mesoscopic phenomena in long-range correlated systems could rely on disordered photonic network~\cite{gaio2019nanophotonic} -- an optical implementation of random graphs~\cite{janson2011random} --, wherein light propagates through 1D waveguides and is scattered at the waveguide vertices. The waveguide lengths and vertices connectivity thus take the role of structural correlations. The network is a low-dimensional medium embedded in three-dimensional space, and allow to design light transport and the optical modes. Complex networks with finely-controlled parameters can be fabricated by self-assembly~\cite{gaio2019nanophotonic}, by standard lithography techniques or implemented on macroscopic systems~\cite{lepri2017complex}.


\subsection{Towards novel applications}

The sensitivity of the LDOS to the local environment discussed in Sec.~\ref{sec:6-LDOS-fluctuations} makes quantum emitters interesting optical probes of nanostructured materials~\cite{pelton2015modified}. Many studies have reported the dramatic impact of the local morphology of a complex medium on the spontaneous emission statistics from neighboring fluorescent molecules or quantum dots~\cite{krachmalnicoff2010fluctuations, sapienza2011long, birowosuto2010observation, de2016fluctuations, riboli2017tailoring, granchi2022near}. A key question to address will be whether optical measurements mediated by near-field probes could reveal statistical information on an unknown morphology, which could be extremely interesting for the remote monitoring of structural phases deep inside a 3D volume~\cite{de2016fluctuations}. This would require a deep understanding on the relation between subwavelength-scale correlations and near-field phenomena. In addition to LDOS measurements, measuring the cross spectral density of states (CDOS)~\cite{caze2013spatial}, which describes mode connectivity in structured media and could be obtained from coherence measurements on the light emitted from two classical or quantum dipole sources~\cite{canaguier2019cross}, may bring precious additional information. Spatially-resolved \textit{intensity} correlations in the optical regime have recently been measured in the bulk of a disordered medium using pairs of emitters separated by distances controlled via DNA strings~\cite{leonetti2021spatial}, unveiling already a very rich physics. First CDOS measurements have recently been realized in the microwave regime~\cite{rustomji2021complete}. 

The coherent control of light waves in disordered media is an important branch of research in multiple light scattering that has been powered in recent years by wavefront shaping techniques~\cite{rotter2017light}. We have seen in Sec.~\ref{sec:6} that structural correlations can result into strong spectral variations of scattering, transport and localization, suggesting that correlated disorder could yield higher degrees of spectral and spatial control, with applications in optical imaging. Disordered media are also being exploited as an unconventional platform for quantum walks and quantum state engineering~\cite{defienne2016two, leedumrongwatthanakun2020programmable}. These are very delicate experiments that require lossless materials, so far attempted in multimode fibers, but which could benefit in the future from correlated disordered media for multiplexing and spectral resolution.

The design of visual appearance is another aspect of research on correlated disorder that has considerably grown in importance in recent years. As beautifully illustrated by many diverse examples in the living world, the interplay or order and disorder has a direct impact on appearance at macroscopic scales, yielding increased transparency or whiteness, iridescent or non-iridescent colors [Sec.~\ref{sec:7-visual-appearance}]. The numerical modelling of \textit{realistic} correlated materials, considering for instance local imperfections~\cite{chung2012flexible} and large-scale random variations~\cite{chan2019visual} in certain ordered systems, will be an essential development of the field in the near future. Our perception of objects indeed relies on many attributes of visual appearance~\cite{hunter1987measurement} -- not only color, but also gloss, haze, translucency, texture, etc. --, which are affected by multiple scattering and are rarely considered in full. Understanding how optical properties created by structural correlations at the microscopic scale translate into visual effects at the macroscopic scale is a great and exciting challenge for the coming years, which could be enabled by merging concepts and techniques from coherent light scattering and computer graphics~\cite{musbach2013full, guo2021beyond, vynck2022visual}. Beyond appearance, correlated disordered media will play an important role in thermal management, for example for radiative cooling~\cite{wang2020dependent}, where broadband light control is needed from inexpensive self-assembled media.  Correlated disordered materials could be used to realize multifunctional materials, where optical (visual) properties could be combined with desired thermal, electrical, mechanical or tribological functionalities. Last but not least, efforts should be amplified to develop and promote low-carbon footprint, ecologically-responsible material synthesis, which can be best achieved in disordered assemblies as those discussed in this review.

\section*{Acknowledgments}

We are grateful to Eugene d'Eon (NVIDIA, New Zealand), Aristide Dogariu (CREOL, University of Central Florida, USA), Akhlesh Lakhtakia (Penn State University, USA) and Lorenzo Pattelli (INRiM, Italy) for fruitful discussions and precious feedbacks on our original manuscript.
We additionally thank M\'{e}lanie M. Bay (University of Cambridge, UK) for preparing Fig.~\ref{fig:classes_fabrication} and Johannes H. Haataja (University of Cambridge, UK) for preparing the cover image.
KV acknowledges funding from the french National Agency for Research (ANR) via the projects ``NanoMiX'' (ANR-16-CE30-0008) and ``Nano-Appearance'' (ANR-19-CE09-0014-01).
This work has received support under the program ``Investissements d'Avenir'' launched by the French Government.
This research was supported by the Swiss National Science Foundation through project numbers 169074, 188494 (FS) and 197146 (LSFP). RS and SV acknowledge funding by the Engineering and Physical Sciences Research Council (EPSRC). SV acknowledges the European Research Council (ERC-2014-STG H2020 639088).
 
\appendix


\section{Green functions in Fourier space} \label{A_Green}

\subsection{Dyadic Green tensor} \label{GFourier}

We consider a statistically homogeneous and translationally-invariant medium. The dyadic Green tensor in a uniform background medium with auxiliary permittivity $\epsilon_\text{b}$ is given by Eq.~\eqref{GFsing} and related to its Fourier transform as
\be
    \Gg_\text{b}(\r-\r') = \frac{1}{(2\pi)^3} \int \Gg_\text{b}(\k) e^{\im \k \cdot (\r-\r')} d\k.
\ee
The Green tensor in Fourier space is given by
\be
    \Gg_\text{b}(\k) &=& \left[ k^2 \Ig - \k \otimes \k - k_\text{b}^2 \Ig \right]^{-1} \\
    &=& \left[ -k_\text{b}^2 \frac{\k \otimes \k}{k^2}  +\left(k^2-k_\text{b}^2\right) \left(\Ig - \frac{\k \otimes \k}{k^2} \right)\right]^{-1} \\
    &=& \frac{1}{k_\text{b}^2} \left[  - \frac{\k \otimes \k}{k^2} + \frac{k_\text{b}^2}{k^2-(k_\text{b}+\im 0)^2} \left(\Ig - \frac{\k \otimes \k}{k^2} \right) \right]. \nonumber \\
\ee
The small imaginary part $\im 0$ introduced here is relevant in integrals involving $\Gg_\text{b}(\k)$. Using Eq.~\eqref{eq:th_distribution_relation}, the Green tensor can finally be rewritten as
\be
    \Gg_\text{b}(\k) &=& \frac{1}{k_\text{b}^2} \left[ - \frac{\k \otimes \k}{k^2} +\text{PV}\left\{ \frac{k_\text{b}^2}{k^2-k_\text{b}^2} \right\}
\left(\Ig - \frac{\k \otimes \k}{k^2} \right) \right] \nonumber \\
&& + \im \pi  \delta(k^2-k_\text{b}^2) 
\left(\Ig - \frac{\k \otimes \k}{k^2} \right). \label{ImG}
\ee

\subsection{Dressed Green tensor} \label{App:Dressed_Green}

Following Eq.~\eqref{Def_Lorentz_Green_1}, the Green tensor can be decomposed into local and non-local terms, the latter being also known as the Lorentz propagator. In Fourier space, we thus have
\be
    \bm{g}_\text{b} (\k) & = &  -\frac{1}{3 k_\text{b}^2} +  \int_{r < a}  \Gg_\text{b} (\r) e^{-\im \k \cdot \r} d\r, \\
    \tilde{\Gg}_\text{b}(\k) & \equiv & \Gg_\text{b}(\k) - \bm{g}_\text{b}(\k).
\ee
Specific expressions for arbitrary values of the radius $a$ are provided by~\citet{bedeaux1973critical}. For $k_\text{b}a \ll 1$, we have
\be
    \bm{g}_\text{b}(\r)= -\frac{1}{3k_\text{b}^2} \delta(\r-\r') \Ig,
\ee
which simply leads to
\be \label{eqS:Gb_tilde}
    \tilde{\Gg}_\text{b}(\k) & \equiv & \Gg_\text{b}(\k) + \frac{\Ig}{3 k_\text{b}^2}.
\ee

Let us now consider a medium composed of small volume elements within the Maxwell-Garnett approximation. Setting the auxiliary permittivity to that of the host medium, $\epsilon_\text{b} = \epsilon_\text{h}$, the average transition operator describing scattering by a single volume element is then given by

\be 
    \bra \tilde{\Tg} (\k) \ket = k_\text{h}^2  \rho \alpha_0 \Ig,
\ee
with $\alpha_0$ the polarizability. The Fourier transform of the propagator $\Hcal$ defined in Eq.~\eqref{Hcaldef} then reads
\be
    \hat{\Gg}_\text{h} (\k) &=& \left[ \Ig - k_\text{h}^2 \rho \alpha_0 \tilde{\Gg}_\text{h} (\k) \right]^{-1} \tilde{\Gg}_\text{h} (\k) \\
    &=&  \left[ \Ig -  \frac{\rho \alpha_0}{3} \Ig - \rho \alpha_0   k_\text{h}^2 \Gg_\text{h} (\k) \right]^{-1} \left(\Gg_\text{h} (\k) +\frac{\Ig}{3 k_\text{h}^2}\right). \nonumber \\
\ee
Taking into account the definition of the Maxwell-Garnett permittivity $\epsilon_\text{MG}$ in Eq.~\eqref{MG_epsilon},
the dressed propagator can eventually be written as
\be
    \hat{\Gg}_\text{h} (\k) &=& \left(\frac{\epsilon_\text{MG} +2\epsilon_\text{h}}{3 \epsilon_\text{h}}\right)^2 \left( 
\Gg_\text{MG}(\k) + \frac{\epsilon_\text{MG}} {\epsilon_\text{MG} +2 \epsilon_\text{h}}
\frac{\Ig}{k_\text{MG}^2} \right),  \nonumber \\ \label{Hk} \ee
where $\Gg_\text{MG}(\k)$ is the Green tensor with $k_\text{h}$ replaced by the Maxwell-Garnett wave number $k_\text{MG} = k_0 \sqrt{\epsilon_\text{MG}}$.

In absence of absorption, the imaginary part of $\hat{\Gg}_\text{h}$ is given by
\be \label{Im_Dressed_G}
    \text{Im} \hat{\Gg}_\text{h} (\k) & = & \pi  \left(\frac{\epsilon_\text{MG} + 2 \epsilon_\text{h}}{ 3 \epsilon_\text{h}}\right)^2   \delta(k^2-k_\text{MG}^2) \left(\Ig - \frac{\k \otimes \k}{k^2} \right),  \nonumber \\ 
\ee
leading to Eq.~\eqref{eq:Im-Gh}.

\section{Derivation of Eq.~\eqref{eq:th_bethe_salpeter_qk}} \label{App:BS-eq-Fourier}

In the Fourier domain and by making use of the average Green function
\begin{equation}
   \bra\bm{G}(\bm{k},\omega)\ket=\left[k^2\bm{P}(\hat{\bm{k}})-k_b^2\bm{1}-\bm{\Sigma}(\bm{k})\right]^{-1},
\end{equation}
we obtain
\begin{multline}\label{eq:th_bethe_salpeter_vector}
   \left\{\bm{1}\otimes\left[k'^2\bm{P}(\hat{\bm{k'}})-\bm{\Sigma}^*(\bm{k}')\right]
      -\left[k^2\bm{P}(\hat{\bm{k}})-\bm{\Sigma}(\bm{k})\right]\otimes\bm{1}\right\}
\\\cdot
      \bm{C}(\bm{k},\bm{k}')
   =\left[\bra\bm{G}(\bm{k})\ket\otimes\bm{1}-\bm{1}\otimes\bra\bm{G}^*(\bm{k}')\ket\right]
\\\cdot
      \int \bm{\Gamma}(\bm{k},\bm{\upkappa},\bm{k}',\bm{\upkappa}')
      \cdot\bm{C}(\bm{\upkappa},\bm{\upkappa}')
      \frac{\ud\bm{\upkappa}}{8\pi^3}\frac{\ud\bm{\upkappa}'}{8\pi^3}
\end{multline}
where we have neglected the source term $\bra\bm{E}\ket \otimes \bra\bm{E}^*\ket$ and used the tensorial relation
\begin{equation}
   (\bm{1}\otimes\bm{B}-\bm{A}\otimes\bm{1})^{-1}
      \cdot(\bm{A}^{-1}\otimes\bm{1}-\bm{1}\otimes\bm{B}^{-1})=\bm{A}^{-1}\otimes\bm{B}^{-1}.
\end{equation}
$\bm{1}$ is the identity tensor. Since we are dealing with dilute media and since the longitudinal part of the Green
tensor is irrelevent regarding light transport, we now consider the transverse approximation which consists in taking
the transverse part of all operators involved in
Eq.~\eqref{eq:th_bethe_salpeter_vector}~\cite{barabanenkov1995poyonting,cherroret2016induced}.
Using the definitions
\begin{align}\nonumber
   \bm{\Gamma}_{\perp}(\bm{k},\bm{\upkappa},\bm{k}',\bm{\upkappa}')
      & = \bm{P}(\bm{u})\otimes\bm{P}(\bm{u}')
          \cdot\bm{\Gamma}(\bm{k},\bm{\upkappa},\bm{k}',\bm{\upkappa}'),
\\\nonumber
   \bm{C}_{\perp}(\bm{k},\bm{k}')
      & = \bm{P}(\bm{u})\otimes\bm{P}(\bm{u}') \cdot\bm{C}(\bm{k},\bm{k}'),
\end{align}
we obtain
\begin{multline}
   \left[k'^2-k^2-\Sigma_{\perp}^*(\bm{k}')+\Sigma_{\perp}(\bm{k})\right]
      \bm{C}_{\perp}(\bm{k},\bm{k}')
\\
   =\left[\bra G_{\perp}(\bm{k})\ket-\bra G_{\perp}^*(\bm{k}')\ket\right]
\\\times
      \int \bm{\Gamma}_{\perp}(\bm{k},\bm{\upkappa},\bm{k}',\bm{\upkappa}')
      \cdot\bm{C}_{\perp}(\bm{\upkappa},\bm{\upkappa}')
      \frac{\ud\bm{\upkappa}}{8\pi^3}\frac{\ud\bm{\upkappa}'}{8\pi^3}.
\end{multline} 
Still considering that we have statistical homogeneity for the disordered medium, we have 
$\bm{\Gamma}(\bm{r}',\bm{r}'',\bm{\uprho}',\bm{\uprho}'')
=\bm{\Gamma}(\bm{r}'+\Delta\bm{r},\bm{r}''+\Delta\bm{r},\bm{\uprho}'+\Delta\bm{r},\bm{\uprho}''+\Delta\bm{r})$ which
leads to
\begin{multline}
   \bm{\Gamma}_{\perp}(\bm{k},\bm{\upkappa},\bm{k}',\bm{\upkappa}')
      =8\pi^3\delta(\bm{k}-\bm{k}'-\bm{\upkappa}+\bm{\upkappa}')
\\
      \times\bar{\bm{\Gamma}}_{\perp}(\bm{k},\bm{\upkappa},\bm{k}',\bm{\upkappa}')
\end{multline}
in the Fourier space. By a change of variable, we also now define the correlation
\begin{equation}
   \bm{L}_{\perp}(\bm{q},\bm{k})\equiv\bm{C}_{\perp}\left(\bm{k}+\frac{\bm{q}}{2},\bm{k}-\frac{\bm{q}}{2}\right)
\end{equation}
which leads to this new form of the Bethe-Salpeter equation in the Fourier domain
\begin{multline}\label{eq:th_bethe_salpeter_qk_app}
   \left[\left(\bm{k}-\frac{\bm{q}}{2}\right)^2-\left(\bm{k}+\frac{\bm{q}}{2}\right)^2
      -\Sigma_{\perp}^*\left(\bm{k}-\frac{\bm{q}}{2}\right)+\Sigma_{\perp}\left(\bm{k}+\frac{\bm{q}}{2}\right)\right]
\\\times
      \bm{L}_{\perp}(\bm{q},\bm{k})
   =\left[\bra G_{\perp}\left(\bm{k}+\frac{\bm{q}}{2}\right)\ket
         -\bra G_{\perp}^*\left(\bm{k}-\frac{\bm{q}}{2}\right)\ket\right]
\\\times
      \int
      \bar{\bm{\Gamma}}_{\perp}\left(\bm{k}+\frac{\bm{q}}{2},\bm{k}'+\frac{\bm{q}}{2},\bm{k}-\frac{\bm{q}}{2},\bm{k}'-\frac{\bm{q}}{2}\right)
      \cdot\bm{L}_{\perp}\left(\bm{k}',\bm{q}\right)
      \frac{\ud\bm{k}'}{(2\pi)^3}
\end{multline}
which is Eq.~\eqref{eq:th_bethe_salpeter_qk}.

\section{Configurational average for statistically homogeneous systems}  \label{App:ConfAv}
\subsection{Particle correlation functions} \label{subsec:part_corr_fn}

We consider a random ensemble of $N$ particles circumscribed in a volume $V$ and centered at positions $\Rg = \left[\Rg_1,\Rg_2,\dots \Rg_N \right]$. A specific configuration is described by a normalized probability distribution $P^{(N)}$ such that
\be
P^{(N)}(\Rg_1, \cdots ,\Rg_N) \dd \Rg_1 \cdots \dd \Rg_N
\ee
is the probability of finding a configuration in which particle $j$ is centered between $\Rg_j$ and $\Rg_j + \dd \Rg_j$.

Assuming that the particles are spherically symmetric (otherwise, the distribution whould also include orientational variables, $\Omega_j$) and identical, the distribution is symmetric in the labels $1,\dots,N$ and we can define the $M$-particle density $\rho^{(M)}(\Rg_1,\cdots,\Rg_M)$ as the probability of finding a configuration of $M$ particles whatever the configuration of the remaining $N-M$ particles
\be
    && \rho^{(M)} (\Rg_1,\cdots,\Rg_M) = \frac{N!}{(N-M)!}\nonumber \\
    & \times & \int P^{(N)}(\Rg_{1}, \cdots ,\Rg_N) \dd \Rg_{M+1}\cdots \dd \Rg_N  \nonumber \\
\ee

The instantaneous particle number density $\rho(\r)$ for a given configuration of particles, $\Rg = \left[\Rg_1, \cdots ,\Rg_N \right]$, is defined as
\be
    \rho(\r) & \equiv & \sum_{j=1}^N  \delta(\r-\R_j),
\ee
and its configurational average $\langle \rho(\r) \rangle$ is given by
\be
    \langle \rho(\r) \rangle &=& \bra \sum_{j} \delta(\r-\R_j) \ket \\
    &=& N \int \cdots \int P^{(N)}(\r, \Rg_2,\cdots ,\Rg_N) \dd \Rg_2 \cdots \dd \Rg_N. \nonumber \\
\ee
In the limit of infinite system size and assuming a statistically homogeneous and isotropic medium (i.e., all properties are statistically invariant by translation and rotation), both the number of particles and the volume of the material tend to infinity (i.e., $\{N,V\} \to \infty$), and one can define an average particle number density, $\rho \equiv \bra \rho(\r) \ket = N/V$, that is constant.

One can then define the $n$-particle probably density functions $\rho_n (\r_1,\r_2,...,\r_n)$ as
\be
    \rho_n(\r_1,\cdots, \r_n) \equiv \bra \sum_{\substack{j_1,\cdots j_s =1 \\ j_1\ne j_2 \cdots \ne j_s}}^N \delta(\r_1-\R_{j_1}) \cdots \delta(\r_s-\R_{j_s}) \ket, \nonumber \\ \label{g_s2}
\ee
as well as the $n$-particle correlation function
\be
    g_n(\r_1,\r_2,...,\r_n) \equiv \frac{\rho_n(\r_1,\r_2,...,\r_n)}{\rho^n}.
\ee
The important quantity in most systems is the pair correlation function $g_2(\r_1,\r_2)$, that describes the conditional probability of finding a particle at $\r_2$ given at particle fixed at $\r_1$. For isotropic media, $g_2$ only depends on the radial distance $r_{12}=|\r_1-\r_2|$. The total correlation function $h_2(\r)$ defined as
\be
    h_2(\r) \equiv g_2(\r)-1,
\ee
has the benefit of converging to zero at separation distances larger than a correlation length $\ell_\text{c}$.

\subsection{Fluctuations of the number of particles in a volume} \label{subsec:fluctNumber}
	   
The fluctuations of the number of particles $N$ in a given volume $v$ can be defined as
\be
    \delta_{N} &\equiv & \frac{1}{\rho v } \left( \bra  N^2 \ket - \bra  N \ket^2 \right) \nonumber \\
     &=& \frac{1}{v} \int_v \frac{\bra \Delta \rho(\r) \Delta \rho(\r') \ket}{\rho} d\r d\r',
\ee
with
\begin{multline} \label{def_S_r}
    \frac{\bra \Delta \rho(\r_1) \Delta \rho(\r_2) \ket}{\rho} = \frac{\bra \rho(\r_1) \rho(\r_2) \ket -\rho^2}{\rho}
\\
    = \frac{1}{\rho} \Bigg[ \bra \sum_{a=1}^N 
	 \delta(\r_1-\R_a)  \delta(\r_2-\R_a) \ket
\\
	+ \bra \sum_{a} \sum_{ b \ne a} \delta(\r_1-\R_a)  \delta(\r_2-\R_b) \ket  - \rho^2 \Bigg]
\\
    = \delta(\r_1-\r_2) + \rho \left(\frac{\rho_2(\r_1-\r_2)}{\rho^2} -1 \right)
\\
     \equiv \delta(\r_1-\r_2) + \rho h_2(\r_1-\r_2). 
\end{multline}
If $v$ is a sphere of radius $R_s$, one gets~\cite{van1977v,torquato2003local}
\be
    \delta_N & = & 1 \nonumber \\
    &+&\rho \int h_2(r_{12})  \left[ 1- \frac{3}{4} \frac{r}{R_s} + \frac{1}{16} \left( \frac{r}{R_s}\right)^3\right] 4\pi r^2 dr_{12} \nonumber \\
    & \sim & 1 +\rho \int h_2(r_{12}) 4\pi r^2 dr_{12},
\ee
where, in the last step, we assumed that $R_s$ is larger than the correlation length $\ell_\text{c}$ of $h_2(r)$. Expressions exist also for two-dimensional systems and non-spherical excluded volumes~\cite{torquato2003local}.

The static structure factor $S(\k)$ is related to the Fourier transform of $h_2(\r)$ via the expression
\be
    S(\k) &\equiv& 1 + \rho h_2(\k),
\ee
with $h_2(\k) = \int h_2(\r) e^{-i\k \cdot \r} d\r$.

\section{Local density of states and quasinormal modes} \label{App:QNM}

The local density of states (LDOS) is defined from the projected LDOS, Eq.~\eqref{eq:projected-LDOS}, as
\begin{equation}\label{eq:LDOS}
\rho_e(\mathbf{r},\omega) = \frac{2 \omega}{\pi c^2} \text{Im}\left[ \text{Tr} \mathbf{G}(\mathbf{r},\mathbf{r},\omega)\right].
\end{equation}
We will now express this quantity in terms of the eigenmodes of the system.

The eigenmodes of non-conservative (non-Hermitian) systems, known as quasinormal modes (QNMs), are described by complex frequencies $\tilde{\omega}_m=\omega_m - i \gamma_m/2$ and normalized fields $\tilde{\mathbf{E}}_m (\mathbf{r})$, the non-zero imaginary part steming from leakage. QNMs have a long history~\cite{baum1976singularity, ching1998quasinormal} and are receiving considerable attention from the photonics community since a few years~\cite{lalanne2018light}. Following a recent QNM formalism~\cite{sauvan2013theory, sauvan2014modal, yan2018rigorous}, we can write the field $\mathbf{E}$ generated by a dipole emitter in terms of QNMs as
\begin{equation}\label{eq:QNM-expansion-total-field}
\mathbf{E}(\mathbf{r},\omega) = \sum_m \alpha_m (\omega) \tilde{\mathbf{E}}_m (\mathbf{r}),
\end{equation}
with $\alpha_m$ the excitation coefficients, defined as
\begin{equation}\label{eq:QNM-excitation}
\alpha_m (\omega) = - \frac{\omega}{2 \epsilon_0 \left( \omega - \tilde{\omega}_m \right)} \mathbf{p} \cdot \tilde{\mathbf{E}}_m (\mathbf{r}).
\end{equation}
Quite expectedly, the efficiency of excitation of a mode depends on the amplitude of the QNM field at the dipole position and the spectral distance with the resonance frequency. Having further that $\mathbf{E}(\mathbf{r},\omega) = \mu_0 \omega^2 \mathbf{G}(\mathbf{r},\mathbf{r}',\omega) \mathbf{p}$, one arrives to a modal decomposition of the dyadic Green function
\begin{equation}\label{eq:QNM-expansion-Green-tensor}
\mathbf{G}(\mathbf{r},\mathbf{r}',\omega) = -\frac{c^2}{2 \omega} \sum_m \frac{\tilde{\mathbf{E}}_m (\mathbf{r}) \otimes \tilde{\mathbf{E}}_m (\mathbf{r}')}{\omega - \tilde{\omega}_m}.
\end{equation}
Note that Eq.~(\ref{eq:QNM-expansion-Green-tensor}) can also be obtained from the Mittag-Leffler theorem which introduces the residues of the Green tensor at the QNM complex frequencies~\cite{muljarov2016exact}. Inserting Eq.~(\ref{eq:QNM-expansion-Green-tensor}) in Eq.~(\ref{eq:LDOS}) finally leads to
\begin{equation}\label{eq:QNM-expansion-LDOS}
\rho_e(\mathbf{r},\omega) = -\frac{1}{\pi} \text{Im} \left[ \sum_m \frac{\text{Tr} \left[ \tilde{\mathbf{E}}_m(\mathbf{r}) \otimes \tilde{\mathbf{E}}_m(\mathbf{r}) \right]}{\omega - \tilde{\omega}_m} \right].
\end{equation}
The LDOS is now explicitly expressed as a sum over resonant modes. To further convince ourselves, we can take the limit of vanishing leakage, in which case both $\mathbf{E}_m$ and $\tilde{\omega}_m$ tend to become real. Using $\lim_{\eta \rightarrow 0^+} \frac{1}{x+i\eta} = \text{PV} \left[\frac{1}{x} \right] -i \pi \delta(x)$, we arrive to the well-known expression of the LDOS for conservative systems~\cite{novotny2012principles, carminati2015electromagnetic}
\begin{equation}
\rho_e(\mathbf{r},\omega) = \sum_m |\tilde{\mathbf{E}}_m(\mathbf{r})|^2 \delta(\omega-\omega_m).
\end{equation}

\end{document}